# Preparing Fermilab to Carry Out the P5 Plan

*An independent review of Fermi Research Alliance (FRA)*

by

A Group of Whistleblowers from Fermilab

July 15, 2024

1st Revision July 30, 2024

Urgent Notification November 27, 2024

# Preface to Urgent Notification – How DOE inaction led Fermilab to go rogue, from a hostile environment to criminal behavior

We were elated that our ArXiv publication produced the **news impact** that it did. On August 14, 2024, an article appeared on Science's website, followed the next day by one on PhysicsWorld, and one on WTTF NEWS on August 16. The Wikipedia article at https://en.wikipedia.org/wiki/Fermilab was updated with a section on the "2024 Whistleblower Report", and a video about it by Sabine Hossenfelder appeared in her YouTube series. The associated web links are below:

https://www.science.org/content/article/trouble-and-strife-deepen-famed-u-s-particle-physics-lab;
https://physicsworld.com/a/fermilab-is-doomed-without-management-overhaul-claims-whistleblower-report/;
https://news.wttw.com/2024/08/16/fermilab-whistleblower-report-alleges-sexual-harassment-loaded-gun-and-chaotic-finances;
https://www.youtube.com/watch?v=JxTCeyG3OvM.

Several reddit threads were also started as shown here below:
https://www.reddit.com/r/Physics/comments/1eymw88/trouble_and_strife_deepen_at_famed_us_particle/;
https://www.reddit.com/r/Physics/comments/1ebv37t/preparing_fermilab_to_carry_out_the_p5_plan_an/;
https://www.reddit.com/r/ParticlePhysics/comments/1ehdr1b/preparing_fermilab_to_carry_out_the_p5_plan/.

On the other hand, the **response of Fermilab's director** has been at the very least underwhelming, such as: "*The document asserts various challenges at Fermilab, some of which are inaccurate, and others of which Fermi Research Alliance (FRA) has been working hard to address for some time. It is our intent to deliver on DOE's mission in the most effective manner and maintain an environment that is worthy of the Fermilab community that FRA serves.*" In reference to the negative media coverage, the response was even more disappointing: "*I am writing to you today in light of the recent negative media coverage of the lab and to express the leadership team's commitment to doing everything we can to support and uplift the wonderful work you all do every day that pushes the frontiers of science. […] The last few weeks have seen us bear the brunt of a lot of criticism. We take these issues seriously, are working to address many of the issues, and invite you to provide input about ways we can continue to improve.*"

In fact, while these hollow statements were being made, the opposite reality was happening in the organization, in that since the publication of this paper, we have received dozens more of horrific testimonials by both employees and users. Some **examples of micromanagement and abuse of authority under FRA** are the following.
1. An esteemed female colleague of ours was further discriminated and retaliated against after being denied once again career advancements and promotions in favor of less qualified (both younger and older) white men! Her story will be found in Appendix I of our next edition.

2. Here is what regularly happened to employees working in what is currently called the Infrastructure Services Division (ISD):
   - Some were required to repeatedly cover rotating shifts for other favorite employees. The latter were managers' buddies, who didn't do shifts because "they couldn't climb a ladder". At the same time, they would spend their free time hunting, fishing and riding motorcycles. The overexploited employees would instead develop apnea and several other body harms as a consequence.
   - Even when injured, the lab would typically refuse to approve employees' claims of workers compensation insurance, and injured employees needed to find a lawyer to fight for their rights. In the meantime, they were forced by the lab to be at work, while injured!
   - During these events, supervisors would get their hands on their employees' private medical records from the lab's Doctor Office. They would sometimes fake their identity, or hire investigators presenting themselves as the patient, to retrieve medical documents from the employees' actual doctor offices!
   - Several employees reported to us that FNAL managers would personally drive to their houses for video surveillance expeditions. More than once, these managers were so engrossed into stalking injured employees at their homes that they were about to run over other people on or around their premises.
   - Lab buddies would bring their rifles on FNAL site at night and shoot at the High Voltage boxes of the Main Injector for fun. Employees who complained about this were intimidated and threatened by their management.

3. Bonnie Fleming personally directed Fermilab security guards to inspect the house of a foreign postdoc who was visiting and living in the village. Afterwards, the guards aggressively questioned him late at work, with their phone surreptitiously connected with Fleming's. The next day, guards showed up at his home from the back door and forced him to Fleming's office. There she expelled him for 6 months from the lab, on the basis of allegedly confiscated drugs paraphernalia from his room, without specifying or showing any of the items in question. When the postdoc denied owning any illegal drugs, and tried to stand up for his rights by refusing to sign the letter where she wanted him to acknowledge ownership of the alleged items, the lab called the cops accusing him of "federal trespassing". Half a dozen cops from the Kane County Sheriff Office <u>immediately</u> showed up, towed his car, handcuffed and dragged him to jail (the Kane County Judicial Center), where he spent the night. Note that this is the same Office whose Sheriff showed up at 3:45 pm when in 2022 the two employees from Appendix H reported a gun on site at 11:15 am. Also, nobody at the police station seemed to question the serious accusation of trespassing, which can make one a felon. And note that the security team that had no qualms encroaching in the foreign postdoc's private quarters is from the same Security Division team that refused to inspect the vehicle of a potential murderer, as previously described in our Appendix H.

4. Lia Merminga personally intimidates employees on the phone, bypassing the whole org chart, when she wants something from them that is not aligned with their managers.

**By October 1, 2024, the DOE reinstated FRA, albeit under a different name, i.e. Fermi Forward Discovery Group, LLC, as core of the Management & Operating (M&O) contractor of Fermilab.** It is of note that in the week of May 13, 2024, the University of Chicago had a celebration party in Chicago with their partners. This party was attended by several FNAL senior leaders and other FNAL managers. By mid-July 2024, the US HEP community too was buzzing about FRA winning the M&O bid together with a couple of for-profit companies.

<u>In the meantime, because of a lab's budget hole of more than $20M caused by inconsiderate hiring by Merminga et al.,</u> a first buyout, called "Voluntary Separation Program", was enacted at Fermilab by end of May 2024. This was followed by a furlough, called Limited Operations and Maintenance Outage or LOMO, in the week of August. 26, 2024. On September 12, 2024, this is how Merminga addressed the staff: "*Let me start by saying **thank you** to all the staff that planned and executed the LOMO the week of Aug 26. The LOMO was a success as we safely completed all the high impact maintenance we had planned for the week. Let me also say **thank you** to all our employees for working together this year to help align our laboratory spending with our funding. With your help, and the support of DOE and FRA, we have achieved our budget goals and will close out fiscal year 2024 within budget.*" Then she proceeded to explain how the annual merit increase would have to be postponed from October to January. <u>In short, the lab has fallen so low that now furloughs and talent leaving the lab are both considered a success</u>. Consistently, a second buyout was announced by Merminga on September 17, 2024.

At around the same time in September, in preparation to the infamous October announcement of a failed organization winning the lab bid because favored by the DOE, Paul Alivisatos, President of the University of Chicago and Merminga's countryman, published his own announcement here
[https://president.uchicago.edu/from-the-president/announcements/240409-regarding-the-office-of-sing](https://president.uchicago.edu/from-the-president/announcements/240409-regarding-the-office-of-sing).
"*<u>John (Jack) Anderson, Jr. is joining the University to provide interim leadership to the teams that steward the University's relationships with the National Labs, effective September 16.</u> Jack is a senior advisor at Brookhaven National Laboratory, where he previously served as deputy laboratory director for operations and as chief operating officer. Prior to his term at Brookhaven, he acted as chief operating officer for Fermi Research Alliance, LLC for a year, so he brings to this role a strong understanding of our regional labs. […]*"

On September 15 we sent the following to Geraldine Richmond, once again without any acknowledgement of receipt or reply:
"*Dear Madame Undersecretary for Science and Innovation:
We understand that the only reaction by DOE to our investigative report is that of coaching the University of Chicago's (UC) "teams that steward the university's* *relationships with the National Labs, effective September 16". By doing so, the DOE is once again showing that it is for the status-quo, however corrupt it is. We are wondering what kind of leadership will be provided to UC to make it change for the better. When there is corruption, one needs to clean the house and reset the culture completely*."

It did not take long for the lab's employees to see a major example of leadership coaching by UC and Fermi Forward Discovery Group. On November 12, 2024, the lab personnel received the following memo by Merminga: "*Our laboratory is undergoing major changes. We are executing the largest project in the DOE Office of Science, and we will soon operate one of the largest experiments in High Energy Physics. […] To realize these objectives, we need to make difficult decisions. After careful deliberation, <u>**lab leadership has decided that approximately 50 existing positions will need to be eliminated**</u> to align our workforce with the lab's priorities and budget. These reductions will occur throughout the laboratory and are independent of the ongoing contract transition. <u>Please be assured that we have assessed each area deeply and diligently and have followed a rigorous process that focused on positions, not individuals</u>. All employees impacted by this workforce reduction will be notified today*."

We have the whole list of the 55 individuals that were eliminated and are analyzing the statistical data as we speak. **A majority of this group has received no specific reasons as for why their work was not aligned with the lab mission**. At some point during the day of November 12, they were called to a meeting with their management and a HR representative. They were read the same statement and informed of the timeline to sign a General Release and Waiver as condition to receive their small several packages. They were told that they would receive pay for two-weeks in lieu of notice. Their questions on the criteria for their selection were not answered. Their badge was immediately removed, as well as any computing access. This prevented them to access their emails from that point on. They lost all of their business contacts and their calendar. They lost their health insurance. The cost for new coverage for those ineligible for Medicare is not affordable without a salary, and some of the eligible ones had surgeries planned that are not always covered by Medicare. Those that were dismissed on site were escorted out of their buildings and to their cars by the Kane County Sheriff. **We received reports of employees who were watching that are still traumatized by seeing their colleagues dragged out like criminals**. <u>This group was clearly treated as outcasts, and certainly completely differently than those who "voluntarily" separated, who received farewell parties and many of them are back on site</u>. **Instead, those that were laid off and had other academic affiliations are being refused their applications as Fermilab Users**.

Despite all of the above, this is how Bonnie Fleming described this process in one of her slides, together with Glenzinski and Unruh, on November 15, 2024 at a Monthly Spokepersons Meeting: "<u>*This was not a decision made lightly. Every effort was made to do this in an objective, fair and respectful manner.*</u>"

What makes the above even worse is that most of the people who were dismissed were competent hard-working professionals. In our ArXiv paper, we recommended, as part of the solutions to improve the lab cost-effectiveness, to eliminate the ballast in management, where also the highest salaries are found. Instead, great care was paid in this lay-off not to have a single manager in this list. Even the few people with manager sounding titles had no direct reports.

Several of the dismissed employees are superiorly qualified, and extremely active in their field. There are also twelve Scientists in this list. Some of them brought millions of dollars to the lab and/or are Principal Investigators of important grants. **How can the work of employees with HEP grants not be aligned with FNAL mission?** Did Merminga even realize that when, once again, she sent out her thanks and claims of success to the rest of the staff, she was congratulating herself for a massacre of talent? **In short, this administration stayed true to their toxic and unethical behavior by doing the opposite of what would benefit the lab, i.e. keeping high salary ballast managers and firing talented, extremely qualified, hardworking professionals.**

Fortunately for the victims of this "position elimination", it is not correct that the lab has followed a "rigorous process". **Lab management made at least two critical unpardonable errors: while being ignorant themselves on their own lab's policies, they also relied on the same incompetent lawyers that we denounced in our original ArXiv paper. Scientist and Senior Scientist positions are tenured positions and there is a very specific process to follow when terminating them.** The process is especially detailed in the case of terminations for a reduction in force, or RIF, as opposed to terminations due to performance. First, the DOE has to agree to the initiation of a RIF by the lab. Then a detailed process starts, with an analysis of the reasons for the selection of a specific Scientist, based on job function, performance, knowledge, skills, and abilities needed for the future mission of the Laboratory. The proposal is then reviewed and has to be approved by several entities and committees, one after the other. There are very few cases, if any, of scientist being dismissed as RIF, and this is because the scientists inherently fill the lab mission, and are adaptable to many different areas in physics. In the rare instances that a scientist is dismissed, the length of time between notice of intent and termination is normally 180 calendar days. **This process was established for the specific reason of preventing a corrupt buddy system to act against the lab's, and therefore the tax payers', best interests!**

Based on the above and our preliminary analysis that already shows concerning illegal trends, we predict a non-negligible number of lawsuits. The recommendations that we had provided in our ArXiv publication to reduce the risk of lawsuits were certainly not followed.

**Stay tuned.** In our next edition we are going to show how a buddy system inevitably converges to mediocrity in science. We have also been investigating several different types of collusion between the lab and other entities, and will present our results in a future Appendix J.

# TABLE OF CONTENTS

## Forewords

I am happy to introduce to our scientific community this white paper. After so many years at Fermilab I have developed a deep sentimental involvement with the laboratory, and I sense a diffused lack of confidence in our future.

The data of the past 15 years show that responding to demands for a change by delaying any incisive action is not productive. It is not leading to a rousing vision for the future of HEP in the US. We must protect the well-being of our students and employees for them to thrive, and for the lab to be able to meet the P5 plans. A move in this direction is advocated by the proponents of this welcomed document.

A transparent and honest approach to the change will substantially improve a productive dialogue with DOE as our funding agency, and strengthen a much-needed mutual trust. Steady, consistent support is important to the success of large, long-term initiatives in particle physics. Trust and mutual appreciation between the lab and DOE are substantial to navigate together the challenging US government's budgeting process, which sometimes results in unforeseen changes in the level of support for projects. Historically, one extreme example of that problem was the abrupt cancellation in 1993 of the Superconducting Super Collider, the SSC, after large investments in that project had been made over several years. By default, that decision ceded leadership at the energy frontier to the Large Hadron Collider of CERN. We do not want that to happen with DUNE. There is the chance of starting a long-term program of research and development of advanced muon facilities at Fermilab, possibly leading to a 10 TeV Muon Collider in the US. The time to act is now.

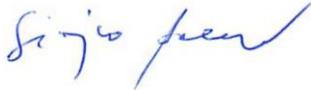

Giorgio Bellettini

I join with Giorgio Bellettini in recommending deep introspection concerning the issues raised by the authors of this document. Like Giorgio, I have had a long and productive association with Fermilab, during which period I have also witnessed a gradual erosion of trust and confidence within the Laboratory staff. In achieving the vision projected by P5, the high-energy physics community in the United States faces both profound challenges, but also great opportunities. Their successful realization demands dedicated, highly-motivated scientific, technical, and administrative professionals unburdened by past deficiencies. To repeat Giorgio's exhortation, "the time to act is now."

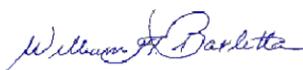

William Barletta

# Preface

Fermi National Accelerator Laboratory (Fermilab) is critical to understanding many aspects of our universe and is famous for producing discoveries and hosting award-winning scientists. Unfortunately, in the past several years it has suffered a number of troublesome setbacks. In some cases, failures to respond effectively have made things worse, resulting in the current crisis atmosphere at the laboratory. This is reflected only in part in the poorly rated performance appraisals that the DOE (Performance Evaluation and Measurement Plan, or PEMP) provides to the lab every year. In this White Paper we review some of the issues, evidence for what might be causing them, and recommend corrective actions. <u>With testimonials, survey results, and reviews available, we hope that this document will provide a motivation for the DOE to authorize a comprehensive review of Fermilab to ensure that it fulfills its potential</u>.

Three entities are currently responsible for supporting, managing, operating, and leading the lab: the Department of Energy (DOE), Fermi Research Alliance (FRA), and the Laboratory Director's Office.

Two recent events provide opportunities for Fermilab to excel: the release in December 2023 of the P5 Report and the DOE's decision to rebid the Management & Operating (M&O) contract. ([https://science.osti.gov/Acquisition-Management/M-and-O-Competitions/Procurement-Information/Request-for-Proposals?utm_medium=email&utm_source=govdelivery](https://science.osti.gov/Acquisition-Management/M-and-O-Competitions/Procurement-Information/Request-for-Proposals?utm_medium=email&utm_source=govdelivery))

The P5 Report ([https://www.usparticlephysics.org/2023-p5-report/](https://www.usparticlephysics.org/2023-p5-report/)), a decadal update of recommendations for US particle physics, includes an ambitious and positive roadmap for projects and activities at Fermilab. Carrying out the P5 plan successfully, however, will require addressing many of the problems highlighted in this White Paper.

<u>The selection of a new M&O contractor is under the responsibility of the Deputy Director for Operations of the DOE Office of Science Dr. Juston Fontaine.</u> A new M&O Contractor would replace FRA and name a new Director effective January 1, 2025. A new management team has the opportunity to solve existing problems and implement the necessary culture and policy changes that will allow all Fermilab employees and users to thrive.

Stimulated by these recent developments, the authors of this document, all present or past employees, convened confidentially in order to collect testimonials on a series of events symptomatic of the lab problems, provide critical analyses and offer possible solutions. The paper is extensive because we tried to document issues, as objectively as possible, in several areas of the lab, drawing on staff with direct experience of each area. <u>For those with little time, we provide here below an Executive Summary, and recommend reading at least appendices E and H.</u> We did our best to remain factual, based on gathered evidence for our statements. <u>We wrote this document as private citizens and not as representatives of Fermilab or FRA</u>.

## Executive Summary

1. We started our think tank to help Fermilab address its Physics and mission impasses. When surveying employees outside of our group, we found evidence (see below) that <u>a large fraction of the staff at the lab are unhappy with management and want change</u>. Regardless of the source of this employee dissatisfaction, the fact of it alone requires attention from leaders. The climate at the lab, as assessed in the latest third-party survey, has worsened further since the National Business Research Institute evaluated it at the 61% percentile in a 2019 survey. The outcome of the latest climate survey points to poor communication between management and employees and a decline of trust in management. <u>Ironically, we cannot provide a reference or actual results of this latest survey because FNAL management has decided to keep this too a secret</u>!

2. While studying the various aspects of the lab difficulties, we uncovered evidence of problems with culture and behavior. Testimonials from employees and visitors often describe a hostile work environment, where constructive criticism is most often ignored and retaliated against. Examples of reports stemming from interactions with the outdated HR/General Counsel system, under FRA purview, are as follows:
    - The alleged sexual assault of the NOvA and DUNE collaborator who filed and won her 2022 lawsuit in the UK was dismissed by FNAL staff; this blunder led to a young woman missing years of her life in solving the problem herself, and she eventually left the field after her postdoc.
    - A cover-up of a case of guns on site in 2023, with promotion of the perpetrator and pretextual firing of the witness.
    - A cover-up of an attempt by an obsessed male employee at badly hurting a female electrician with an industrial vehicle.
    - A cover-up of Beryllium windows blasting, with subsequent promotion of the person in charge.
    - The previous case led to a successful EEOC lawsuit for retaliation against the employee who had predicted and warned management of the risk of those windows failing.

    Former and current employees and visitors willing to come forward provided detailed testimonials in Appendices E and H. A most telling example of the culture of cover-ups is in the latter. <u>FRA's preoccupation with controlling the narrative with DOE at all costs has led to more than a decade of unethical conduct and sometimes conduct of questionable legality</u>. After 15+ years of such data, the perception is that opportunities to acknowledge errors and implement effective policy changes to improve the Fermilab experience have been neglected. FRA's

management has responded to demands for change by offering just the appearance of change.

3. In replacing the M&O contractor, 98% of the staff must remain the same. However, it is the top 2% in the leadership that makes or breaks an organization. A successful organization requires leaders who lead, inspire and teach by example. They have the courage to accept their mistakes, learn from them, and implement necessary changes. In the last decade, the average seniority, i.e. knowledge, experience and maturity, as well as scientific reputation of Fermilab's managers shrank to a fault. We think that hiring, evaluation, and promotion methods at FNAL need external review to make sure that the latest research and best practices from business and social sciences are utilized in the selection and support of FNAL leaders at all levels to eliminate micromanagement and maximize effectiveness. Previously, internal committees and/or task forces composed of mature and prominent scientists had a say in how operations were conducted at the lab. In contrast, because of the large percentage of early career members without sufficient experience, knowledge and insight in administrative positions of responsibility and in critical scientific and technical committees, sufficient checks and balances have evaporated.

4. Consistently with all of the above, what could have become a rousing vision for the future of HEP in the US has suffered, and has not been motivating enough for hundreds of potential PhD students who have left our field for more inspiring endeavors. Similarly, many lab's scientists, engineers and users have already opted to transfer their expertise to activities at other facilities. The consequent loss of skills and experience has already been destructive for the lab and will jeopardize the lab's capabilities to answer to the P5 report.

5. FRA's lack of state-of-the-art processes clearly extends to business, finances and procurement. Several employees involved in these endeavors invoke the lack of modern and effective processes. In these past 15+ years, FRA has extensively demonstrated not to have the knowledge and capability to implement effective procedures for business. Such management ineffectiveness has progressively caused, among others, the following self-inflicted problems:
    - Disruptively long vacancies in important leadership positions, both scientific and administrative.
    - Serious budget overruns and delays in LBNF/DUNE, which required a revision of the CD1 cost point estimate and range.
    - PIP-II contingency overruns within one year from CD3 approval.
    - The perception of lack of knowledge and effective administrators in business/ finance/ procurement.
    - Administrative obstacles both in safety and general business that slow down and sometimes stop scientific productivity of the experiments.
    - Lack of vision/focus in technological areas, and too little to no oversight.
    - Lack of focus on ensuring the quality of life of employees.

6. A consequence from point 5. is budget insolvency, with the lab continually threatened with hiring freezes, furloughs, or layoffs to support its existing staff. A major financial burden on salaries are benefits and indirect costs (overhead). The actual cost of employees to the lab is approximately 3 times their salary. <u>The overhead is correlated to the ratio of the cost of administrative employees to that of non-administrative ones.</u> In any mature organization, it can be difficult to ensure that every hire and every decision is to further the founding mission. It takes sustained commitment, attention, action, assessment, and adjustment. FNAL needs an enforceable pledge that every hire and personnel decision advance the scientific mission of the lab. In particular, the number of bureaucrats and administrators on the lab payroll has to be monitored and closely controlled. Implementation of automated processes and solutions in projects' operations is also necessary to help reduce the number of redundant administrators. This is critical for the lab not to lose itself in heavy bureaucracy and forget its Physics vision!

7. A laboratory's M&O must prioritize that lab's interests. As the University of Chicago (UC) is a major scientific research institution in its own right, conflicts of interest can result from the fact that <u>a major partner in Fermilab's M&O manager is also its potential competitor. This potential conflict of interest is exacerbated by the UC being also the M&O contractor of ANL, where several Fermilab's competing groups collaborate directly with the UC.</u> Section L.12 ORGANIZATIONAL CONFLICT OF INTEREST (OCI) of the RFP specifies that "A contract shall not be awarded to any Offeror having an unresolved OCI" (Appendix C, pg. 76). Additionally, it is our opinion that <u>voting power in any board where members have a potential OCI due to the research activity carried on at their institution should be carefully evaluated</u>.

While comparing Fermilab relative to the other DOE labs, we observed that there is only another lab, Thomas Jefferson National Accelerator Facility (JLab), whose grades dropped a few years ago and currently remain below the acceptable threshold. Consistently with DOE regulations, the DOE is rebidding JLab's M&O contract too. Testimonials concerning events occurring at FNAL and JLab, not just to employees but also to users, <u>show a correlation between a hostile environment at a lab and a poor PEMP rating at the same lab</u>. This correlation echoes historical precedents in Western societies ["Rules for Whistleblowers", book by Stephen Martin Kohn, ESQ] that show that treating all employees well is actually good for business. <u>Ignoring and/or covering up unethical and sometimes illegal behavior is NOT good business</u>.

In June 2024, most of the above information was conveyed in separate letters to the Deputy Director for Operations of the DOE Office of Science Dr. Juston Fontaine; the DOE Under Secretary for Science and Innovation Geraldine Richmond; and Congressman Bill Foster.

# I. Introduction

Fermi National Accelerator Laboratory is a national treasure and a valuable international resource. At Fermilab, as at other labs, the Office of Science of the Department of Energy provides resources and facilities for particle physics that are too large to locate at individual universities. Founded in 1967 as the National Accelerator Laboratory, Fermilab reached the energy frontier of particle physics via its 200 GeV Main Ring Accelerator. Following the vision of its founding Director, Robert R. Wilson, it achieved twice the Main Ring energy with its Tevatron superconducting synchrotron. After Europe's CERN laboratory established a new record center-of-mass energy with a proton-antiproton collider based on its SPS, Fermilab regained the energy frontier by transforming the Tevatron into a similar proton-antiproton collider. The collision energy at the center of mass reached 1.96 TeV in 2001. The Tevatron ceased operations on September 30, 2011, after ceding leadership at the energy frontier in 2008 to CERN's Large Hadron Collider. The lab collaborates actively with CERN on particle physics experiments and in developing accelerator technology critical to exploring the energy frontier.

Because the statistical significance of particle physics experiments depends upon the number of events detected, Fermilab aspires to lead the particle physics field at the intensity frontier. Its flagship project, in that regard, is the Deep Underground Neutrino Experiment at the Long-Baseline Neutrino Facility, or LBNF/DUNE.

In addition, Fermilab performs world-class research in astrophysics, quantum physics, superconducting technology development, and related instrumentation. In each of these endeavors, collaboration with U.S. and international laboratories and universities is key to the success of Fermilab's research activities.

Fermilab is managed by a Management & Operating contractor (M&O), currently the Fermi Research Alliance, LLC (FRA), which is a partnership between the University of Chicago and Universities Research Association (URA). Fermilab Director is the President of FRA. FRA's performance-based M&O contract requires accomplishing specified DOE missions and programs in an efficient, safe and secure environment, while effectively managing all business operations. <u>The M&O contractor has full responsibility and accountability</u> for total performance under the contract, including determining the specific methods for accomplishing the work effort, performing quality control, and assuring that all operations are consistent with DOE, EPA, OSHA, and EEOC regulations and guidelines. DOE is responsible for the oversight of all activities conducted under the contract and for assuring that funds are properly and effectively utilized.

In the last 5-6 fiscal years, FRA management of Fermilab has received poor performance reviews (see Appendix B) and has been responsible for large cost overruns in LBNF/DUNE and its associated accelerator project PIP-II.

Due to several safety accidents, accelerator operations have been unexpectedly closed for several months.

With respect to the management of its workforce, FRA has acquired a reputation for developing an unsupportive work environment among many of its present and former employees. We have gathered testimonials to the fact that challenges have often been ignored and negative consequences covered up. Employees and visitors who submit constructive criticism are too often ignored, or in some cases, retaliated against.

The healing and improvement process MUST begin soon if Fermilab is ever to return to its former glory and be the lab its employees are proud to work for.

## I.A  Looking to the Future

The DOE has recently published the report of the Particle Physics Project Prioritization Panel (P5), a subpanel of the High Energy Physics Advisory Panel (HEPAP). This report constitutes a ten-year strategic plan for particle physics in the U.S. (https://www.usparticlephysics.org/2023-p5-report/) It offers an aggressive, optimistic vision for the future of Fermilab, including LBNF/DUNE at the intensity frontier and a "Muon Shot" as a possible path back to the energy frontier. The following quotes exemplify P5's expansive plans for future development of those major accelerator facilities.

"DUNE is the centerpiece of a decades-long program to reveal the mysteries of elusive neutrinos. This US-hosted international project will exploit a unique underground laboratory (now nearing completion) and neutrino beams produced at Fermi National Accelerator Laboratory (Fermilab). The Rubin Observatory Legacy Survey of Space and Time (LSST) anchors an ongoing program of cosmic surveys. Realizing the full scientific potential of these and other ongoing projects is our highest priority." The completion of DUNE on the projected time scale without further cost growth is an essential DOE Goal for the Office of High Energy Physics.

"The panel recommends dedicated R&D to explore a suite of promising future projects. One of the most ambitious is a future collider concept: a 10 TeV parton center-of-momentum (pCM) collider […]. As part of this initiative, we recommend targeted collider R&D to establish the feasibility of a 10 TeV pCM muon collider. A key milestone on this path is to design a muon collider demonstrator facility. If favorably reviewed by the collider panel, such a facility would open the door to building facilities at Fermilab that test muon collider design elements while producing exceptionally bright muon and neutrino beams. By taking up this challenge, the US blazes a trail toward a new future by advancing critical R&D that can benefit multiple science drivers and ultimately bring an unparalleled global facility to US soil." (Ref: p xi P5)

"Although we do not know if a muon collider is ultimately feasible, the road toward it leads from current Fermilab strengths and capabilities to a series of proton beam improvements and neutrino beam facilities, each producing world-class science while performing critical R&D towards a muon collider. At the end of the path is an unparalleled global facility on US soil. This is our Muon Shot." (Ref p23 P5)

Those quotes are only a part of the comprehensive vision for Fermilab outlined in the report, as the following quote demonstrates: "DOE National Laboratories are critical research infrastructure that must be maintained and enhanced based on the needs of the particle physics community. This is particularly true for Fermilab as the only dedicated US laboratory for particle physics." (Ref: p23 P5)

The P5 Report also advocates a balanced portfolio of small, medium, and large experiments. This recommendation takes on added importance for Fermilab because the largest experiment, DUNE, is not located near the lab campus.

More than a decade ago, the DOE decided to emphasize projects over general R&D. This changed the lab's fund distribution and increased the administrative efforts needed to manage larger projects. In somewhat of a reversal, Chapter 6 of the current P5 Report, entitled "Investing in the Future of Science and Technology", emphasizes that besides projects, substantial support for R&D initiatives is essential to the health of the field.

These ambitious goals would be challenging to achieve even under ideal conditions. In reality, however, the present state of the lab is far from ideal. Articles in newspapers and scientific journals have described some of the problems. (https://www.science.org/content/article/major-shake-coming-fermilab-troubled-u-s-particle-physics-center) Recent performance appraisals of the lab by the DOE (Performance Evaluation and Measurement Plan, or PEMP) have also found significant shortcomings in several areas of the lab (See Appendix B). FNAL needs to improve to achieve the goals proposed in the P5 plan.

Due to the lab's poor performance, the DOE has initiated a Request for Proposal (RFP) for bids to manage and operate (M&O) the lab starting on January 1, 2025. In response, several organizations have filed public Expressions of Interest to be Fermilab's M&O contractor. The deadline for bidding was March 4, 2024.

The P5 Report and DOE's rebidding of the M&O contract create a great opportunity to rethink how things are currently being done at Fermilab. In response, the authors of this white paper collected information about issues relevant to achieving the vision of P5 and, where possible, to mitigate potential impediments to progress toward those goals. The information and advice presented here is meant to be a call to action for those individuals and institutions that will be responsible for overseeing the success of Fermilab.

Though there are many challenges ahead, there is also plenty of discovery science happening now and planned for the future at the lab. We therefore remain optimistic for the lab's potential with the right vision, experienced management team, and responsiveness to the scientific community. The areas where instead the lab does well already can be seen in Appendix B, which summarizes in some detail the PEMP results from FY 2019 to FY 2023. However, <u>the purpose of this document is to start the conversation, to suggest possible causes of the problems at the lab, and to propose recommendations on how to do better.</u> The next step is for the DOE to organize a thorough external review of Fermilab to ensure that it reached its highest of excellence.

In the next section we present observations about problems and applicable solutions.

## II. Observations

"This was a tough year", said Lia Merminga, Fermilab Director, addressing Fermilab employees and users at an All-Hands Meeting on December 18, 2023, thereby acknowledging what most people in her audience already knew. At the same meeting, a member of lab management presented some results of a recent climate survey that confirmed anecdotal reports of widespread low morale at Fermilab. Morale problems are to be expected when troublesome issues proliferate. This document attempts to explore the underlying root causes, identifies lessons learned, and recommends methods for improvement.

Tracing the root cause and its time of origin of particular problems over the past dozen years is not always possible. But we believe that it is important to make the attempt. Highly qualified employees have retired from the lab and more may do so, or seek employment elsewhere, if they do not feel valued. The consequence is a loss of skills and experience critical to the future success of the lab. Similarly, many scientific users may opt to transfer their expertise to activities at other facilities. Fermilab will be able to achieve the bright future that P5 has envisioned for it only by reversing these concerning trends.

We also want to ensure that the University Research Associates (URA) and the University of Chicago (UC) as M&O keep prioritizing the best interest of the lab, since members of their Physics departments are involved in the research carried out at the laboratory. As the University of Chicago[1] (UC) is a major scientific research institution in its own right, it needs to manage conflicts of interest if <u>Fermilab's M&O</u>

---

[1] First filed on January 9, 2022 in Illinois federal court, a lawsuit against the UC alleges that the University was a part of a "price-fixing cartel" that used a shared methodology for calculating financial need that limited financial aid for admitted students. The UC has agreed to pay $13.5M to students after settling a claim that it conspired with other universities to intentionally limit financial aid offers for students. The latest lawsuit blasts UC cancer site contractors after two workers fell over 80 feet, which is eerily similar to what happened at Fermilab in May 2023.

manager is also its potential competitor. This may be exacerbated by the UC being also the M&O contractor of ANL, where several Fermilab's competing groups collaborate directly with the UC. Section L.12 ORGANIZATIONAL CONFLICT OF INTEREST (OCI) of the RFP specifies that "A contract shall not be awarded to any Offeror having an unresolved OCI" (Appendix C, pg. 76). We note here that our group worked on this document off site, on personal time and with our own personal devices, as we were prohibited from discussing FRA's performance based on alleged conflict of interest. Additionally, it is our opinion that voting power in any board where members have a potential OCI due to the research activity carried on at their institution should be carefully evaluated.

## II.A      Selection of the Leadership

Hiring Fermilab's director is a responsibility of the M&O contractor. In December 2021, during FRA's hiring process for the lab director, one of us talked to FRA liaison Juan De Pablo and asked him questions about the training that the search committee members would receive; whether they had measurable criteria to assess the ability of the candidates; and whether they had a set of best practices, questions, role playing procedures and processes to reduce errors. (https://collection.cloudinary.com/apsphysics/62487d369d42a47b79d67d14f9097f7f? Scroll down to "gazette-2022-Spring.") His answer was that "the search is proceeding well, with a good array of outstanding individuals having expressed interest in the position."  We advocate for a transparent process, with input from all constituencies, and the application of best practices when hiring leaders.

Within months from his starting date in Fall of 2013, there was concern that the new Lab Director had removed former high-level managers, such as the Associate Director for Accelerators and the LBNE Director, who both left the lab. The position of Associate Director for Accelerator was eliminated from the org chart, and a former Navy Admiral was hired as LBNE Director. A few years later, the beam physics expert and academic in charge of the Accelerator Division was replaced with a particle physicist. By the end of his tenure, the Lab Director was perceived to have promoted people he openly favored[2] to positions of Senior Scientists as early as 5 years from hiring, and assigned them to positions of highest responsibilities. The appearance of such micro-management (i.e. not trusting and fully using the organization chart to delegate responsibilities according to professional qualifications) and the need for transparent processes, including for instance who should be recipient of Laboratory Directed Research and Development (LDRD) grants, have also contributed to morale issues.

It is our opinion that in the last decade, the appearance of favoritism and associated micro-management has often produced managers that do not have the confidence of those who report to them, resulting in stunted growth. Very often management positions get filled without any job opening and search, which prevents many qualified people from applying. In the rare instances that a job opening is created, it is most often all just smoke and mirrors.

---

[2] See all-hands presentations by N. Lockyer over the years.

The climate at the lab, as assessed in the latest third-party survey, has worsened further since the National Business Research Institute evaluated it at the 61% percentile in a 2019 survey.
https://indico.fnal.gov/event/21064/contributions/60683/attachments/37970/46109/Climate_Survey_Details.pdf
Whereas the official claim was that this was on par with the industry, Fermilab must aspire to be better than that. The outcome of the latest climate survey[3] points to poor communication between management and employees and a decline of trust in management. This finding is not surprising, since secrecy and communication of misinformation are demoralizing for those who aspire to scientific and human integrity. Former internal committees and/or task forces of mature and prominent scientists used to have a say on how operations were conducted at the lab. Because of the presently large percentage of early career members without sufficient experience, knowledge and insight in these committees, they have become nearly powerless in advocating and enacting positive change.

To succeed an organization requires leaders who lead, inspire and teach by example. We note here that Contract Clause I.9 – FAR 52.203-13 CONTRACTOR CODE OF BUSINESS ETHICS AND CONDUCT is consistent with this basic concept when requiring in I.9.(c), a "Business ethics awareness and compliance program and internal control system" (pg. 13). The Internal control system section I.9.(c)(2) includes timely discovery of improper conduct, prompt corrective measures, adequate resources to endure effectiveness of the internal control system, effort not to use as managers individuals whom due diligence would have exposed as having engaged in conduct that is in conflict with the code of business ethics and conduct, periodic evaluations of the internal control system, periodic assessment of the risk of criminal conduct (i.e. such as sexual assault or bringing a lethal weapon on site as described later on), positive enforcement of reports, disciplinary action for either improper conduct or its enabling, and finally, <u>timely disclosure to the OIG</u>.

In our opinion, Fermilab needs a leadership team consisting of people with long term proven scientific expertise, with the highest integrity as an inspiration to others, who own their mistakes, value productivity and excellence over the appearance of it, and put facts and fair play above politics and convenience. Unfortunately, the current administration has not focused on this foundation. This includes:
1. Not replacing inept/inexpert managers inherited from the former administration pipeline with more productive ones;

---

[3] Ironically, we cannot provide a reference or actual results of the latest climate survey because FNAL management has decided to keep this too a secret. The answer that employees receive when inquiring about a reference or a link to the latest climate survey is the following: "Unfortunately, the information is not available at this time, and there are no plans to make it accessible in the future."

2. Sometimes even promoting these ineffective managers to increased levels of responsibility;

3. Hiring inexperienced people as managers, or people with a poor track of accomplishments;

4. Overpaying for the so-called "executive positions" and therefore attracting money hungry people and ladder climbers for high responsibility roles;

5. Not reverting executive salaries of former managers to scientific salaries when managers are done with their executive service.

For instance, ALD positions at national labs typically require mature and distinguished leaders in their field, capable of developing and accomplishing a scientific vision in the areas of their directorates. At Fermilab, the present role of the lab's multiple Associate Lab Directors (ALDs) sometimes resembles that of secretaries to the Director and CRO. <u>Fermilab reputation suffers from having in highly visible positions inexperienced employees with sometime very few scientific accomplishments.</u> To identify effective leaders, searches should ideally follow best practices as developed by Industrial-Organizational Psychology (I/O Psychology).
https://onlinelibrary.wiley.com/doi/10.1111/j.1744-6570.1976.tb00404.x
Some are described in Section II.A.1 below.

**II.A.1 – Best Practices**
<u>Best practices aim to match desired behavior to the values of an organization through measurable processes</u>. To reduce hiring errors, the first step of the quantitative process is called "job analysis" that identifies the characteristics needed in the candidate for a successful performance. A rubric, based on the identified specifications should be agreed upon before examining applications, wherein characteristics are prioritized and a normalized scoring procedure is created. Only steps that add measurable value to the process should be used, sometimes iteratively. The process should be as transparent (and defensible in case of litigation) as possible with respect to the selection committee and the stakeholders. When stakeholders are invited to provide feedback, the search committee obtains valuable information that reduces the risk that unexpected and potentially embarrassing information arrives after a new leader is announced. Behavioral research shows that interviewing candidates using a structured and consistent process across applicants, including personality evaluations, integrity and reliability tests, work simulations (both situational and behavioral), is two to three times more reliable than non-structured processes.

As for performance evaluations, a majority of employees are disappointed and disillusioned about the existing Employee's Performance Review process, which is primarily used for managing raises rather than for employees' development.

**RECOMMENDATIONS**

- All Division leaders and their deputies, once they are given formal authority and responsibility, as well as much larger executive salaries, should be made accountable for their Division's budget vs. productivity with clear, measurable criteria that define scientific, technical and administrative accomplishments. The Division budget and spending plans should be presented regularly to the Division staff to ensure transparency and cost-effective use of the money.

- As part of their duties and responsibilities it should be required and mandatory for Division leaders and their deputies to establish non-subservient relationships with DOE mid-level managers. We note here that as part of the Goal 3 (i.e. Science and Technology Project/Program Management) evaluation by the HEP Office, the 2021 PEMP criticized the "codependent attitude" (Appendix B, pg. 45) of some project managers in their communication with DOE.

- To offset any negative bias that might exist toward Fermilab's HEP teams, Fermilab line managers should undertake as a crucial duty an effective marketing/ advertising of their team's talents and capabilities based on actual and impactful results. Prompt disciplinary action should ensue if the opposite is done.

- Clear, measurable, and explicit criteria must be defined to evaluate individuals' accomplishments and justify promotions in any category. This principle is especially urgent for promotions in the Scientific category, that are known to have been subject to pressure from top management, with cases of "favorites" being promoted to Senior Scientists within 5-8 years from PhD compared to 15 to 20 years average at other SC laboratories. An improved Employee's Performance process should be employed and we propose a 360 review process in Appendix A.

- Membership of influential committees such as promotion committees, scientific advisory committees, LDRD committees, etc. should be selected among the most knowledgeable and accomplished scientists and engineers both inside and outside the lab, if needed. Members should be free to offer constructive criticism without fear of retaliation by superiors, to evaluate scientific proposals without pressure, and without bias due to possible conflict of interest with their own research and professional agenda.

## II.B       The Flagship Projects

The 2014 P5 panel, on which both current Fermilab Director and her Deputy Lab Director served, had recommended the following major initiatives for Fermilab: the Deep Underground Neutrino Experiment (DUNE), the Mu2e Experiment, the PIP-II linear accelerator, and Fermilab participation in the LHC at CERN.

DUNE, at the Sanford Underground Research Facility (SURF) in Lead, South Dakota, was originally envisioned by DOE as a $1B project. The world's highest-intensity

neutrino beam is to be produced at Fermilab with the Long Baseline Neutrino Facility (LBNF) and sent 800 miles away to SURF. The term "baseline" refers to the distance between the neutrino source and a neutrino detector. The farthest and largest detector is at SURF and is composed of four modules of instrumented liquid argon with a fiducial volume of 10 kilotons each. In the 2016 Conceptual Design Report, the first two modules were expected to be complete in 2024, with the beam operational in 2026. The final modules were planned to be operational in 2027. By 2022, the cost for the excavation and outfitting of the caverns in the South Dakota mine, two far detector modules, the construction of the new neutrino beamline and near detector hall and Near Detector, had risen to about $3B. DOE then decided to phase the project, with two modules in Phase I, to be completed by 2028-2029, and the beamline by 2032; with the remaining two far detector modules for a Phase II, yet to be planned. (The current P5 Report revises some of these recommendations.)

The DOE $1B budget for DUNE was originally intended just for the far and near detectors, with NSF carrying the brunt of the cost of the LBNF infrastructure. This included the construction of an underground lab, called DUSEL at the time. Unfortunately, the NSF decided not to initiate construction of the DUSEL facility in 2011. When the 2014 P5 panel pushed nevertheless for DUNE, it was apparently done with the implicit assumption of financially engaging the international community. By 2020 or so, this approach was recognized to have only partially worked, securing some international contribution, but requiring a much larger investment on the part of the U.S. to build the actual facility, i.e. LBNL. An unrealistic budget cap forced the IPT (Integrated Project Team) to make a series of ill-founded decisions between 2015 (CD1-R) and 2022 (CD-1RR).

Only by 2022 the project was split into several sub-projects, recognizing the different levels of design maturity and progression of the various components of the project: conventional facilities, including underground access, one-of-a-kind particle detectors, and a new beamline.

On FRA side, the Project Team was not up to the standard required by the execution of a DOE Order 413 project, between at least 2015 and 2022, both for the overall project and its sub-areas. This was often reported in DOE Lehman's reviews, but went unaddressed for a long time. Rigorous adherence to Order 413 requirements was downplayed with consequent lack of transparency in the interactions with all the stakeholders.

The Proton Improvement Plan-II (PIP-II) aims to deliver 1.2 MW of proton beam power from the main Injector to the LBNF production target at 120 GeV. The plan will also support the current 8 GeV experiments including Mu2e, and other short-baseline neutrino experiments that require a Linac that injects into the Booster at 800 MeV instead of 400 MeV. PIP-II received CD3 approval in 2022 and, within a year, incurred in significant cost overruns that will most likely require to re-baseline the project. This coincidentally happened after its former project director became Fermilab director, and most of the former PIP-II management team disbanded.

The Mu2e experiment was to produce results in 2020, but is now delayed until 2026.

In our opinion, it is obvious that a lack of knowledgeable leadership; familiarity with how the field and the U.S. system work; a mature, realistic and unbiased view of the problems that the field faces, have brought Fermilab and FRA to act reactively to a series of self-inflicted problems.

**RECOMMENDATIONS**

- Oversight between 2014 and 2021 on the part of all stakeholders should have been stronger and more forceful in identifying the inadequacy of the proposed budget and in selecting capable person-power to support such a large enterprise. This was partially improved only starting in 2021, and has to be further improved.

- To execute the newly released P5 plan and achieve measurably excellent results, the laboratory should seek to fill leadership positions with people with proven knowledge and genuine investment and stake in solving the laboratory issues in a manner that will ensure a robust HEP future.

## II.C  DOE Performance Evaluation and Measurement Plan (PEMP)

Since fiscal year 2006, DOE uses a homogenous appraisal process for the performance of the 11 labs and institutions under its Office of Science (SC).
(https://www.energy.gov/science/office-science-lab-appraisal-process)
These evaluations provide the basis for determining annual performance fees and the possibility of winning additional years on the contract through a 3-year "Award Term" extension. They also serve to inform the decisions that DOE makes regarding whether to extend or to compete the management and operating contracts when they expire.

The Performance Evaluation and Measurement Plan (PEMP) is structured around eight Performance Goals, and emphasizes the importance of delivering the science and technology necessary to meet the missions of DOE; of operating the Laboratories in a safe, secure, responsible, and cost-effective way; and of recognizing the leadership, stewardship and value-added provided by contractor managing the Laboratory. The eight Performance Goals are:
1. Mission Accomplishment (Delivery of Science & Technology), (30%)
2. Design, Construction and Operation of Research Facilities, (45%)
3. Science and Technology Project/Program Management, (25%)
4. Leadership and Stewardship of the Laboratory.
5. Integrated Environment, Safety and Health Protection, [30%]
6. Business Systems, [30%]
7. Facilities Maintenance and Infrastructure, [25%]
8. Security and Emergency Management, [15%]

Each Performance Goal is comprised of a small number of Objectives. As can be seen in more detail in Appendix B, the Objectives usually do not change much from one year to the next. Within each Objective, the SC HEP and Site Office can further identify a small number of Notable Outcomes that illustrate or amplify important features of the laboratory's performance for the coming year. The Performance Goals, Objectives, and Notable Outcomes are documented at the beginning of each year in the lab PEMP. <u>Goals 1 to 3 are evaluated by the HEP Office</u> and are associated to what is called the "Initial S&T Score" when using as weights the percentages shown in round parenthesis in the goal list above. <u>Goals 5 to 8 are graded by the FSO</u> and are associated to what is called the "Initial M&O Score" when using as weights the percentages shown in square parenthesis in the goal list above. Goal 4 is assessed at the SC level. The final S&T and M&O scores are calculated by weighing 75% of each initial scores with 25% of the score given to Goal 4. Evaluations are performed by considering the lab's performance against the Notable Outcomes, as well as other sources of performance information that becomes available to DOE throughout the year.

Appendix B summarizes in some detail the PEMP results from FY 2019 and FY 2021. To put it in the context of other DOE labs, anything lower than a B+ (or 3.1) is considered failure. The aim is to get as many A/A- (i.e. ≥ 3.5) as possible, and several other labs manage to do that. An example is LBNL that had all (i.e. 8) A/A- for 3 years straight, or ANL and BNL that had between 6 and 7 A/A- each over the last 3 years. Fermilab has received the lowest grades among the national laboratories since at least FY2018. For the FY 2023 PEMP, Fermilab was again at the bottom, this time very close to JLAB. In addition, FRA received a $1M fine, which brought the effective rating down from B+ to C+. In 2021, a rare C grade was assigned for Goal 3.0, i.e. Science and Technology Program Management project management, reflective of the delays and cost overruns (See Appendix B). In an article in the journal Science, James Decker, who was principal deputy director of DOE's Office of Science from 1973 to 2007, stated that the performance evaluation for 2021 was "one of the most scathing I have seen." (https://en.wikipedia.org/wiki/Fermilab)

As can be appreciated from the detailed tables in Appendix B, the PEMP goals encompass most of the responsibilities required in the RFP for M&O bidders. Appendix C shows the table of content of the RFP. To achieve the S&T goals, it is critical that the M&O goals be achieved. The first one of the latter, i.e. Goal 5, relies on both workers health and safety and environmental protection under the CSO. This is covered in the Section II.D on safety below. Goal 6 requires productive Finance, Procurement, HR, and Quality control systems, under the COO. Business, Financial, and Accounting for Operational Support are covered in Section II.E. Both HR and General Counsel systems are covered in Section II.F.

**II.C.1 – The Accident**
On May 25, 2023, a contractor fell 23 ft while attempting to secure reinforcing bars on a wall for the new PIP II project site.

(https://www.chicagotribune.com/business/ct-biz-fermilab-proton-construction-accident-investigation-20230914-yj5c6gnhfnejtaixtm4qsktla4-story.html).

The PIP-II Project had been led by current lab director Lia Merminga until April 2022. The contractor was air-lifted to the hospital after an accident considered the worst on site in decades.
(https://www.chicagotribune.com/suburbs/aurora-beacon-news/ct-abn-fermilab-injury-st-0526-20230525-fhcqfqi4b5arzcknvy57y4iu2u-story.html )

The DOE- Accident Investigation Board concluded that the incident was preventable and "recommended a long list of managerial and safety controls needed to prevent a recurrence of such an accident."
 (https://www.energy.gov/ehss/articles/accident-investigation-may-25-2023-ironworker-fall-injury-fermi-national-accelerator)
Some conclusions included:
1. Unclear and non-specific definition of work activities/responsibilities.
2. No methodology to communicate a clear flow down of requirements.
3. Oversight failures at multiple management levels.
4. Systemic lack of attention; etc.

As a result of the accident, the $1B PIP II project that is crucial for the success of the DUNE Experiment has been delayed. The DOE report on the May 2023 PIP-II accident is very much representative of the way that business is generally conducted at the lab.

**II.C.2 – The Shutting Down of the Beam**
On Sept. 1, 2023, the Chief Research Officer announced that the Fermilab accelerator system was temporarily shut down for safety reasons. On Sept. 9, 2022, DOE had issued order DOE O 420.2D entitled "Safety of Accelerators". This document establishes the accelerator-specific safety requirements for DOE-funded accelerators and their operations. When pressed for answers by one of us, Douglas Glenzinski wrote in a December 2023 email that "the criteria for restarting the beam operations are captured in DOE O 420.2D
https://www.directives.doe.gov/directives-browse#b_start=0."
He also stated that "There are a series of readiness reviews planned. As each of those is successfully completed, we'll be able to restart a portion of the complex." The current earliest estimate for resuming the beam is March, 2024. However, no answer was provided to the question "Where can one find info on why the beam was stopped in the first place?"

In the following we expound on the results of our investigations. In the fall of 2021, two chapters of the Safety Assessment Document (SAD) were modified to include any changes to the neutrino muon beamline and the experimental enclosure to support a new experiment, E1039/SpinQuest.  Review of the chapters were delayed for nearly a year as the chair of the SAD committee wished to review and comment before another review committee sent it to the Lab Director for approval. In the fall of 2022, the new chapters were approved by the Director and sent to the DOE Site Office. That document was rejected with 39 comments just before the Christmas holiday. In the meantime, DOE order 420.2D was approved requiring external reviews of the SAD and Accelerator Safety Envelope (ASE)[4]. Response to the Site Office comments were submitted in April of 2023 and roughly half of the responses accepted. Whereas responses to the comments were discussed in meetings which were attended by members of the DOE Site Office, little/no useful input was available about how best to word the document.

In early 2023, a committee of over 50 people was formed to re-write the SAD and ASE. Such a large committee made it difficult to obtain consensus on how best to proceed based on the little guidance obtained on what was required in a new version of the document. The group was disbanded in the late Spring of 2023. Outside contractors were hired to assist in re-writing the SAD similarly to other national labs. An intense effort over several weeks ensued to prepare both documents for review in September of 2023. The hastily compiled documents were reviewed with comments from the external reviewers that indicated that whereas the requirements of DOE 420.2D were satisfied, there was still some work to be done. Further comments from the DOE Site Office pointed to additional concerns about the wording of the documents. At this time the Director decided to postpone accelerator startup until a new document could be written. We note that other labs were running while re-working the 420.2D requirements. Since then, a program to rewrite chapters of the SAD/ASE and turn on the main accelerator complex has been underway with the LINAC recently having their chapter approved and granted approval to run[5].

### II.C.3 – Site Access Safety/ Security Problems

Fermilab was founded in 1967 as an open-access laboratory, and, to this day, does not host classified research. For 50 years, both scientists and the public could easily access the site for research, educational activities, arts programs, and recreation.

In the late 2010s and early 2020s, the management of Fermilab began to introduce severe restrictions on access to the Fermilab site by the public and by scientists. By spring 2023, the restrictions had become so onerous that more than 2500 physicists and visitors to the laboratory signed an "open petition to elected representatives to reopen Fermilab." (https://www.reopenfermilab.com/) The petition stated that: "The access policy changes undermine critical aspects of the scientific process as well as the basic functioning of Fermilab. Hosting research meetings, interviewing prospective employees, collaborating with scientists outside the lab, and enacting our famously impactful education programs have all been hindered." With respect to the general public, the petition stated: "Today, the general public is only permitted to access the main road, and with ID requirements that are becoming increasingly stringent, soon its doors will be closed to tourists and even to some immigrants. We can no longer drive or bike around the premises freely. The dog park, Wilson Hall with its exhibits on the top floor, and other areas are no longer generally accessible. Fishing and other activities open to the public have been canceled." The petition emphatically requested that access policies be reverted to the open laboratory model that governed the laboratory prior to 2020.

---

[4] The SAD is a general overarching document summarizing the safety systems and protocols implemented at a DOE laboratory. This document is approved by the director of the laboratory and sent to the DOE Site Office but does not require their approval. However, the ASE summarizes key components of the SAD and does require approval of the DOE Site Office.

[5] We note that the test accelerators FAST/IOTA have been allowed to continue running under their existing safety chapters.

In May 2023, the Lab Director posted a response to the petition on the Fermilab website, ([https://news.fnal.gov/2023/05/from-director-lia-merminga-accessing-fermilabs-batavia-site/](https://news.fnal.gov/2023/05/from-director-lia-merminga-accessing-fermilabs-batavia-site/)) noting that some areas on site remain open to the public during specific hours with ID access requirements. Merminga's response justifies the new restrictions because the lab "manage[s] a large amount of non-public information", which conflicts with the petition that points out that the lab is fully tax-payer funded, does no classified research, and has a government mandate to publish all of its scientific results. Further coverage of the petition and the management response appeared in the magazines Physics Today ([https://pubs.aip.org/physicstoday/online/42386/](https://pubs.aip.org/physicstoday/online/42386/)) and Physics World ([https://physicsworld.com/a/fermilab-faces-protest-over-visitor-restrictions/](https://physicsworld.com/a/fermilab-faces-protest-over-visitor-restrictions/)).

The FSO decided that service buildings' doors have to be locked at all times. In the Accelerator Directorate building managers have to check 200 doors, three times a week to make sure that they are locked. Access gates at the Linac, Tevatron A0, and Tevatron C0 roads were installed at the FSO request. Interim COO Marc Clay was heard stating that he worked at Los Alamos for 20 years and Fermilab is far worse as far as what is being required by its Site Office.

**RECOMMENDATIONS**

- Negotiate with the DOE the opportunity for Fermilab to be assigned an ODSA, similarly to other DOE labs. Some DOE labs have an ODSA, which is the Officially Designated Security Authority for the site and is a position appointed by DOE, and specifically by Operations under the Office of Science. For instance, at LBL the ODSA developed the system they now have in place by which everyone badges onto the site and again to most buildings. It had to satisfy both the Site Office's requirements for control of everybody on site and maintaining secure areas at the Lab, like the Joint Genome Institute and NERSC. Conversely, at labs that do have an ODSA, the FSO is the site ODFSA (Federal added). As much as formally the ODFSA might have 51% of the authority with respect to the ODSA at 49%, a FNAL employee covering this latter role would presumably improve mutual communication.

- In reference to security, the lab needs an effective Research Compliance Office on the related issue of research security, which is also getting new oversight from Washington. This office would be in charge of research integrity, conflicts of interest and conflicts of commitment, research security and export control. The office head has to have good connections and communication with the research ALDs and the other Directors to help them manage through the new research environment, with special attention on international aspects.

## II.D     The Safety Matter

The physical and psychological safety of an organization's workforce is of paramount importance. Organizations lacking in one or both of these areas will have difficulty succeeding in their missions. In this section, we will analyze Fermilab's recent performance when it comes to physical safety, and attempt to determine possible solutions to the problems discussed.

The following definitions apply:
https://www.energy.gov/sites/default/files/2024-02/OPEX%20Awareness%20-%20Understanding%20TRC%20and%20DART%20Rates%20-%20FINAL%2002.06.2024_0.pdf

- Total Recordable Cases (TRC) is the total number of work-related injuries or illnesses that result in death; days away from work; job transfer or restrictions; or other recordable cases as identified in columns G, H, I and J of the OSHA Form 300.

- Days Away, Restricted or Transferred (DART) is a subset of the TRC events and is the total number of work-related injuries or illnesses that result in the most serious outcome of the cases involving days away from work, or days of restricted work activity or job transfer, or both.

- Occurrences are events or conditions that adversely affect, or may adversely affect, DOE or contractor personnel, the public, property, the environment, or the DOE mission. Occurrences are to be reported into the Occurrence Reporting and Processing System (ORPS) database.
  https://www.energy.gov/ehss/occurrence-reporting-and-processing-system

Reportable occurrences at the lab are plotted per year in Fig. 1 (https://cdv-fdashp.fnal.gov/Reports/mobilereport/Management%20Systems%20-%20MS/ESH/Reports/ORPS%20NTS).

The total TRC and DART counts from FY20 to FY24 are shown in Fig. 2 (https://cdv-fdashp.fnal.gov/Reports/mobilereport/Management%20Systems%20-%20MS/ESH/Reports/Injury%20Illness%20Charts).

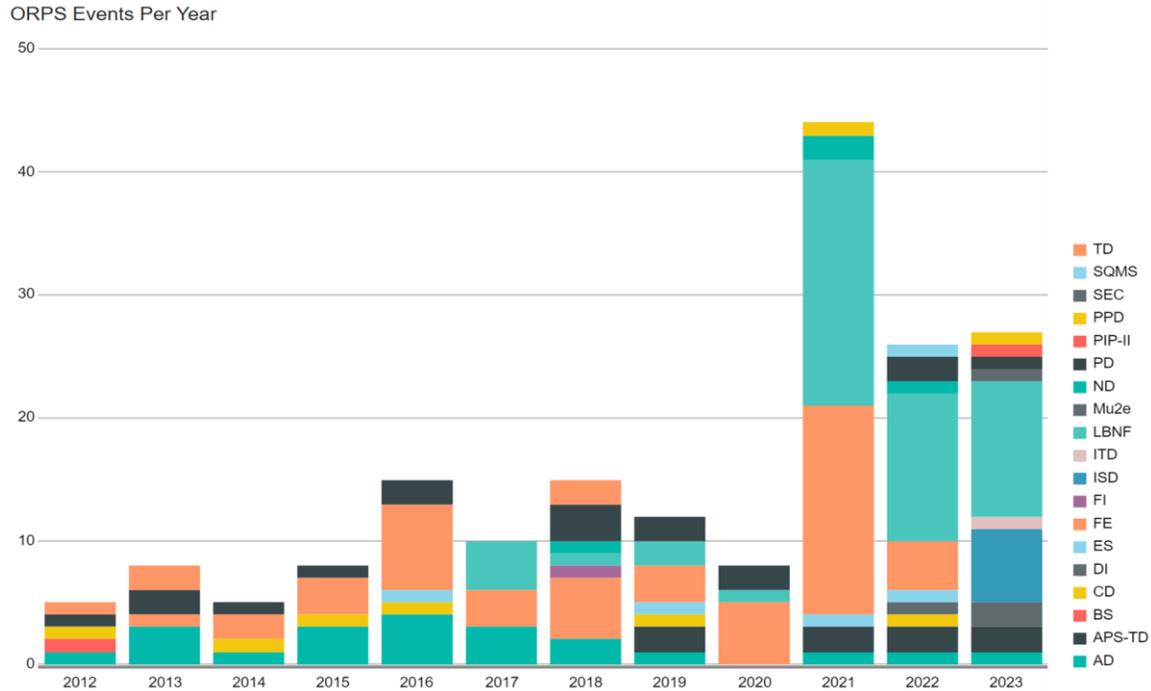
**Fig. 1.** ORPS events per year from 2012 to 2023.

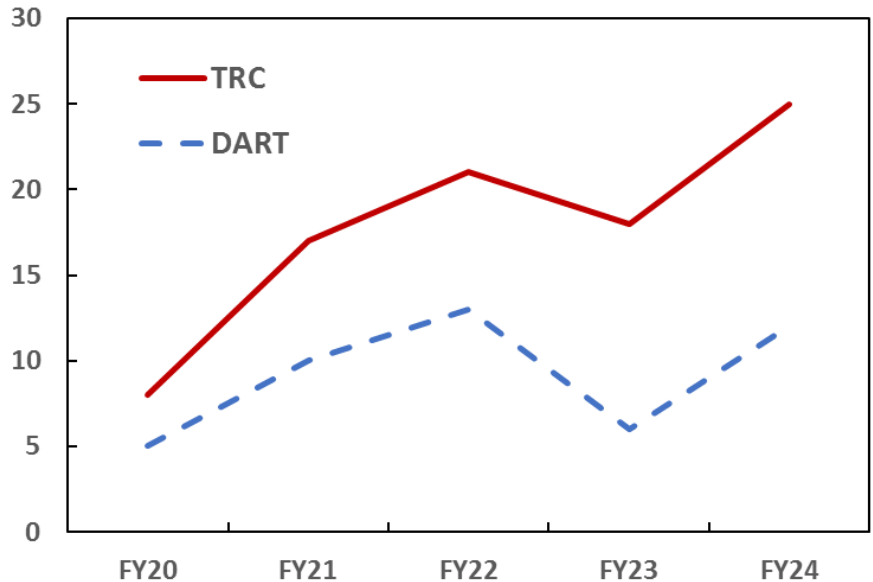
**Fig. 2.** Total TRC and DART counts, including employees, users and subcontractors, from FY20 to FY24 (ongoing).

As can be seen, safety incidents increased significantly over the last few years. More extensive and detailed information on safety problems can be found in Appendix B in the FY 2021 and FY 2023 PEMPs.

In an attempt to solicit feedback on its performance, Fermilab's Environment, Safety, and Health (ES&H) Division conducted a Customer Satisfaction Survey in 2022. About half of the respondents classified their overall satisfaction with support received from ES&H as "moderate" or worse. Similarly, about half also reported their confidence in the expertise provided by ES&H personnel as "moderate" or below. These results strongly suggest that Fermilab staff do not have an adequate level of confidence and trust in ES&H.

The origins of this lack of confidence can, in part, be traced to the centralization of ES&H in 2016, as alluded to in several comments from the survey and in discussions leading to the writing of this document. Prior to 2016, each Division at Fermilab had its own embedded Safety organization, encompassing both industrial and radiation safety. These Division-level Safety departments were separate from the ES&H Division (then the ES&H Section), and their personnel reported directly to the head of the Division that their department was embedded within. This direct-report organizational structure empowered each Division Safety Officer (DSO), who had the advantage of having the same priorities and goals as the Division Head. People within the Division knew this and took the role of the DSO more seriously. An additional advantage to this arrangement was that there was little to no confusion on whom to speak with about a particular safety issue in one's Division. Long-standing relationships were developed, which in turn led to increased trust.

In the time leading up to the reorganization, it was felt by ES&H Section management that some of the Division Safety organizations were not being adequately cooperative with the Section-level management. To address this concern, it was decided by Senior Management to centralize all of the Division Safety departments into the ES&H Section, including the physical relocation of personnel away from their former Divisions. While it is true that communication and cooperation between these Division-level Safety organizations, and between them and the ES&H Section, could have been improved, this could have been accomplished without a complete reorganization, which soon proved to be disruptive to operations and damaging to the lab's safety culture.

The reorganization resulted in confusion over responsibilities, leaving people unsure of whom to contact about a given issue. Exacerbating the situation, some people were moved around to other groups or departments in a seemingly haphazard fashion. As a result, at times even members of ES&H were unaware who was responsible for what. Like other ill-conceived management decisions, the reorganization led to some key personnel leaving the laboratory, retiring or finding work elsewhere. <u>Perceptions of cronyism and age discrimination within ES&H became abundant, with many wondering why management positions were handed to early-career employees with little or no experience and with fewer professional qualifications than others</u>. To date, little effort has been made by ES&H management to transfer institutional knowledge between new and existing staff and to foster a common ES&H vision.

Best business practices call for appropriately involving personnel who will be affected by overarching decisions. The centralization of safety personnel was a large-scale example of FNAL non following best practices. A smaller-scale example of this phenomenon is the attempted movement of people into other job positions without first consulting them—this has happened numerous times and has resulted in the affected employees, who were not interested in the position, leaving the lab. In other cases, employees felt pressured into taking the offered position, leading to increased stress and decreased job satisfaction. Despite repeated feedback from staff, management has continued this pattern of decision-making without discussion.

<u>Another major issue pointed to by the 2022 Customer Satisfaction Survey (but well-known to ES&H staff prior to it) is the ongoing understaffing of the Division</u>. Again, this may be also true in other parts of the laboratory, but it has been particularly acute in certain areas of ES&H. For example, the Radiation Physics Operations (RPO) Department has been almost continuously understaffed since the 2016 reorganization—this department in particular was called out quite often in the survey results, likely because it interfaces with other Divisions more regularly than other parts of ES&H. In this case, Senior Management under the current Director responded to the issue and gave the go-ahead for RPO to hire a large number of additional staff members. However, other areas of ES&H remain understaffed, and promised new job positions have been put on hold.

Meanwhile, additional demands made by the Fermilab Site Office (FSO) have continued to increase the workload of an ES&H staff already spread too thin. For example, it is well understood that ionizing radiation exposures above natural background levels are unavoidable in certain areas at the lab. Historically, signs and fences have been used to keep unauthorized persons away, thus keeping doses As Low As Reasonably Achievable (ALARA). A key assumption in these controls is the maximum amount of time that members of the public are presumed to potentially be exposed to such sources. A common conservative assumption at the lab since its inception has been one hour per day per working week, or 250 hours per year. FSO has recently insisted on assuming continuous occupancy (24 hours per day for 365 days per year, or 8,760 hours per year) for members of the public; being the highest possible value, this directly conflicts with the "reasonable" part of the ALARA concept. As a consequence, Radiation Safety personnel had to divert significant time and attention away from their daily work to devise additional controls to limit these inflated potential dose rates to members of the public. Over the past few years, numerous ES&H staff hours have been diverted to meeting similar new requirements from FSO, a situation that cannot be described as ideal when it comes to oversight of actual safety.

While the significant expansion of RPO has helped alleviate the delays in getting work accomplished as well as the lack of communication described in the survey results, there is a concern that it may introduce new issues, and these concerns also apply to any department which gains a large number of new employees. All of the new hires, including all but one of the new managers, came from outside Fermilab, meaning only a small fraction of RPO staff have any knowledge about the department's operational history and practices. Outside hires who don't have experience with accelerators, may bring with them preconceived notions about how to address issues, which could lead to a reactionary management style and further friction between departments. This is even more of a risk if new managers and/or staff already knew each other prior to being hired into a particular department.

Another finding of the Customer Satisfaction Survey was that many at Fermilab no longer think that ES&H staff are knowledgeable or helpful. To quote a typical comment provided to the authors of this paper, "ES&H people have no idea of the bigger picture and fail to understand how to apply safety in the context of physics. They are just blindly applying what they have been taught is safe without knowing how things work in practice." Many survey comments suggest that ES&H is perceived as being overly focused on compliance, poor at communication, reactionary rather than proactive, and just there to hinder one's work. Similar to the other findings, this is a situation that can lead an organization to failure. When one looks at the results of the survey, it is not surprising that the lab has had accidents like the serious injury at the PIP-II construction site, for which management failures were identified as a major contributing factor.

Similarly, management shortcomings were also found to be partially responsible for the Proton Source Test Stand unplanned dose event that took place in May 2022. The lab's investigation of this incident found that several communications and oversight failures within the Radiological Control Organization contributed to a worker apparently receiving a significant unplanned dose as measured by their dosimetry badge. This event could have been avoided or mitigated had there been oversight by more qualified and knowledgeable staff. Additional details of this event are included in Appendix D.

Psychological safety is another area where Fermilab comes up short, despite proclamations of its importance. Section II.F and Appendices E and H provide sobering examples of the laboratory's disregard for the psychological well-being of its staff. Similar stories seem to abound; many concerned employees have been horrified to hear of the unwillingness of HR to do anything about the perpetrators of terrible acts, whether physical, verbal, or otherwise. This issue is further addressed in Section II.F, but it is important to note here that a psychologically toxic environment is conducive to potential safety violations of all kinds. Certainly, when one is in a state of heightened stress and fear at work due to the actions and behaviors of a fellow employee, that person is not going to be able to pay as close attention to their work as someone who feels psychologically safe on the job, a situation that can and will lead to mistakes and possibly accidents and injuries.

**RECOMMENDATIONS**

- An obvious potential solution to the issues described above may be to once again decentralize portions of the ES&H Division. There is precedence for this, as early in the lab's history the Safety Section was centralized, and Senior Management at the time made the decision to decentralize it. However, concerns are valid that such a major move could further exacerbate certain issues rather than alleviate them, especially if the other management issues are not addressed in parallel.

Furthermore, in the past year or so, improvements have been noted in ES&H. The Division Safety Officers have gotten more integrated into their assigned Divisions, and although their teams are still smaller than they were prior to the centralization, Division Safety Specialists continue to be added to assist the DSOs. The improvements to RPO were already noted above. These positive changes must persist and be built upon, bringing us to our next point:

- An alternate solution is to make ES&H a major area of accountability for ALL line managers. A central ES&H division can supply Division Safety Coordinators as experts matrixed to line managers of relevant operations. Employees need to feel that they have real authority to STOP WORK. Employees need to be able to stop work without being subject to retaliation.

- Additional ES&H staff is desperately needed to reduce workloads. Job positions that have been put on hold should be reviewed and reopened. Particular focus needs to be on hiring for non-management positions, as the Division (and the lab as a whole) is perceived to be top-heavy, with too many managers and too few non-managerial workers.

- Workers should not be shuffled around from position to position as they were after the centralization. This has improved recently but must continue if trust in ES&H is to be rebuilt. Rebuilding trust will take time, and breaking relationships with personnel in other Divisions by moving staff to different positions will only serve to reset that clock.

- Until ES&H understaffing is not solved, Senior Management has to have the capabilities to negotiate priorities for the demands made by the Fermilab Site Office (FSO).

- Since the centralization, various ES&H departments have been relocated to isolated "islands" at the Training Center, Feynman Computing Center, the Linac Loft, etc. This has had the effect of reducing interaction between staff members. Serious consideration should be given to co-location or devising methods to increase interdepartmental interaction, cooperation, knowledge transfer, etc.

- Train scientists to impress on them that safety is real and necessary for their work to succeed. For example, a 'Safety for Managers' course aimed specifically at scientists, engineers, and project managers could be developed to help them inspect field work in which they are engaged for well-known error precursors and to identify and abate problems before they occur.

## II.E      Business, Financial, Accounting – Operational Support

The business/financial side of the laboratory is beset with a large chaotic mix of problems involving personnel issues, severe lack of depth in subject matter expertise in many mission critical functions, accounting system software deficiencies, government accounting complexities, frequent invasive audits, etc.

Personnel problems start at the top, where the COO, CFO, and Comptroller departed in early 2024 with no qualified replacements promptly available to replace them. The stress is so intense that personnel retention is a challenge from top to bottom; stressful situations result from the many issues on the operations side of the lab.

Government accounting rules and regulations are unique and quite complicated, so that <u>accountants need specialized skills and experience to function effectively out of the gate</u>. That in turn means that departing long-time employees/subject matter experts are hard to replace, and the temptation is to replace them with people who are not fully qualified – either by education and/or by experience. Additionally, the government pay scale and recent events/changes negatively impacting the lab's overall reputation have also contributed to the inability to compete with the current job market. The lab has reacted recently by recruiting candidates from other national labs to fill senior leadership positions. Though that is a partial solution, the newcomers still take time to familiarize themselves with the Fermilab culture, DOE Site Office, policies and procedures. <u>The compensation packages offered of late have created extreme inequity among existing/long time employees, resulting in further decline in morale and desire to stay.</u> In addition, despite the lucrative compensation packages offered, the lack of transparency, communication, and growing toxic environment have resulted in considerable loss to the lab and FRA upon their choice to leave soon after being brought on board.

Investments to maintain accounting software systems up-to-date have been inadequate for a long time, resulting in inefficient workflow and the ensuing need for extensive and time-consuming manual intervention. <u>Investment in automation solutions is desperately needed to improve productivity, efficiency, accuracy, and ensure overall compliance with all federally imposed regulations (e.g. DOE, OIG, DCAA, DEAR, FAR, FTR, etc.).</u> These infrastructure problems were exacerbated when the lab was appointed host of the international LBNF/DUNE project. Although the scientific side of the lab was prepared for such a complex endeavor and role, there was inadequate consideration and assessment made of the state of the operational support side of the lab to accommodate and support the increased and complex needs of such a dramatic change in mission focus.

As already expounded on, a main problem for FNAL is budget insolvency, with the lab being very much in the red even just because of existing salaries. A major financial burden for salaries is the indirect cost, or overhead, i.e. the actual cost of an employee to the lab is approximately 3 times their salary due to this indirect cost. <u>The overhead is correlated to the ratio of the cost of administrative employees to that of non-</u>

administrative ones. It clearly follows that the number of bureaucrats and administrators on the lab payroll has to be optimized and closely controlled. Implementation of automated processes and solutions in Operations is necessary also to reduce the number of redundant administrators. This is critical for the lab not to lose itself in heavy bureaucracy and forget its Physics vision! In the meantime, this year the lab is forcing nearly everybody to take a 2-week mandatory vacation in a specific period before the end of the fiscal year.

In addition to administrators, the overhead includes the cost of scientists in management positions over the whole lab. Very often these employees are ballast and are not managing anything. Another example of waste is establishing so many ALD positions and other directorates, which has introduced a supplementary layer in the org chart. This has resulted in a much larger number of employees in so-called "Executive positions" with "executive", i.e. much higher, salaries. For instance, whereas before we had a dozen of Division Heads and Deputies (the threshold for an executive salary is Deputy Division Head), now we have about 5 times this number.

Another dire example of managerial cost ineffectiveness is for the Chief of Equity, Diversity & Inclusion to currently have 178 direct reports! In the Summer season this number includes several interns, students and teachers. However, dozens are staff. One of their tasks is to closely monitor the meetings of so-called resource groups, which splits people into the African-American/Black Association, the Fermilab Asian Pacific Association, the Hispanic/Latino Forum, the Spectrum (LGBTQ+) Community, the Veterans Group, the Women's Initiative, the Fermilab Young Professionals, the Fermi Accessibility Communities, the Inclusivity Journal Club, the Fermilab Society of Women Engineers, the Fermilab Society of Hispanic Prof. Engineers. Each group has meetings where the scope is very specific and narrowly defined. When attending these meetings, bringing up themes of ampler interest is often discouraged. Clearly, with this divide and conquer approach, the lab reduces the risk of people effectively communicating with each other, and therefore educating each other on what is going on at a larger scale. For instance, even the chairs of the Fermilab Society of Women Engineers did not know anything about the legal troubles of the lab associated with their own female engineer constituent! On par with the above, several of the sponsors for these groups are known, from testimonials, to have been key players in cover-ups. See more in Section II.F and Appendices E and H.

The DOE subjects the lab to frequent financial audits. As a result of the recent OIG decision to transition from a cooperative audit strategy to an independent audit strategy, the Department of Defense Contracting Audit Agency (DCAA), whose auditors are experienced at assessing the allowability, allocability and reasonableness of costs associated with the manufacture of ships and weapons, but not necessarily high energy particle beam targets, has assumed responsibility of conducting the lab's annual cost allowability audits. The resulting need to educate another federal agency to the ways and means of DOE has been extremely time consuming and has negatively impacted the overall productivity and efficiency of an already overstretched staff. Additionally, the need to respond to audits of prior years has also resulted in time-consuming examinations of past history. (As in "We think we may have found a small problem in 2018; please examine your records to find out whether a similar problem occurred in 2019, 2020, 2021, 2022, or 2023.") Time wasted on such efforts would be much better spent on forward-looking activities.

DOE imposed site access restrictions and changes to the lab's long standing onsite housing policy for users, affiliates and guests have caused considerable hardship for this group and another bureaucratic morass for employees. It has also resulted in users, affiliates and guests seeking out participation at other sister labs as opposed to Fermilab thereby further impacting the lab's overall national and international reputation.

There are deficiencies also in basic hiring processes. An example is the following. A few months after the start of her tenure in 2022, the present Director's team had completed the search for a PIP-II Project Director, which is the position that Lia Merminga had left vacant when she became Fermilab's Director. A candidate had been selected for the job. However, two months later, their hiring had not been finalized (we do not know what was the bottleneck.) When the lab was finally ready to finalize it, the candidate rejected the position.

Another example of basic deficiency is the signature by the Director or a designated official of standard commitment letters for grant applications. This used to be a non-automated process, with the applicant sending a request to an appropriate office, including a letter draft containing required information. The letter would come back edited as required and signed in less than a month. A recent attempt at using a new portal for the same purpose had us wait nearly 3 months because of micromanagement. This does not help the lab's participation as a partner of European projects and all its associated benefits.

**II.E.1 – Procurement**

Procurement is part of PEMP Objective 6.2, i.e. Provide an Efficient, Effective, and Responsive Acquisition Management System and Property Management System. As can be appreciated in Appendix B, this is an area that has consistently underperformed since at least FY 2019. <u>Cost-effective and timely procurement of quality packages is key to abide by deadlines and milestones of Projects and Programs</u>. Procurement deficiencies therefore hinder the lab's ability to deliver and enable its Science Mission.

Procurement encompasses goods, works and services. According to Fermilab's FSO Roger Snyder, <u>all National Labs are operated under the same basic set of terms and regulations for Procurement (in the contract, in regulations, in Orders, etc.), but each lab defines its own system/procedures</u>.  There are substantive differences in implementing processes – for example, most labs have pre-negotiated subcontracts in place allowing for quick task orders for facility repairs – FNAL does not yet; most labs do not have as much university subcontracting as FNAL so that is a greater focus/challenge. Pre-negotiated contracts for facility repairs are sorely needed at the lab, as is well-known.

Most labs have greater internal authority without DOE review – but that level is set based on system maturity, etc. For instance, ANL has also had procurement issues resulting in unallowable costs but is substantially more mature than FRA's. Still according to Fermilab's FSO, FRA's procurement system has not been approved without contingencies or conditions for years – issues have been cited in IG reviews, HQ assessments, field assessments, legal proceedings, etc.

The lab incurs in losses for its mission due to vendors that are reluctant to keep doing business with Fermilab because of lack of responsiveness for offers; delayed awards; delayed payments; and more, such as the following:
1. Lack of effective communication between both the Procurement and Accounts Payable sides with the scientists/engineers invested in the process. <u>Personal relationships and warm communication are key to getting productive business done and should not be sacrificed in the name of automation either.</u>
2. <u>High turn-over rate in Procurement</u>. This could include non-clear/unreasonable procedures, responsibility and ownership non-aligned with management levels, lack of customer service, etc.
3. <u>Some unique high-tech companies, such as Avnet, that produce highly engineered and customized products, have been banned from the lab's list of allowed vendors</u>. The alleged reason invoked by the bureaucrats is that Avnet does not follow the buying terms and regulations established in the DOE contract. Avnet has had a productive 30+ years relationship with the lab. For instance, the electronic QICK system for QIS, that has become a world standard, is based on specialized hardware developed by Avnet. No other company offers the same cost effectiveness and quality. Avnet is 100% American, has a $5B net worth, with multiple offices in 25 states, Canada and South America, produces its own R&D and sells to Universities, the DoD and NASA. <u>Because the same company can sell their products to other DOE labs too, such as Argonne National Laboratory (ANL), clearly there must be a way to use this company for Fermilab too</u>.
4. Some companies that had productive business with Fermilab for decades do not bid anymore when jobs are offered because sometimes Accounts Payable "does not pay their bills".
5. Some Divisions use web requisition systems for approvals, but then the information has to be entered into a different Fermi REQ system by hand. Such an approach is time consuming and prone to errors.

Consistently with the above, the 2021 PEMP criticized the acquisition strategy/planning, the engagement, market research, the selection of subcontract type, the evaluation criteria, subcontract options, and subcontract administration, as well as the quality and substance of procurement packages (more details are in Appendix B). The 2021 PEMP also states that "the laboratory has had difficulties complying with their DOE approved procurement policies/procedures and DOE requirements". Despite a B grade, for Objective 6.2 the 2023 PEMP mentions a FY 2023 Procurement Evaluation and Reengineering Team (PERT) Report to which the lab was responsive. The hiring of 27 new Procurement employees, thanks to whom a backlog of requisitions was greatly reduced, was praised. However, with implementation of automated processes and solutions, the number of new hires could have been presumably reduced accordingly.

**RECOMMENDATIONS**

- An out-of-the-box potential solution to some of these financial problems would be for the lab to request that the DOE send an experienced team of their own accountants to assume responsibility for reorganizing the operational support side of the lab, at least until the major issues are resolved. Alternatively, the DOE could select an M&O contractor who is experienced, knowledgeable, motivated and capable of investing in solutions to these current accounting challenges and deficiencies.

- One root cause for the Procurement issues is efficient handling of the complex regulations. A legal review by an outside party of the lab Procurement's boiler plate materials, policies, and procedures could be performed. The review panel should include experts from other labs, business systems consultants, contract writing experts, etc. The purpose of the review should be to establish cost-effective processes within the present DOE rules and regulations.

- Vendors do not just interact with the Procurement teams in the Procurement Office, but apparently also with the Office of General Counsel. This is the same Office that continues to show ineffectiveness in handling serious complaints (see Section II.F), which sometimes escalate in lawsuits. They are presently doing a similarly ineffective job when interpreting regulations, and negotiating terms and conditions with vendors. <u>The former General Counsel left the lab a year ago and there is still time to replace him with an effective leader</u>.

- It may be instructive to check out the Procurement protocols and models followed within the European Union, as described here [https://acer.europa.eu/sites/default/files/documents/en/The_agency/Public_Procurement/OLAF_DOCS/Doc%2014%20Vademecum-public-procurement-Nov%202015-11_updated%20Feb%202016.pdf](https://acer.europa.eu/sites/default/files/documents/en/The_agency/Public_Procurement/OLAF_DOCS/Doc%2014%20Vademecum-public-procurement-Nov%202015-11_updated%20Feb%202016.pdf)

- <u>To reduce the rate of indirect cost, the number of bureaucrats and administrators on the lab payroll has to be optimized and closely controlled through implementation of automated processes and solutions in Operations</u>. This is critical for the lab not to lose itself in heavy bureaucracy and lose track of its Physics vision!

- Also, <u>the number of management positions have to be reduced over the whole lab</u> to reduce the indirect cost. For instance, the costly extra layer of ALDs should be replaced with a maximum of two Associate Lab Directors, one combining "Accelerators and APS-TD" and the other encompassing "Particle Physics", "Computational Science/AI" and "Emerging Technologies".

## II.F       A More Modern HR/ General Counsel System

As mentioned above, the climate at the lab was assessed as extremely poor by the latest survey conducted by an independent organization.[6] Effective leadership would recognize the problem quickly and move beyond legal compliance to address the culture of the lab. <u>The "culture" of any given group or organization is defined by the set of human values that the group selects as a foundation for their behavior</u>.

It is a fundamental concept that no organization can ever succeed with unproductive/inept leaders. State-of-the-art hiring processes, as described in Section II.A, are critical to maintain knowledge, experience and human quality of employees promoted to manager positions. In addition to providing competent and productive leaders, transparent and merit-based promotion and hiring processes improve an organization's effectiveness by building the necessary trust. When ineffective management is compounded with old-school HR/General Counsel partisanship to the institution, this is conducive to harassment/abuse of power and retaliation, as well as complaints filed with the EEOC. Some of these result in lawsuits. As succinctly quoted in the testimonial from Appendix E, "FRA actions have left the impression on myself and others that they prioritize maintaining a positive "reputation" by trying to make problems go away at the expense of victims over cultivating a safe working environment for Fermilab employees and Users."

We expound in the following on how to reform the whole HR/ General Counsel system towards more modern processes that reduce lawsuit risk while improving the culture and climate of the organization.

### II.F.1 – The Culture of an Organization

An equitable and just community does not spontaneously occur without effort. (https://arxiv.org/ftp/arxiv/papers/2209/2209.06755.pdf) In order to implement ethical procedures, it is necessary to understand the mechanisms by which discrimination occurs. The most prevalent form of discrimination is gender discrimination and retaliation. Since 2008, the Equal Employment Opportunity Commission (EEOC) has reported that retaliation is the most common discrimination finding in federal sector cases (https://www.eeoc.gov/retaliation-making-it-personal). The most powerful deterrent of discrimination is <u>organizational climate— the degree to which those in the organization perceive that misconduct of any form is or is not tolerated</u>.

---

[6] We cannot provide a reference or actual results of the latest survey because FNAL management has decided to keep this a secret. The answer that employees receive when inquiring about a reference or a link to the latest climate survey is the following: "Unfortunately, the information is not available at this time, and there are no plans to make it accessible in the future."

Institutions can take concrete steps to reduce discrimination of any sort by making systemic changes that demonstrate how seriously they take this issue and that reflect that they are listening to those who speak up. This is in contrast with the policies and procedures that aim (and often ineffectively) to protect the liability of the institution but are not effective in preventing misconduct. Hierarchical power structures with strong dependencies on those at higher levels are more likely to foster and sustain sexual harassment, bullying, and other forms of discrimination. This is exacerbated in the aforementioned "rigid organizational structures," where power is concentrated in single individuals. "Organizations that foster a climate of aggression and bullying are more likely to have managers who abuse power and retaliate when claims are made. Other organizational factors that influence retaliation are: a lack of administrative policies discouraging retaliation; an authoritarian management culture; overly hierarchical organizations, where rank or organizational level is prized; high levels of task-related conflicts; reward systems and structures that promote competition; and the ability to isolate the accuser."

A textbook case study of all of the above is provided by the 2021 lawsuit described in (https://www.theguardian.com/education/2022/oct/12/former-ucl-academic-to-pay-damages-after-harassing-colleague-for-months), which was settled in favor of the plaintiff by a high UK court. The U.S. female academic had reported an alleged sexual assault by a former UK collaborator to Fermilab HR, since the lab is the U.S. host lab of their joint international collaborations NOVA and DUNE. After no finding of fact was made against him by Fermilab, a long series of retaliatory events occurred, which produced significant distress on the victim that has affected all aspects of her personal and professional life, including leaving the field. The story told by the plaintiff herself is in Appendix E. Her case includes a majority of textbook factors that make retaliation thrive in organizations that have insufficient oversight, such as described above.

Another representative example of the flaws of the current system was exposed in the 2021 lab settlement with the EEOC on a claim that it had "denied promotion to a female engineer in retaliation for complaining about discrimination". The plaintiff had applied for an internal position that was given to a less qualified (even if female) individual. During the multi-year duration of this process, the former General Counsel et al. were not able to de-escalate the process through productive negotiation before the victim filed with the EEOC in 2016. They were unable to do so also after the EEOC successfully filed suit on her behalf. It is worth noting that the same employee who was discriminated against had also been one of the experts warning her division management of the risk of Beryllium windows blasting, due to poor engineering calculations and poor reviews. Through a convoluted process, a new policy was put in place for Beryllium windows' safety, and transferred to the newly promoted manager. When the failure of a Beryllium window indeed occurred shortly thereafter, the new policy was made to disappear. The person in charge was later further promoted, whereas the expert who tried to help the lab is still at the same engineering level.

When approached on these topics in February 2023, the former lab's COO Scott Tingey replied: "I can tell you that the laboratory takes concerns seriously and has worked diligently to improve and enhance its processes and to communicate those to the laboratory community." Merely verbally advocating for a cause or issue to gain attention, support, or profit rather than caring about making a difference in the cause is deleterious to any organization (https://mocostudent.org/2021/11/why-is-performative-activism-a-problem/ ; https://tulanehullabaloo.com/58978/intersections/opinion-performative-dei-initiatives-are-far-from-productive/).

On December 9, 2022, within an all-hands presentation, Deputy Director Bonnie Fleming presented slides (reported in Appendix F) on an allegedly new "Fermilab Concern Reporting System". On March 3, 2023 a critical analysis (reported in Appendix G) of that system was emailed to her. Since then, we have gathered dozens of testimonials that attest to the contrary of most of those performative statements and promises. Some of these testimonials are in Appendix H. In this re-edited version of this paper, we added Section II.F.2 below on how these reporting systems historically work, based on the book "Rules for Whistleblowers" by Stephen Martin Kohn, ESQ. The critical analysis in Appendix G was performed just based on rational thinking and common sense, as it was written well before we learned about the laws and legal precedents from that book.

In Appendix H we relate recent events associated to a series of cover-ups in the Infrastructure Services Division (ISD). These cover-ups encompass, among others, guns on site and serious attempts at hurting female employees.

At the end of calendar year 2023, the lab established a new policy by which any employee, including scientists, can be given a disciplinary leave of absence without pay that most of the time ends in firing. That we know of, this policy was applied to half a dozen people in the first couple of months from its inception. Our first reaction when we learned of this new policy was that of hope, finally, for positive change. Unfortunately, it does not seem that this policy was created to discipline bad actors, as <u>it was used to fire witnesses of cover-ups, as we will show in Appendix H</u>.

**II.F.2 – Internal Reporting Systems**
We summarize here a couple of chapters from the book "Rules for Whistleblowers" by Stephen Martin Kohn, ESQ.

<u>Know the Limits of "Hotlines", pp. 65-69</u>
*"The court held that... Complainant's internal safety and quality complaints "were not protected by the statute." Accordingly, the complaint in this case is DISMISSED.*
*-Final order of the secretary of labor dismissing the whistleblower case of Ronald Goldstein*

Every major employer knows that employee whistleblowers are the single most important source of information in your company, and you need reasonable and effective channels for information to be disclosed and investigated. As a consequence, there has been a worldwide proliferation of internal reporting programs. They usually start with a hotline that urges employees who witness misconduct to place a confidential phone call to a responsible agent. Thereafter, a compliance department supposedly independently investigates the concern.

There's just one hitch. Is the hotline truly independent? Can it really keep callers' identities confidential? Will there be a proper investigation? Is contacting the internal compliance group really the right thing to do? As explained in Rule 11, <u>corporate compliance programs that are managed by attorneys working for the company are completely compromised and can lawfully throw whistleblowers under the bus</u>. But there are other problems inherent whenever a corporation tries to police itself that can land an honest employee in hot waters.

Before you use a company's internal compliance program, make sure that any such contacts are protected under law. Regardless of company propaganda, corporations regularly go into court and argue that internal reporting is NOT a protected activity and that employees who raise concerns to their managers CAN BE FIRED." Clearly, this does not apply to lab's Users, and we have received testimonials of at least one case of User receiving justice.

<u>Don't Let the Lawyers Throw You Under the Bus, pp. 71-75</u>
"When you talk to an attorney for the company, that lawyer does not represent you. They are not your friend. They work for the boss – and the boss may want to throw you under the bus and blame you for the violations they sponsored, planned, or turned their back on.

Corporate lawyers will try to convince you to help the company, even if the company ultimately decides to blame you for a violation or turn you in to the government for prosecution. They will not advise you of your whistleblower right. Corporate counsel's role in compliance investigations can be downright devious. A typical example is to invoke corporate attorney-client privilege not to release the information provided by employees during an internal compliance investigation. In companies whose internal reporting systems report to the company's general counsel, all of its documents, including evidence of fraud, abuse, etc. provided to the company via internal whistleblowers, can be confidential.

The ability for some companies to twist compliance programs to serve their own interest is far greater than simply hiding information. A corporate client can always decide to waive the privilege and release the attorney-client information any time it is to their advantage, for instance to discredit the whistleblower and/or the victims. <u>In short, an internal reporting system can become a prime enabler for corporate crime instead of promoting ethical corporate behavior.</u>"

"CONCLUSION
*Question: Are corporate compliance and ethics programs just window-dressing?*
*Answer: In many companies, probably yes.*
*- Donna Boehme, former Chief Compliance Officer, BP*"

**ADVICE FOR CONTACTING HOTLINES/ INTERNAL REPORTING SYSTEMS**
"Before contacting a corporate compliance program, employees should carefully consider whether the program is INDEPENDENT or whether it can act as a Trojan Horse and be used to undermine your rights. Steps to take are as follows:

1. Ensure the contact is protected under law.
Are the internal reports you want to make protected under law? If not, find another way to raise your concerns.

2. Research the in-house program.
Try to first obtain information about the program's reputation. If the company has a history/reputation of retaliating against whistleblowers, contact the program at your own risk.

3. Document everything.

4. Don't rest on laurels.
Whistleblowers CANNOT ASSUME THE SYSTEM WILL WORK. It is common for workers to first raise a concern with a supervisor, then elevate the concern to an internal compliance department, and finally file a concern with a government agency. But simply relying on a hotline investigation to fix a problem pro provide protection is naive.

5. Don't take legal advice from a compliance officer.
Compliance officers and hotline investigators work for the company; they do not work for the employees. They are under no obligation to provide employees with complete or accurate advice, nor to inform employees of their rights or the law that may protect them.

6. Be skeptical about confidentiality.
Many hotlines promise confidentiality, but it may be easy for the employer to figure out the identity of the whistleblower.

7. <u>Avoid programs managed by corporate attorneys</u>
Whistleblowers should investigate who manages the compliance program. Programs that report directly to a company's chief executive officer or to an independent audit committee tend to have more integrity than programs that report to, or are managed by the company's general counsel. Company attorneys focus on protecting employers from lawsuits, NOT FIXING PROBLEMS. Programs that report to the office of general counsel or other company attorneys SHOULD BE AVOIDED or approached with extreme caution.

8. Don't take advice from the company attorney.
They can use your disclosures to "throw you under the bus" (see in the following).

9. Give them the rope to hang themselves.
The failure of a corporation to properly investigate a hotline concern can constitute evidence of a cover-up and evidence that a company was hostile to the whistleblower. Sometimes hotline records can be obtained as part of pretrial discovery.

10. Blow the whistle on the lack of proper internal controls.
A publicly traded company that lacks effective internal controls necessary to detect securities fraud, ensure accurate corporate disclosure, and detect improper payments may be in violation of securities laws."

## RECOMMENDATIONS (IN ADDITION TO THE ABOVE)

- <u>The culture of an organization can be changed</u>. To be effective, the lab has to move beyond legal compliance, and improve transparency and accountability. When institutions do not have the means or neglect to enforce disciplinary sanctions as necessary, it is called "institutional betrayal," because it causes additional harm to the victims [Carly Parnitzke Smith and Jennifer J. Freyd. "Dangerous safe havens: Institutional betrayal exacerbates sexual trauma. Journal of Traumatic Stress", 26:119–124, 2013], as in the case of Appendix E. To effect positive change, policies have to be written in such a way that the process of dealing with violations/imposing sanctions is clearly defined and enforceable.

- An enforcement policy that includes equitable standards of fact-finding, clear-cut consequences, and adjudication that is unbiased and equal for all is necessary. The recommendations of the National Academy of Sciences (https://doi.org/10.17226/24994) and the EEOC include:
  1. Creating diverse, inclusive, and respectful environments.
  2. Addressing the most common form of sexual harassment: gender harassment.
  3. Moving beyond legal compliance to address culture and climate.
  4. Improving transparency and accountability.
  5. Diffusing the hierarchical and dependent relationship between employees and managers.
  6. Providing support for the victim.
  7. Striving for strong and diverse leadership.
  8. Making the entire community responsible for reducing and preventing harassment and discrimination.

- Because of the demonstrated conflict of interest by the lab HR/General Counsel system, we propose that complaints to HR about abuses (as defined by the National Labor Relations Board) not be handled by the lab, but by an independent entity paid by the DOE Site Office. An option is that its cost would be deducted from the M&O management fee. The independent entity should be a law firm with labor lawyers, with the provision that FNAL lawyers be EXCLUDED from any investigation. Money and issues concerning complaints against the lab already have to legally pass through the DOE Site Office. This is done to take the lab directorate out of the loop until DOE has had independent, third-party fact-finding.

- Alternatively, the DOE could select an M&O contractor who is experienced, knowledgeable, motivated and capable of investing in solutions to these current HR and legal challenges and deficiencies.

## III. Conclusions

Our main conclusion is that without a complete renewal of its human and professional values, the HEP community is destined to die out because hiding problems under the rug has brought our lab to a level of ineffectiveness bordering extinction. It is unlikely that an unethical entity will suddenly become interested in principles. This kind of failings made us realize that no field of science is more important than human dignity and rights, i.e. NOTHING should be worth unethical and/or degrading behavior within a civilized community. <u>Only by exposing the truth Fermilab has a chance to heal and ideally reprise its former glory.</u> The latter is the main reason why we decided to come forward to the taxpayers, against the opposite view that exposing the truth would instead kill the lab. This is a call for action by those in charge now and in the future.

# Appendix A – A 360 Performance Review Process

The current yearly Performance Review process is based on the performance evaluation of each employee by his/her direct supervisor followed by the approval of the supervisor's supervisor. This approach cannot be objective for the following reasons:
1. With a matrix system, the direct supervisor may not be aware of a person's real role and contribution to the program/project if the former is not part of it. They are supposed to ask the program leader but they may not be interested to do it for various reasons;
2. The performance rating is usually averaged within each group and if the group is small and consists of various categories (scientists, engineers, techs, etc.), the rating is even less objective;
3. The performance rating depends significantly on the personal relationship between employee and direct supervisor, and between the two supervisors. It is the weakest point of the present system;
4. A useful case study is described in the following. Upper managers can downgrade, in retaliation, the rating of a yearly performance review against the will of one's direct supervisor. When one of us presented this inconsistency to the general counsel in 2019, "Those are our procedures" was the answer. There is currently no provision to prevent the same people who abuse an employee be in charge of their rating. One has to accept the downgrading of their rating, and then go through an alleged "grievance" process. Those of us who have used this process know that ii is a charade, as all the managers responsible at each of the four review steps of the process side with each other.

<u>The new system should include instead the following:</u>
1. The evaluation should be done at 360 degree, i.e. not just by the supervisors but also by peers and subordinates. Each would answer a specific set of questions, such as ability to communicate, greatest contributions, best qualities, any blunders, weakest points, etc.
2. The evaluation of group leaders, department heads, etc., including division directors and their deputies, should include the opinion of their employees, using a separate set of questions or table.
3. The performance should be reviewed, evaluated and compared for the specific job category (Scientist, ENG, TECH, etc.) within each division.
4. For the scientific category, a special lab committee with the most experienced members should be considered for evaluation of the employee's performance report and supervisor's recommendation. Then, the result has to be approved by the division (directorate) head.
5. The final results of the performance review should be public.
6. All promotions should be based on the results of the performance reviews. The proposed promotion list with brief justifications should also be public and available before any promotion for comments from the community.

# Appendix B – The Lab's Performance Evaluation and Measurement Plan (PEMP)

Areas with insufficient grades are highlighted in yellow below.

## Analysis of 2019 PEMP (B+ Grade)

| ITEM | DESCRIPTION | OWNER | RATING |
|---|---|---|---|
| **Goal 1.0** | **Provide for Efficient and Effective Mission Accomplishment (30%)** | **Director** | **A-** |
| Objective 1.1 | Provide Science and Technology Results with Meaningful Impact on the Field | CRO | B+ |
| Notable Outcome | HEP: By Feb. 2019, the lab should provide specific responses (including action plans as appropriate) to the lab-specific recommendations provided in the 2018 HEP comparative reviews of the General Accelerator R&D, Intensity Frontier, and Theoretical Research programs. | Lykken | Achieved |
| Objective 1.2 | Provide Quality Leadership in Science and Technology Results that Advances Community Goals and DOE Mission Goals | CRO | A- |
| **Goal 2.0** | **Provide for Efficient and Effective Design, Fabrication, Construction and Operations of Research Facilities (45%)** | **Director** | **B+** |
| Objective 2.1 | Provide Effective Facility Design(s) as Required to Support Laboratory Programs (i.e., activities leading up to CD-2) | CPO | A |
| Notable Outcome | HEP: Complete final design for LBNL Far-Site Conventional Facilities, including all drawings, models, and specifications. | Mossey | Achieved |
| Notable Outcome | HEP: Submit a completed international contributions plan for PIP-II to HEP before the CD-2 review. | Merminga | Achieved |
| Notable Outcome | HEP: By Dec. 2018, have a viable plan for the inclusion of the MIP Timing Detector (MTD) in the HL-LHC CMS project, within the funding constraints provided by the program, consistent with the larger international plan for the final design choice, and have the subproject fully staffed with people who have demonstrated both project management and technical expertise or drop the MTD from the project baseline. | Tschirhart | Achieved |
| Objective 2.2 | Provide for the Effective and Efficient Construction of Facilities and/or Fabrication of Components (execution phase, post CD-2 to CD-4). | CPO | B+ |
| Notable Outcome | BES: Execute the assigned LCLS-II project scope in compliance with the technical performance | Tschirhart | Achieved |
| | specifications and within the established DOE performance goals for cost and schedule. Performance will be assessed based on the work planned and accomplished during FY 2019, not on the cumulative performance of the project. | | |
| Objective 2.3 | Provide Efficient and Effective Operation of Facilities | Director | B+ |
| Notable Outcome | HEP: By Jan. 2019, submit transition to operations plans, including externally vetted operations budgets, for Mu2e and SBN. | Frieman/ Brice | Achieved |
| Notable Outcome | HEP: By July 2019, complete the Accelerator and Detector Modernization Reviews and submit the completed reports to DOE. | Lindgren | Achieved |
| Objective 2.4 | Utilization of Facilities to provide impactful S&T Results and Benefits to External User Communities | CRO | |
| **Goal 3.0** | **Provide for Efficient and Effective Science and Technology Program Management (25%)** | **Director** | **B+** |
| Objective 3.1 | Provide Effective and Efficient Strategic Planning and Stewardship of Scientific Capabilities and Program Vision | Director | B+ |
| Notable Outcome | HEP: Working with OHEP, and taking into account Laboratory Optimization guidance, deliver and begin to implement a Cosmic Frontier Strategic Plan by March 2019 that builds on the laboratory's core capabilities in this area. | Frieman | Achieved |
| Objective 3.2 | Provide Effective and Efficient Science and Technology Project/Program/Facilities Management | Director | B+ |
| Notable Outcome | HEP: By June 2019, develop a documented process for determining and prioritizing long-term investments in infrastructure in Batavia and at SURF that provides OHEP with an appropriate long-term and all-inclusive view of coming investments needs. Identify a labwide point of contact to communicate lab priorities for investments to OHEP and SLI. | Meyer | Achieved |
| Notable Outcome | HEP: By Nov. 2018, submit a fully integrated installation and commissioning plan, including a revised budget estimate, for the SBN program. | Brice | Achieved |
| Objective 3.3 | Provide Efficient and Effective Communications and Responsiveness to Headquarter's Needs | Office of Communication | B+ |
| **Goal 4.0** | **Provide Sound and Competent Leadership and Stewardship of the Laboratory** | **Director** | **A-** |
| Objective 4.1 | Leadership and Stewardship of the Laboratory | Director | A- |
| Objective 4.2 | Management and Operation of the Laboratory | COO | B+ |
| Notable Outcome | FSO/SC: FRA must rebuild the FNAL procurement team to support successful project execution, in particular for LBNF/DUNE and PIP-II. | Peoples | Achieved |
| Notable Outcome | FSO/SC: FRA must ensure successful project and procurement management for all projects, in particular LBNF/DUNE and PIP-II. | Tschirhart | Achieved |
| Notable Outcome | FSO/SC: Implement any updated DOE foreign visits and assignments (FV&A) policy requirements, when issued. | Meyer | Achieved |
| Objective 4.3 | Leadership of External Engagements and Partnerships | COO | A- |
| Objective 4.4 | Contractor Value-added | COO | A- |
| **Goal 5.0** | **Sustain Excellence and Enhance Effectiveness of Integrated Safety, Health, and Environmental Protection [30%]** | **CSO** | **B+** |
| Objective 5.1 | Provide an Efficient and Effective Worker Health and Safety Program | CSO | B+ |
| Notable Outcome | FSO: Establish the process map and framework for the Work Planning and Control (WPC) tool and complete development of a functional replacement for the current, hazard-analysis process. | Michels | Achieved |
| Notable Outcome | FSO: Evaluate the implementation of the Human Performance Improvement (HPI) program by assessing how supervisors set expectations for their teams with respect to avoiding critical mistakes through work planning and controls. Include the level of supervisory engagement, recommendations for improvements, and an implementation plan. | Michels | Achieved |
| Notable Outcome | FSO: FRA will develop and implement an assurance plan that demonstrates its ability to assure corrective actions from the 2018 DOE Reviews at SURF, as well as reviews of other high hazard and critical activities and operations at the site, are completed and effective. | Weber | Achieved |
| Objective 5.2 | Provide an Efficient and Effective Environmental Management System | CSO | A- |
| **Goal 6.0** | **Deliver Efficient, Effective, and Responsive Business Systems and Resources that Enable the Successful Achievement of the Laboratory Mission(s) [30%]** | **COO** | **B** |
| Objective 6.1 | Provide an Efficient, Effective, and Responsive Financial Management System | CFO | B |
| Objective 6.2 | Provide an Efficient, Effective, and Responsive Acquisition Management System and Property Management System | CSO | B- |
| 6.2 Part 1 | Acquisition Management System | CFO | |
| Notable Outcome | FSO: Respond to the recommendations identified in the FRA and DOE Procurement Surveys, and perform an effectiveness evaluation of corrective actions. | Peoples | Achieved |
| 6.2 Part 2 | Property Management System | Facilities Eng. Services | |
| Objective 6.3 | Provide an Efficient, Effective, and Responsive HR Management System and Diversity Program | WDRS | B+ |
| Objective 6.4 | Provide Efficient, Effective, and Responsive Contractor Assurance Systems, including internal Audit and Quality | WDRS | B+ |
| 6.4 Part 1 | Contractor Assurance Systems and Quality | COO | |
| 6.4 Part 2 | Internal Audit | FRA Internal Audit | |
| Objective 6.5 | Demonstrate Effective Transfer of Knowledge and Technology and the Commercialization of Intellectual Assets | OPPT | A- |
| **Goal 7.0** | **Sustain Excellence in Operating, Maintaining, and Renewing the Facility and Infrastructure Portfolio to Meet Laboratory Needs [25%]** | **COO** | **A-** |
| Objective 7.1 | Manage Facilities and Infrastructure in an Efficient and Effective Manner that Optimizes Usage, Minimizes Life Cycle Costs and Ensures Site Capability to Meet Mission Needs | Facilities Eng. Services | A- |
| Notable Outcome | FSO: Modify current strategic budget and program interface processes to address emerging infrastructure and operational needs, including development and documentation of appropriate Earned Value Management and oversight procedures for General Plant Projects and Accelerator Improvement Projects. | Tschirhart | Achieved |
| Notable Outcome | FSO/SC-33: Develop a plan with sufficient detail including options and cost to address the over 80 buildings identified as excess. | Ortgiesen | ??? |
| Objective 7.2 | Provide Planning for and Acquire the Facilities and Infrastructure Required to Support the Continuation and Growth of Laboratory Missions and Programs | Office of Campus Strategy & Readiness | B+ |
| **Goal 8.0** | **Sustain and Enhance the Effectiveness of Integrated Safeguards and Security Management (ISSM) and Emergency Management Systems [15%]** | **Director** | **B+** |
| Objective 8.1 | Provide an Efficient and Effective Emergency Management System | CSO | B+ |
| Notable Outcome | FSO: Identify necessary enhancements in the Emergency Operation Center that will improve the | Michels | Achieved |
|  | room's ease of use, reliability and functionality, and implement solutions. |  |  |
| Objective 8.2 | Provide an Efficient and Effective Cyber Security System for the Protection of Classified and Unclassified Information | CIO | B+ |
| Objective 8.3 | Provide an Efficient and Effective Physical Security Program for the Protection of Special Nuclear Materials, Classified Matter, Classified Information, Sensitive Information, and Property | CSO | B |
| 8.3 Part 1 | Part 1: Physical Security Program for the Protection of Special Nuclear Material and Property | CSO |  |
| Notable Outcome | FSO/SC-33: Develop an effective and actionable implementation plan addressing the new design basis threat requirements that is approved by DOE by March 31, 2019. | Meyer/ Michels | Achieved |
| 8.3 Part 2 | Part 2: Classified Matter, Classified Information, and Sensitive Information | CIO |  |

**Goal 6.0: Deliver Efficient, Effective, and Responsive Business Systems and Resources that Enable the Successful Achievement of the Laboratory Mission(s)**

*This Goal evaluates the Contractor's overall success in deploying, implementing, and improving integrated business systems that efficiently and effectively support the mission(s) of the Laboratory.*

**Site Office**
**Score:** 3.0    **Grade:** B
**Goal Evaluation:**
[…] continued focus is essential to consistently deliver quality products which are aligned with all Federal acquisition requirements and address strategic needs.
[…]
Human resources actions to develop and promulgate the climate survey to support management efforts to better understand opportunities for staff needs and to develop Diversity and Inclusion activities to build this program are both laudable initiatives.
[…]

**Objective 6.1: Provide an Efficient, Effective, and Responsive Financial Management System**
**Weight:** 20.0%
**Score:** 3.0    **Grade:** B
**Objective Evaluation:**
Overall, FRA had mixed performance against aspects of this objective. The financial management controls were determined to be satisfactory, however, assessments conducted during the year indicated that some weaknesses still exist. Several deficiencies in FRA's financial systems were identified by the 2019 Financial Management Assessment conducted by an independent review team, which exposed several risks in the Financial Management System. The audit findings identified control weaknesses associated with accounting not consistently adhering to the published General Payables and Disbursements Policies and Accounting procedures to ensure all payments have authorized support and documentation prior to payment. Corrective actions are underway to address the audit findings and efforts should include verification that that goods and services purchased and delivered match the actual contract requirements, invoices are approved in a timely manner as to not violate Prompt Payment Act, and that authorized approval of invoices occurs.
[…]

**Objective 6.2: Provide an Efficient, Effective, and Responsive Acquisition Management System and Property Management System**
**Weight:** 30.0%
**Score:** 2.5    **Grade:** B-
**Objective Evaluation:**
FRA has spent an inordinate amount of management capital to address deficiencies on the acquisition management system to drive improved performance. Improvement plans have been developed, implemented and tracked by all levels of management. FRA is progressing and is about halfway through completion of nearly 100 identified improvement items. Completion and verification of these items should result in improvements. The improvements to date have focused on the fundamental building blocks; personnel, training, processes, procedures, metrics and roles and responsibilities in the procurement organization. These foundational elements are beginning to reap benefits and will be important as a robust acquisition program is put in place. The progress is substantial, but there is a long way to go. The level of complexity of the procurement work and the continued legacy of existing actions that were initiated prior to corrective actions being put in place, continues to result in products that require significant Site Office involvement to meet product quality and project schedule requirements.
Site Office
**(Objective 6.2) Notable 1:** Respond to the recommendations identified in the FRA and DOE Procurement Surveys, and perform an effectiveness evaluation of corrective actions.
*Outcome*: This notable was addressed by a series of actions and corrective action reviews by the Chief Financial Officer. Monthly Status meetings with the Site Office provided reviews of all action categories and documentation of completion for corrective actions. Independent evaluation of the effectiveness of initial actions was conducted by a variety of actions and series of review teams. Actions to close out follow-on actions continue, but actions to date meet the recommendations in the initial Procurement.

**Objective 6.3: Provide an Efficient, Effective, and Responsive Human Resources Management System and Diversity Program**
**Weight:** 20.0%
**Score:** 3.4    **Grade:** B+
**Objective Evaluation:**
[…] One area that has the greatest need for improvement is representation of women in the area of Technical Research Staff.
[…]
FRA's current representation of women in the area of Technical Research Staff is abysmal. FRA needs to make a concerted effort to focus on this as part of its broader efforts.

**Goal 8.0: Sustain and Enhance the Effectiveness of Integrated Safeguards and Security Management (ISSM) and Emergency Management Systems**

*This Goal evaluates the Contractor's overall success in safeguarding and securing Laboratory assets that supports the mission(s) of the Laboratory in an efficient and effective manner and provides an effective emergency management program.*

**Site Office**
**Score:** 3.2    **Grade:** B+
**Goal Evaluation:**
FNAL continues to struggle with a number of legacy weaknesses in security related compliance despite significant improvements in the organization, operation and improved responsiveness of the security section and plans to utilize technology and data analysis to make the most of limited S&S funding.
[…]

**Objective 8.1: Provide an Efficient and Effective Emergency Management System**
**Weight:** 30.0%
**Score:** 3.2    **Grade:** B+
**Objective Evaluation:**
The Emergency Management program strives for continuous improvement to ensure emergency prevention practices are in place and emergency response capabilities are adequate. […]
The Emergency Management team conducted a number of drills to ensure adequate response preparedness. Those efforts included three tabletop EOC personnel training and drills and participation in DOE's quarterly accountability drills. FRA also utilizes real events, such as the recent Tornado warning to activate the EOC as a proactive response to a possible emergency situation. This approach not only allows for the evaluation of EOC capabilities and shortcomings, but also other emergency shelter areas. As a result of this activation, deficiencies were noted within several shelter areas. These deficiencies were documented in an after action report for resolution of those deficiencies.
[…]
Facility conditions in some legacy areas remain a concern for emergency planning and response. ES&H will continue to work with FESS to evaluate facility conditions and identify required improvements.
Site Office
**(Objective 8.1) Notable 1:** Identify necessary enhancements in the Emergency Operations Center that will improve the room's ease of use, reliability and functionality, and implement solutions.
Outcome: FRA finished an evaluation of enhancements necessary to ensure the Emergency Operations Center is able to provide leadership an environment that is reliable, functional and supports on-scene personnel, and completed numerous upgrades to improve conditions including equipment relocation communication system upgrades and modification of process documents and job aids. These efforts included benchmarking three other EOC sites to evaluate observe and understand best practices, lessons learned and the physical layout of other facilities. The actions taken were derived from a combination of the best practices from the other facilities and self-assessment of gaps and needs at the Fermilab EOC.

**Objective 8.2: Provide an Efficient and Effective Cyber Security System for the Protection of Classified and Unclassified Information**
**Weight:** 35.0%
**Score:** 3.4    **Grade:** B+
**Objective Evaluation:**

FRA has worked with DOE to develop needed cyber security infrastructure to allow for the safe transmission and storage of critical scientific data.
[…]
The FRA management continues to be open to advancing the cyber security posture; ensuring employees, visitors and users are able to do their work while not impacting the security of the site. FRA is carefully approaching DOE mandated requirements to add specific cyber technologies to the portfolio. The FRA reputation would benefit from recognizing where risks of noncompliance outweigh other considerations. Legacy systems retirement plans should be developed.

**Objective 8.3: Provide an Efficient and Effective Physical Security Program for the Protection of Special Nuclear Materials, Classified Matter, Classified Information, Sensitive Information, and Property**
**Weight:** 35.0%
**Score:** 3.0   **Grade:** B
**Objective Evaluation:**
FRA continued to take a more proactive, disciplined and reasonable approach to improving security operations in FY19. Continuing modifications to organization and personnel within security operations, FRA continues to improve communications with DOE on security matters. FRA has made incremental improvements; especially through the use of enhanced visual coverage of remote sites and areas of high personnel concentrations and traffic. However, incremental improvement was not enough to make up for significant challenges identified by the DOE S&S Survey. Several issues with DOE Order compliance were identified that will require advancing the improvement process. Additional expansion of the ShotSpotter system is underway enabling security operations to assess and observe areas identified by the ShotSpotter system. Security management continues to optimize and better integrate numerous security, monitoring, and alarm systems to improve efficiency in the communications center.
FRA has continued to demonstrate ineffective oversight of the security subcontract, which has also resulted in the lack of fulfillment of all security requirements. FRA has initiated actions to re-compete the subcontract.
[…]
Site Office
**(Objective 8.3) Notable 1:** Develop an effective and actionable implementation plan addressing the new design basis threat requirements and submit to DOE by March 31, 2019.
<u>Outcome</u>: The Fermi Site Office reviewed the submittal of the Design Basis Threat Implementation Plan and determined it met the intent of the Department's requirements specifically the Plan directly or indirectly; provides a current listing of Departmental assets based on Protection Level (PL) 1 through PL-7 assets, Provides the current status of protective measures; Identifies and addresses potential changes in site missions; Identifies tasks necessary to achieve full implementation; and Establishes realistic and measurable milestones. The strategy defined in the Plan builds on multiple assessments and a comprehensive Hazard Analysis to identify and mitigate risk and ensure continual program improvement. This approach addresses the unique challenges presented by the open nature of the site, the sheer size and location of the laboratory, and the highly cooperative and supportive of the local community, through innovative and technology-based solutions.

# Analysis of 2021 PEMP (B Grade)

| ITEM | DESCRIPTION | OWNER | RATING |
|---|---|---|---|
| **Goal 1.0** | **Provide for Efficient and Effective Mission Accomplishment (30%)** | **Director** | **A** |
| Objective 1.1 | Provide Science and Technology Results with Meaningful Impact on the Field | CRO | A |
| Notable Outcome | HEP: Contribute to establishing the synergistic research program and deliver impactful science from the Fermilab-led QIS Center, as measured by the FY 2021 annual report, research publications and highlights, and participation in periodic conference calls. | Lykken | Achieved |
| Objective 1.2 | Provide Quality Leadership in Science and Technology Results that Advances Community Goals and DOE Mission Goals | CRO | A |
| **Goal 2.0** | **Provide for Efficient and Effective Design, Fabrication, Construction and Operations of Research Facilities (45%)** | **Director** | **B-** |
| Objective 2.1 | Provide Effective Facility Design(s) as Required to Support Laboratory Programs (i.e., activities leading up to CD-2) | CPO | C |
| Notable Outcome | HEP: Complete the Option 1A scope of work and start the LBNF Excavation work by July 31, 2021. | Mossey | Achieved |
| Notable Outcome | HEP: Pass an Independent Project Review for the LBNF/DUNE Project in the second half of FY 2021 with all charge questions satisfactorily answered and a clear set of recommendations to address for CD-2 readiness. | Mossey | Not Achieved |
| Notable Outcome | BES: Effectively manage and execute the assigned LCLS-II project scope in accordance with DOE Order 413.3B, in compliance with the technical performance specifications, and within the established DOE performance goals for cost and schedule. Performance will be assessed based on the work planned and accomplished during FY 2021, not on the cumulative performance of the project. | ? | Achieved |
| Objective 2.2 | Provide for the Effective and Efficient Construction of Facilities and/or Fabrication of Components (execution phase, post CD-2 to CD-4). | CPO | B+ |
| Notable Outcome | BES: Effectively manage and execute the assigned LCLS-II project scope in accordance with DOE Order 413.3B, in compliance with the technical performance specifications, and within the established DOE performance goals for cost and schedule. Performance will be assessed based on the work planned and accomplished during FY | ? | Achieved |
| | 2021, not on the cumulative performance of the project. | | |
| Objective 2.3 | Provide Efficient and Effective Operation of Facilities | Director | A- |
| ~~Objective 2.4~~ | ~~Utilization of Facilities to provide impactful S&T Results and Benefits to External User Communities~~ | ~~CRO~~ | |
| **Goal 3.0** | **Provide for Efficient and Effective Science and Technology Program Management (25%)** | **Director** | **C** |
| Objective 3.1 | Provide Effective and Efficient Strategic Planning and Stewardship of Scientific Capabilities and Program Vision | Director | C+ |
| Objective 3.2 | Provide Effective and Efficient Science and Technology Project/Program/Facilities Management | Director | C |
| Objective 3.3 | Provide Efficient and Effective Communications and Responsiveness to Headquarter's Needs | Office of Communication | C |
| **Goal 4.0** | **Provide Sound and Competent Leadership and Stewardship of the Laboratory** | **Director** | **B** |
| Objective 4.1 | Leadership and Stewardship of the Laboratory | Director | B |
| Notable Outcome | Effectively and efficiently administer the PIP2/LBNF/IERC subcontracts (including KAJV and TMI). | ? | Not Achieved |
| Objective 4.2 | Management and Operation of the Laboratory | COO | B |
| Notable Outcome | The Laboratory must keep senior SC leadership informed of key events (e.g., VIP/protocol visits, news releases, media requests) through timely population of the Science News Dashboard with all the relevant information on such activities and/or through other appropriate mechanisms. | ? | Achieved |
| Objective 4.3 | Leadership of External Engagements and Partnerships | COO | B+ |
| Notable Outcome | Benchmark key FRA business systems/tools and business processes against an agreed set of labs to identify and apply best practices/implementation improvements. Develop and submit 5-year strategic improvement plans (FY2022-2026) for the lab's business support systems (including workforce, infrastructure, business software/tools, and major equipment). Develop and position for initial FY2022 funding and investment required for implementation. | ? | Achieved |
| Objective 4.4 | Contractor Value-added | COO | A- |
| Notable Outcome | The Laboratory and contractor leadership must ensure that all communication with interested stakeholders on DOE/SC program | | Achieved |
| | priorities/objectives are communicated in advance to DOE and aligned with DOE/SC goals, strategies, and guidance. | | |
| **Goal 5.0** | **Sustain Excellence and Enhance Effectiveness of Integrated Safety, Health, and Environmental Protection [30%]** | **CSO** | **B+** |
| Objective 5.1 | Provide an Efficient and Effective Worker Health and Safety Program | CSO | B |
| Notable Outcome | FSO: Complete the implementation review of the Accelerator Division (AD) Control Rooms Formality of Operations Tripartite to include the AD Operations Self-Assessment and field work activities to assure actual accelerator control room operations align with internal procedures and requirements. | ? | Achieved |
| Objective 5.2 | Provide an Efficient and Effective Environmental Management System | CSO | B+ |
| Notable Outcome | FSO: Assure Tritium source migration is identified, and corrective measures analyzed and implemented as agreed. | ? | Achieved |
| **Goal 6.0** | **Deliver Efficient, Effective, and Responsive Business Systems and Resources that Enable the Successful Achievement of the Laboratory Mission(s) [30%]** | **COO** | **B-** |
| Objective 6.1 | Provide an Efficient, Effective, and Responsive Financial Management System | CFO | C+ |
| Objective 6.2 | Provide an Efficient, Effective, and Responsive Acquisition Management System and Property Management System | CSO | C- |
| 6.2 Part 1 | Acquisition Management System | CFO | |
| 6.2 Part 2 | Property Management System | Facilities Eng. Services | |
| Notable Outcome | FSO: Develop and implement a comprehensive property stewardship program that includes an accurate inventory of deployed and stored personal property and expectations for accountability around the stewardship by divisions and sections of this personal property. | ? | Achieved |
| Objective 6.3 | Provide an Efficient, Effective, and Responsive HR Management System and Diversity Program | WDRS | B+ |
| Objective 6.4 | Provide Efficient, Effective, and Responsive Contractor Assurance Systems, including internal Audit and Quality | WDRS | A- |
| 6.4 Part 1 | Contractor Assurance Systems and Quality | COO | |
| 6.4 Part 2 | Internal Audit | FRA Internal Audit | |
| Objective 6.5 | Demonstrate Effective Transfer of Knowledge and Technology and the Commercialization of Intellectual Assets | OPPT | A |
| **Goal 7.0** | **Sustain Excellence in Operating, Maintaining, and Renewing the Facility and Infrastructure Portfolio to Meet Laboratory Needs [25%]** | **COO** | **A-** |
| Objective 7.1 | Manage Facilities and Infrastructure in an Efficient and Effective Manner that Optimizes Usage, Minimizes Life Cycle Costs and Ensures Site Capability to Meet Mission Needs | Facilities Eng. Services | B+ |
| Notable Outcome | FSO: Create, deliver, and implement a three-year plan and budget that defines how the laboratory will increase investment in maintenance each year to stabilize deferred maintenance and reach DOE target levels by FY2024. | ? | Achieved |
| Objective 7.2 | Provide Planning for and Acquire the Facilities and Infrastructure Required to Support the Continuation and Growth of Laboratory Missions and Programs | Office of Campus Strategy & Readiness | A- |
| Notable Outcome | FSO: Re-evaluate GPP processes (including requirement identification, pre-planning, acquisition strategies, status and issues tracking, dependency identification, and communication) with the goal of improving transparency, accountability, and expedited delivery. Leverage this insight to expedite the Sanitary Sewer GPP as practical. | ? | Achieved |
| **Goal 8.0** | **Sustain and Enhance the Effectiveness of Integrated Safeguards and Security Management (ISSM) and Emergency Management Systems [15%]** | **Director** | **B** |
| Objective 8.1 | Provide an Efficient and Effective Emergency Management System | CSO | A- |
| Notable Outcome | FSO: Promptly implement a revised Site Security Plan (SSP) and any additional required updates as approved by DOE. Complete directed security actions per DOE approved timelines. | ? | Achieved |
| Objective 8.2 | Provide an Efficient and Effective Cyber Security System for the Protection of Classified and Unclassified Information | CIO | C+ |
| Objective 8.3 | Provide an Efficient and Effective Physical Security Program for the Protection of Special Nuclear Materials, Classified Matter, Classified Information, Sensitive Information, and Property | CSO | B |
| Notable Outcome | FSO/SC-33: Develop a revised comprehensive ATO package(s) (based on an updated set of risk assessments, security plans and security control assessments) to align Fermilab systems as directed by DOE to the new CSPP and DOE risk | ? | Achieved |

| | expectations. Complete actions per DOE approved timelines. | | |
|---|---|---|---|
| 8.3 Part 1 | Part 1: Physical Security Program for the Protection of Special Nuclear Material and Property | CSO | |
| 8.3 Part 2 | Part 2: Classified Matter, Classified Information, and Sensitive Information | CIO | |

**Goal 2.0 Provide for Efficient and Effective Design, Fabrication, Construction and Operations of Research Facilities**
*The Laboratory provides effective and efficient strategic planning; fabrication, construction and/or operations of Laboratory research facilities; and are responsive to the user community.*

**SC High Energy Physics**
**Score:** 2.7      **Grade:** B
**Goal Evaluation:**
The excellent performance of the PIP II, HL-LHC CMS, and HL-LHC Accelerator Upgrade projects was countered by problems on Mu2e and the Long Baseline Neutrino Facility/Deep Underground Neutrino Experiment (LBNF/DUNE). The very large size of LBNF/DUNE weighs down the performance in this goal.
The accelerator complex returned to operation after the COVID-19 shutdown and performed well including set a new beam intensity record.

**Objective 2.1: Provide Effective Facility Design(s) as Required to Support Laboratory Programs (i.e., activities leading up to CD-2)**
**Weight:** 40.00%      **Score:** 1.8      **Grade:** C

**Objective Evaluation:**
The HL-LHC CMS and LBNF/DUNE projects are both at CD-1, although LBNF/DUNE has a CD-3a that will be graded under 2.2. COVID 19 has continued to impact the HL-LHC project both directly and indirectly. The workforce has not yet fully returned to the labs. There have been increased costs to components due to supply change issues. […]
At the January 2021 review LBNF/DUNE project presented their plans to the review committee. The committee found that the project would likely not stay within 150 percent of the top of the CD-1R cost range and recommended preparing for a CD-1R reaffirmation (CD-1RR). At a follow-up review, the committee found the goal of having a CD-1RR review in October to be unrealistic. The committee called for strengthening the project team and adjustments to how the cost range would be developed. The committee also recommended that the planned Work Breakdown Structure (WBS) reorganization be executed prior to the CD-1RR review. The project is addressing all of these recommendations, but concerns with staffing progress, contingency allocation, and lack of clarity in the Near Detector (ND) scope raises concern with the ability to be ready for the Project's stated goal of a CD-1RR review in April 2022. The liquid nitrogen cryogenic system procurement is behind schedule, and the design deliverable from that contract may no longer be delivered prior to the CD-1RR review which would add additional uncertainty to the cost range.
The large cost of LBNF/DUNE clearly weighs heavily on the overall grade and pull it down despite the good performance of HL-LHC CMS.

SC High Energy Physics

**(Objective 2.1) Notable 1:** Complete the Option 1A scope of work and start the LBNF Excavation work by July 31, 2021.
<u>Outcome</u>: Satisfactorily achieved
**(Objective 2.1) Notable 2:** Pass an Independent Project Review for the LBNF/DUNE Project in the second half of FY 2021 with all charge questions satisfactorily answered and a clear set of recommendations to address for CD-2 readiness.
<u>Outcome</u>: The project did not pass the review and will have to undergo a CD-1 reaffirmation review in CY2022 – Not Achieved

**Goal 3.0 Provide Effective and Efficient Science and Technology Program Management**
*The Laboratory provides effective program vision and leadership; strategic planning and development of initiatives; recruits and retains a quality scientific workforce; and provides outstanding research processes, which improve research productivity.*

**SC High Energy Physics**
**Score:** 2.0     **Grade:** C
**Goal Evaluation:**
Fermilab is the primary HEP lab in the nation leading several flagship domestic projects that are in serious trouble in terms of performance, staffing plans and deliverables. Laboratory management should immediately consider balancing resources to the core mission of the laboratory to allow for the flagship HEP programs and projects to thrive and to benefit at the same time the entire HEP community.

**Objective 3.1: Provide Effective and Efficient Strategic Planning and Stewardship of Scientific Capabilities and Program Vision**
**Weight:** 30.00%     **Score:** 2.2     **Grade:** C+
**Objective Evaluation:**
There are a few areas that need attention.
Fermilab is the lead U.S. laboratory for the nationally coordinated CMS program at the Energy Frontier. The program continues to effectively plan and align research and operations activities with the international CMS collaboration at CERN. Proper awareness to the CMS program, particularly by Fermilab upper management, however, is lacking and needs to strengthen at a level commensurate of a host laboratory for the U.S. program.
Core HEP capabilities and commitments seem to be often neglected in favor of new initiatives where the leadership of a High Energy Physics lab is going to take some time to build up. The laboratory deserves credit for responding rapidly to new opportunities to participate in DOE priority areas and initiatives such as QIS and AI/ML. Very recently they supported an LDRD to collaborate with an ARPA-E project on muon catalyzed fusion that allows the Lab to be part of a clean energy project. However, while being open to new research avenues is commendable, the laboratory is also the primary HEP lab in the nation leading several flagship domestic projects that are in serious trouble in terms of performance, staffing plans and deliverables. Laboratory management should immediately consider balancing resources to the core mission of the laboratory to allow for the flagship HEP programs and projects to thrive and to benefit at the same time the entire HEP community.
At the Cosmic Frontier the lab is actively involved in the Snowmass planning process. The laboratory needs to develop a plan for the DES dark energy group to complete their data processing and analysis in a timely manner in order to be able to participate fully and take on leadership roles in DESC collaboration efforts as well as data analysis, to ensure a strong program going forward. They need to ensure the CMB group stays focused on efforts aligned with the HEP portfolio to ensure leading participation in CMB-S4.

**Objective 3.2: Provide Effective and Efficient Science and Technology Project/Program/Facilities Management**
**Weight:** 45.00%    **Score:** 1.9    **Grade:** C
**Objective Evaluation:**
As for Objective 3.2, there are some areas of concern.

Although improved from prior years, additional care is needed by both Fermilab staff and lab's legal counsel in preparing high-quality written instruments such as international Cooperative Research and Development Agreements (CRADAs), including the supporting annexes to these agreements, which are subsequently sent to DOE for processing. Oftentimes provisions in such instruments appear to be copied and pasted from other project or technical documents, leading to structural format and language inconsistencies contemplated for an agreement.

At the Cosmic Frontier, individual experiments are responsive to needs and questions by Cosmic Frontier program manager. There is concern however in how the lab management is developing the Cosmic Frontier efforts for the future to ensure the lab's program makes strong contributions and carries out leadership roles for LSST DESC and on CMB-S4. This is also the case for planning the dark matter experiments for the future and for the lab's role on DarkSide.

**Objective 3.3: Provide Efficient and Effective Communications and Responsiveness to Headquarters Needs**
**Weight:** 25.00%    **Score:** 1.9    **Grade:** C
**Objective Evaluation:**
As for Goal 3.3, there are some serious areas of concern.

The Cosmic Frontier, program manager at HEP has regular meetings with the laboratory Cosmic Frontier and Physics Department management. However, there is some disconnect. The laboratory needs to be more proactive in reporting on how they're carrying out current roles as well as planning for the future. Regular presentations or writeup to the Cosmic Frontier program manager would help (as is done by the other labs). The laboratory program doesn't seem to have an overall Cosmic Frontier plan that is presented to HEP. In addition, although there are biweekly HEP and Fermilab calls, it is unclear the importance or priority of the Cosmic Frontier efforts are within the overall laboratory management since even when a presentation is made in the HEP call, it is left with waiting for HEP to respond and make decisions rather than lay out what the laboratory plans to do. It is not up to HEP to lay out the laboratory's program. HEP has also had to do a lot of the legwork in developing the DarkSide plan and iCRADA, which should be done by the laboratory.

In the area of projects, with the exception of HL-LHC AUP and HL-LHC CMS, other projects carried out at the laboratory show a worrisome lack of transparency in their communication with DOE and often assume an antagonistic and/or codependent attitude. As mentioned above for the Cosmic Frontier program, here too HEP often finds itself observing the projects waiting for HEP to respond to arising issues and make decisions rather than lay out what the laboratory and/or the project plans to do. This often results in lack of pro-activity and unexpected surprises to the stakeholders, particularly international collaborators, and funding agencies. This seems to be a cultural issue common to several projects management teams and the laboratory directorate.

The Chief Project Officer doesn't seem to be empowered enough to oversee some of the most complex projects. Where he is actively overseeing (ex: CMS or AUP) his contribution is very good. This is an issue that needs immediate attention to be resolved in order for flagship projects to be able to deliver successfully on their commitments.
[…]

**Goal 4.0 Provide Sound and Competent Leadership and Stewardship of the Laboratory**

*This Goal evaluates the Contractor's Leadership capabilities in leading the direction of the overall Laboratory, the responsiveness of the Contractor to issues and opportunities for continuous improvement, and corporate office involvement/commitment to the overall success of the Laboratory.*

**Score:** 3.0       **Grade:** B

**Goal Evaluation:**

Laboratory leadership has created an exciting vision and strategy for the future of Fermilab by expanding its expertise in neutrino science, elemental particle physics, accelerator science and moving into breakthroughs in quantum science and technology. The laboratory's leadership team should seek opportunities to add more formality and visibility into important programs to ensure ongoing actions to correct identified deficiencies are accurately captured and addressed. Causal analyses and corrective actions are not always comprehensive and require FSO direction and input to ensure issues are adequately addressed.

The challenges with LBNF/DUNE are serious and glaring and reflect poorly on laboratory management (e.g., procurement, independent project review). In addition, laboratory management has expressed strong support of the P5 plan, and that has been strongly in evidence for the LBNF/DUNE and PIP II projects. However, there seems to be relative neglect of the LHC program which is the highest priority in the field, and we almost never hear from senior management about the HL-LHC projects, and the HL-LHC Accelerator Upgrade Project has struggled to get the resources it needs.

The laboratory has put in extensive work on developing international partnerships that are a great benefit to the program, but then turns in draft agreements with a variety of problems that have to be corrected by DOE slowing the process and increasing DOE's workload.
[…]

**Objective 4.1: Leadership and Stewardship of the Laboratory**

**Weight:** 32.00%       **Score:** 2.8       **Grade:** B

**Objective Evaluation:**

High Energy Physics

The laboratory management has expressed strong support of the P5 plan, and that has been strongly in evidence for the LBNF/DUNE and PIP II projects. The problem has been relative neglect of the LHC program which is the highest priority on the field. We almost never hear from senior management about the HL-LHC projects, and the HL-LHC Accelerator Upgrade Project has struggled to get the resources it needs.

Office of Science

[…]

Laboratory leadership participated in a revision to the strategic vision and partnership commitment approved by the Fermi Site Office (FSO), DOE High Energy Physics Program (HEP), and FRA. This vision set expectations for a commitment to foster safe, diverse, equitable and inclusive work, research, and funding environments that value mutual respect and personal integrity.

[…]

Laboratory leadership should set the expectation that deliverables to the Office of Science and FSO are of high quality and require little to no rework.

FRA's laboratory leadership team should seek opportunities to add more formality and visibility into important programs to ensure ongoing actions to correct identified deficiencies are accurately captured and addressed.

**(Objective 4.1) Notable 1:** Effectively and efficiently administer the PIP2/LBNF/IERC subcontracts (including KAJV and TMI).

<u>Outcome:</u> This notable required the effective and efficient administration of PIP-II, LBNF DUNE, and IERC subcontracts. To aid in this effort, DOE and the laboratory collaborated to identify a scope of procurements for reporting and evaluation. The scope of procurements included all active PIP-II, Integrated Engineering Research Center (IERC), and LBNF DUNE procurements that met or exceeded the $5 Million DOE Site Office review threshold, or any procurement on one of the project's critical paths irrespective of dollar value. The quarterly reports were submitted by the laboratory, as agreed, with few errors and required some engagement by DOE. With respect to effective and efficient subcontract management, there were no major issues with the PIP-II and IERC subcontracts. While laboratory efforts to have Kiewit-Alberici Joint Venture (KAJV) solicit, award, and administer the LBNF Far Site Excavation bid package proved unsuccessful in FY 2020, the laboratory had to assume ownership and the associated risks of administration of the Thyssen Mining, Inc (TMI) subcontract in FY 2021. Efforts to accomplish the three-party assignment required substantial involvement by DOE. In order to avoid delays in the excavation phase of the LBNF project, significant DOE involvement was required for consent to purchase Builders Risk Insurance (BRI). Systemic procurement issues (as detailed in Goal 6, Objective 6.2) also contributed to the laboratory's failure to achieve this notable in FY 2021. - Not Achieved

==**Objective 4.2: Management and Operation of the Laboratory**==
**Weight:** 32.00%          **Score:** 2.8     **Grade:** B
**Objective Evaluation:**
<u>High Energy Physics</u>
The laboratory's biggest initiative is struggling. There are a variety of issues such as procurement. Many procurement packages struggle to get through DOE approval. LBNF's performance on Independent Project review has been mixed, with general evidence of progress but always short of DOE expectations.
<u>Office of Science</u>
[…]
The new Services Oversight Group (SOG) forum, which includes both key laboratory and DOE members, provided for a broad view of laboratory performance issues, and has facilitated the establishment of new metrics to track performance. The new metrics have been established for all major business and operations functions and compare planned versus actual performance. To date, improvements as a result of this effort include reduction in division and section procurement errors, reduction in overdue iTrack items, redirection of effort to key facilities metrics, reduction in expired badges, and other areas, as well as in-depth discussions regarding lab risks, cyber security, and financial Fermi National Accelerator Laboratory Evaluation FY 2021 21 data. DOE is eager to see how this group can continue to drive proactive improved laboratory performance in several functional areas, while reducing risk of future activities to aid in meeting the laboratory's strategic vision.

Throughout FY 2021, numerous challenges have arisen, both new and recurring, related to project performance, procurement and safety, and cyber security. In some instance laboratory responses to Office of Science and FSO feedback on contractor performance are defensive and dismissive, resulting in repeat findings and observations.

Causal analyses and corrective actions are not always comprehensive and often require FSO direction and input to ensure issues are adequately addressed. Additionally, the analyses take an extended time to complete and would benefit from the line organization leading the charge on the causal analyses and corrective action plans. While it is understood there is a need to balance thoroughness with timeliness, these extensive delays can result in a lack of thorough recall of events, and unchecked safety concerns which require mitigation. A lab-wide emphasis by management is needed to drive involvement, timeliness, and accountability.

**(Objective 4.2) Notable 1:** The Laboratory must keep senior SC leadership informed of key events (e.g., VIP/protocol visits, news releases, media requests) through timely population of the Science News Dashboard with all the relevant information on such activities and/or through other appropriate mechanisms.

<u>Outcome</u>: FRA provides weekly updates on laboratory activities to DOE/FSO leadership. The SC-1 dashboard is updated every Thursday/Friday and includes information on VIP/protocol visits, agreement signing, meetings/engagements with international funding agencies, news releases, and media requests.

**Goal 5.0 Sustain Excellence and Enhance Effectiveness of Integrated Safety, Health, and Environmental Protection**
*This Goal evaluates the Contractor's overall success in deploying, implementing, and improving integrated ES&H systems that efficiently and effectively support the mission(s) of the Laboratory.*

**Fermi Site Office**
**Score:** 3.2    **Grade:** B+
**Goal Evaluation:**
The laboratory has sustained their focus to address issues and make improvements with their Work Planning and Control (WP&C) process. While challenges in this area remain, the laboratory continues to make progress in its commitment to programmatic improvements.
[…]
The laboratory has experienced significant issues with subcontractor implementation of safe excavation measures. FSO observed numerous examples of unsafe excavations and shared these observations with FRA. A letter was transmitted to FRA directing a formal causal analysis be performed and scheduled monthly meetings with the Site Office to discuss actions and progress toward completion.
[…]
A DOE assessment was performed of the Fermilab Radiation Protection Program, resulting in two level one findings and thirteen level three findings. This review identified programmatic gaps and operational shortcomings. Due to the severity and the quantity of findings, a letter was transmitted to FRA directing a formal causal analysis be performed and scheduled monthly meetings with the Site Office to discuss actions and progress toward completion.

**Objective 5.1: Provide an Efficient and Effective Worker Health and Safety Program**
**Weight:** 60.00%    **Score:** 3.0    **Grade:** B
**Objective Evaluation:**
<u>COVID-19 Response</u>
[…]
<u>Subcontractor/Construction Safety</u>
The Laboratory has been challenged by numerous ES&H related activities and events this year at both the Batavia campus and the LBNF project work at the Sanford Underground Research Facility (SURF) facility in Lead, South Dakota. FSO oversight of these activities identified improvements in the hazard analysis documentation for construction work in general; however additional efforts are needed to maintain recent improvement with hazard analysis documentation for subcontractor construction activities. The laboratory has implemented several actions to bolster their subcontractor safety program including:
[...]

While the laboratory continues to reinforce and maintain the subcontractor communication improvements, there are areas of the program that remain of concern. Many of the above noted improvements were initiated due to the continued FSO on-site observations as well as the occurrence of significant ES&H events. In the first half of FY 2021, various excavation deficiencies were identified on active worksites including instances of improperly classified soil, improper cave-in protection systems, the use of benching in Type C soil (not allowed by OSHA), and unsuitable access/egress into trenches/excavations. The laboratory responded to these initial incidents by discussing the issues with the task managers, construction coordinators, and subcontractors at the Batavia site. In the second half of FY 2021, there were several instances of utility strikes and one instance of wet saw work performed within a "no dig" radiation boundary.

It was noted by FSO construction safety staff that neither DOE nor the laboratory ES&H were being allowed full unencumbered access to the IERC project site due to perceived ES&H concerns. Although the issue was eventually resolved, it is FSO's expectation that unless a hazard to DOE personnel exists that is not able to be adequately controlled, access to conduct safety oversight will be allowed.

==Due to frequency and severity of these events, a stand down of FESS-directed excavations was issued at the end of July requiring subcontractors to perform a safety review of their programs prior to the resumption of excavation work.== The laboratory leadership reemphasized safety expectations during this time. The ES&H Section has implemented increased presence of their construction oversight personnel to assure safe excavation operations and conformance with applicable standards and best practices. During this increased laboratory oversight, another incident of a subcontractor employee working in an excavation without a proper cave-in protective system in place was identified. The work was immediately paused by a Fermilab ES&H staff member, and a formal Stop-Work Order was issued to the subcontractor.

Given the aforementioned concerns, FSO issued a letter to the laboratory directing the completion of root cause analyses of these events and a recurring monthly meeting with FSO to discuss the fundamental causes of poor contract performance in this area.

<u>Work Planning and Control (WP&C)</u>

The laboratory's hazard analyses (HAs) have been stored on several platforms, each with different document formats, not allowing for consistent review and implementation across the site. To address this concern, the laboratory developed and implemented the IMPACT tool to standardize the digital location and format of the HAs. In January 2021, the laboratory decommissioned all other HA tools and in March they utilized the IMPACT tool for their first e-permit. An internal WP&C tripartite is currently assessing the rollout of the IMPACT tool by interviewing 29 different groups of employees at various levels of the organization. The IMPACT tool is a promising application that facilitates collaboration between various work groups and is an overall positive change to laboratory WP&C activities onsite.

The laboratory's HA documents for subcontractor construction work had recurring observations that did not meet expectations. While on-going construction inspections performed by FSO staff identified gradual improvements of the quality of HA documentation, observations were made noting that some HA documentation was not available on-site as required or did not adequately address specific work activities or provide for a description of the work activities being performed.

A Tripartite review of the internal subcontractor WP&C process revealed that ES&H needs to have a process to identify on-site contractor work and its locations, a threshold for when several roles are required for projects and recognized several FES&HM chapters that require revision. The assessment highlighted the need for a follow-on review focusing on subcontractor service work. To date, the results of the assessment findings have not been fully implemented.

[…]

The Laboratory has maintained a significant focus on effective and documented work planning and control processes. While increased reporting demonstrates a mature reporting culture, the potential severity and frequency of incidents creates concern and may be an indicator of more serious programmatic issues.

Issues Management

The laboratory continues to document construction walkthrough observations using their Predictive software. The laboratory's ES&H Section has demonstrated some improvement by being more proactive in interpreting data and recognizing health and safety trends to avoid/minimize issues. The Laboratory should look for opportunities to add more formality and visibility into important programs to ensure ongoing actions to correct identified deficiencies are accurately captured and addressed, and causal analyses and corrective actions are comprehensive and thoroughly documented.

The laboratory's iTrack system is a centralized database to capture observations and open corrective actions. iTrack items are primarily items that are significant in nature and may take longer periods of time to effectively address. This system is not being fully utilized for effective issues management and several opportunities for improvement exist to improve its usefulness. These opportunities include increasing responsibility and accountability for assuring corrective actions are completed and documented in a timely manner within the system.

ES&H Data Analysis

For data available in FY 2021, the laboratory averaged approximately 155,255 construction hours worked per quarter showing an increase in construction hours worked per quarter by approximately 25% from the previous year's representative sample. This indicates a significant increase in construction activities in FY 2021, mostly attributable to the LBNF and PIP-II work.

There was an increase in ORPS events over FY 2021 with 33 events occurring as compared to eight in FY 2020. One reason for the large number of reports is due to the LBNF Far Site personnel nitrogen dioxide overexposures which represent 13 of the ORPS events. These numbers will be reduced in FY 2022 due to an ORPS reporting equivalency approved by DOE that will change individual exceedance to a quarterly report.

The Non-compliance Tracking System (NTS) events remained similar to the previous years with six worker safety and health events, with no nuclear events occurring this fiscal year. The TRC and DART rates were the highest since 2016 and 2012 with rates of 1.6 and 1.1 respectively. The first aid rate stayed below the previous several years with a rate of 0.66.

Work at Sanford Underground Research Facility (SURF)

Performance during the year was generally sound with a great deal of work performed by the laboratory to assist subcontractors in effectively investigating incidents to assure causes are identified and addressed. The laboratory's work planning and control processes have been effective as the laboratory has transitioned from Reliability Projects into the main cavern excavation activities.

The Laboratory worked closely with DOE to develop a variance to OSHA 800(k)(3) ventilation requirements which was successfully submitted to the Office of Science to be forwarded to DOE-AU for approval. Additionally, an Abatement Plan and ORPS equivalency were developed to ensure worker safety due to the project's inability to meet 2016 American Conference of Governmental Industrial Hygienists (ACGIH) Threshold Limit Values (TLVs) for NO2. The Laboratory worked collaboratively with DOE to effectively develop the required documentation which was successfully submitted and ultimately approved by DOE.

There were also recent incidents of Silica TLV exceedances during underground construction. The laboratory response to the Silica TLV exceedances was swift and effective with a focus on protecting worker health and limiting exposures. Sampling processes allowed the laboratory to identify the issues and refocus their response which included enacting a new sampling plan in conjunction with KAJV and TMI to ensure a proper understanding of the hazards and the ability to measure efficacy of controls.

Radiation Safety Program Compliance

The laboratory 's Radiation Protection Program (RPP) was reviewed by an external DOE team in April 2021. The review identified two Level 1 findings, thirteen Level 3 findings and one Best Practice. The Level 1 findings were associated with deficiencies in the Lab's Environmental RPP (ERPP) and overall compliance with DOE O 458.1 Radiation Protection of the Public and the Environment, in particular with the laboratory's radiological material release and clearance program. This assessment confirmed many of the concerns and issues identified during FSO oversight activities. The results of the RPP review, coupled with FSO oversight results identified significant gaps in the Radiation Protection Program effectiveness, primarily due to a general informality of operations with many procedures/processes not formally documented.

Several RPP related HPI reviews were performed throughout the course of the fiscal year:

Main Injector-30 re-lamping incident. While interviews and root cause analysis were productive, FSO input was required to ensure the investigation focused on issues pertaining to 10 CFR 835 vice almost entirely on WP&C.

Saw cutting activities performed in a "no dig zone" in the 8GeV/Booster area while beam was present. The HPI team did a thorough job interviewing participants and discussing extent of condition, causal analysis, and root causes of the event.

Additionally, numerous events violating Fermilab's Radiation Safety Program occurred during FY 2021. DOE is concerned with the laboratory compliance in implementing two areas of radiation protection: 10CFR835 Subpart L, Radioactive Contamination Control and 10 CFR 835 Subpart F, Entry Control Program. The following events described below highlight programmatic issues in these areas of concern.

10 CFR 835 Subpart L Radioactive Contamination Control

In May 2021, a personnel contamination event occurred when a high frisker reading alerted Radiation Safety to respond to the MT6.1 enclosure in the Fermilab Test Beam Facility (FTBF). The affected individual's fingertips were reading about 400 cpm above background using a frisker. This corresponds to about 4,000 dpm and exceeds the 10 CFR 835 Appendix D Surface Contamination Value of 1,000 dpm for removable beta-gamma emitters.

Operations were stopped at Muon Test Area/Irradiation Test Area (MTA/ITA) and MT6.1 area while the event was investigated, and decontamination efforts were performed. Interim measures were implemented including updating the Radiation Work Permit and issuing a new procedure for handling contaminated parts, and operations resumed. A full investigation and causal analysis into the personnel contamination event remains ongoing.

In July 2021, a failure of a planned design change in the NuMI air handling unit (AHU) condensate collection system resulted in approximately 100 mL of fluid containing tritium to leak onto the floor causing contamination in the MI65 Fire Command Room. An operational readiness review of the system was completed and did not identify the potential for this event to occur. The Laboratory initiated an HPI to determine the cause of the design failure that resulted in a new contamination area.

While investigating the MI65 tritium spill, condensation was found accumulating on piping in the area and in the utility room of the service building. The condensation was sampled and found to exceed the 100 pCi/mL limit for personnel exposure. The areas are posted as contamination areas and additional controls are needed for work in the areas.

10 CFR 835 Subpart F Entry Control

In July, technicians entered an area posted as a Controlled and Radioactive Materials Area to perform work. Once inside the service building, the technicians noticed an enclosure door leading to a posted Radiation Area was not fully closed or locked as required. The technicians immediately called the Main Control Room to report it.

Employees of the Facility Engineering Services Section performed work in a MI30 High Radiation Area (HRA) located in the Accelerator Division without proper authorization and coverage. A job specific Radiation Work Permit (RWP) and RCT coverage is required for work in HRAs due to the potential for individuals to be exposed to greater than 100 mrem/hr dose rates per 10 CFR 835.502 High and Very High Radiation Areas. Initial discussions with the laboratory indicated that since the individuals only received 10-15 mrem dose, and an investigation would be conducted while shutdown, work could continue as normal. After further review of the event and FSO expressing concern about the potential consequences of the incident, the laboratory's Senior Radiation Safety Officer (SRSO) issued a stop work in all high radiation areas with compensatory measures in place to release work on a cases-by-case basis until an investigation could be completed. The investigation took over 3 months to complete and a report to be issued.

Disregard of Radiation Boundaries and Postings

FSO oversight has observed several instances where radiation boundaries, signs or postings have been moved by unauthorized personnel.

During a walkthrough at Minos, a Tritium Contamination Area posting stanchion was moved by unauthorized personnel to accommodate work.

The MC7 enclosure door was removed without following proper radiation safety procedures. Non-Radiation Safety personnel removed radiological postings from the original door in order to replace the door with a new one. The original door was intended to be released for recycling but did not immediately go through required release and clearance surveys to verify the door was not activated. Radiation Safety tracked the door down in order to survey it while it was still onsite.

A second event at the MC7 area occurred that involved hand digging in an area that would have required a Joint Utility Locating Information for Excavators (JULIE) to determine the proximity of the excavation activity to required radiation shielding. Following requirements of the Accelerator Safety Program, an Unreviewed Safety Issue Determination (USID) was performed to determine if an Accelerator Safety Envelope (ASE) violation occurred as required when potential impacts to shielding are discovered. At the time of the digging, critical devices controlling the beamline were locked off and no beam was present. It was determined not to be a violation of the ASE (negative USID).

A subcontractor working on the Accelerator Campus Utility Corridor (ACUC) Project performed wet saw cutting on the existing asphalt in the South Booster Road intersection and entrance to the Central Utility Building. The saw cutting occurred while the accelerator was operating and within an established and marked "No-Dig Zone." The cutting was observed by a Fermilab Radiation Safety Technician, who notified Radiation Safety management. Upon notification of the incident, the Accelerator Division de-energized the beam to the Booster complex and all beamlines downstream. A USID was issued, and the change in shielding as a result of cutting in the "No-Dig Zone" was determined not to be a violation of the ASE.

Eating/Drinking/Smoking/Cosmetics Policy in Radioactive Material Area

During FY 2021, cigarettes and cigarette smoke were discovered in a Radioactive Material Area (RMA) along the F Sector Service Buildings during a Highly Protective Risk Inspection. Additionally, during a walkthrough of Particle Physics Division spaces, an individual working in an RMA was observed to have food and beverages. Eating, drinking, smoking, and applying cosmetics in an RMA presents a risk of exposure and intake through inhalation or ingestion pathways. During monthly Radiation Safety Subcommittee meetings, FSO has expressed concern that the laboratory does not have a policy forbidding personnel from eating, drinking, or applying cosmetics in an RMA. This was also noted during the recent DOE RPP Assessment.

Laser Safety

During FY 2021, several new laser systems were installed which is indicative of a trend in the use of higher-powered lasers being used in accelerator operations. The principles of ANSI Z136.1, Safe Use of Lasers, are fully implemented in laser labs around the complex. The Laser Safety Program at the laboratory remains robust and effective at preventing unwanted outcomes. No laser safety incidents were reported in FY 2021.

Accelerator Safety

The Accelerator Safety Program remained effective and successful at preventing unwanted outcomes during FY 2021. No violations of the ASE were reported, and the laboratory tracked multiple potential unreviewed safety issues through the USID process. Improvements and enhancements to the Accelerator Safety Program were made by continuing to develop formal processes when determining the level of Accelerator Readiness Reviews (ARRs) for a new or modified accelerator. The new process is integrated into the USID procedure and describes selecting an ARR review team and type of review based on a graded approach. Formally documenting processes such as this creates a more robust and cohesive Accelerator Safety Program.

New Safety Assessment Documents (SAD) Chapter updates were published for Minos and the Short Baseline Neutrino Experimental Areas (i.e., ICARUS, SBND, and MicroBooNe Experiments) in the Neutrino Division and the Proton Area and the Tevatron Main Ring of the Accelerator Division. Additionally, several USIDs were conducted for planned and discovered unreviewed safety issues. Some of the planned changes to shielding that required a USID include the Main Injector berm shielding reorganization for LBNF site preparation, extra shielding was added for the MC7 Nova test beam, and the Muon Campus' installation of additional shielding in the delivery ring to prepare for future Mu2e beamline. Two negative USIDs were issued during FY 2021 following unplanned excavation activities into accelerator beamline shielding at MC7 and 8GeV/Booster areas.

The laboratory completed several actions including a qualitative beam-on survey of the entire accelerator complex due to concerns from the Department. Ongoing evaluation is being performed to ensure all postings and surveys are complete.

**(Objective 5.1) Notable 1:** Complete the implementation review of the Accelerator Division (AD) Control Rooms Formality of Operations Tripartite to include the AD Operations Self-Assessment and field work activities to assure actual accelerator control room operations align with internal procedures and requirements.

Outcome: During FY 2021, the laboratory finished the remaining portion of FY 2020 PEMP Notable 5.1.1, Complete the implementation review of the Accelerator Division (AD) Control Rooms Formality of Operations Tripartite to include the AD Operations self-assessment and field work activities to assure actual accelerator control room operations align with internal procedures and requirements. The laboratory completed the second part of the Tripartite Assessment during FY 2021 by performing field observations of operations in the Main Control Room and the g-2 Remote Control room. The team identified six recommendations and two best practices during this assessment. Some of the recommendations focus on clarifying and establishing additional operating procedures in the g-2 Control Room. Additionally, the MCR and Remote-Control Rooms deployed strict protocols during operations to address concerns related to the COVID-19 Pandemic including limiting capacity, requiring people to call ahead before picking up enclosure keys, and installing physical barriers inside the MCR to decrease the chances of transmitting the virus.

**Goal 6.0 Deliver Efficient, Effective, and Responsive Business Systems and Resources that Enable the Successful Achievement of the Laboratory Mission(s)**

*This Goal evaluates the Contractor's overall success in deploying, implementing, and improving integrated business systems that efficiently and effectively support the mission(s) of the Laboratory.*

**Fermi Site Office**
**Score:** 2.5     **Grade:** B
**Goal Evaluation:**
The laboratory continues to have challenges in Financial Management and Acquisition Management. Audits repeatedly highlight the same deficiencies and control failures year after year. Any corrective actions implemented have resulted in little to no progress. Significant procurement issues have hindered the laboratory's ability to successfully deliver efficient and effective business systems/resources to enable the Science Mission. Substantial concerns remain regarding the ability to expend Government funds in an effective, efficient, and compliant manner.
[…]
The Laboratory organizational structure does not facilitate inclusion of key staff to resolve major operational challenges, nor has the structure provided a framework to flexibly adapt to needed staffing, mission, and project changes.
The laboratory has made great strides in strengthening its Contractor Assurance System (CAS) program by implementing recommendations/best practices and integrating performance metrics for key laboratory support areas into the oversight process. Additionally, the laboratory formed a Human Performance Improvement (HPI) Subcommittee and developed an assessment program.
Fermilab's Internal Audit (IA) activity continues to be effective and has met all contract deliverables for FY 2021 and is on track to complete the Annual Internal Audit for FY 2021. The Laboratory should continue efforts to resolve communication breakdowns between IA, management, and compliance.
[…]

**Objective 6.1: Provide an Efficient, Effective, and Responsive Financial Management System**
**Weight:** 20.00%     **Score:** 2.4     **Grade:** C+
**Objective Evaluation:**
Scoring reflects the laboratory's continued challenges with financial functions related to capability, compliance, and quality. Integration with procurement is still problematic. The laboratory's financial management system has still not reached maturity. Additionally, the laboratory has failed to make adequate progress in determining the root cause of repeated audit findings and, as of the date this report was written, have yet to submit formal comprehensive corrective action plans for DOE consideration.
Results of the FY 2021 OMB Circular A-123 audit findings are concerning and highlighted repeated control failures for three consecutive years (FY 2019-FY 2021) in the areas of payables management and project cost management. Fermi National Accelerator Laboratory Evaluation FY 2021 36 Specifically, the OMB Circular No. A-123 audit resulted in 20 control failures from the Finance Section which inclued Accounting and Procurement. Of the 20 control failures, eight control failures are repeated failures from prior years FY 2019 and FY 2020, eight control failures are repeated failures from FY 2020, and four new controls failed in FY 2021.

Of the seven audits conducted by FRA's Internal Audit in FY 2021, there were 13 findings of which seven were related to compliance and/or internal control failures in finance and accounting. Audit report FNAL2021-02 highlighted control failures in accounting and finance related to reimbursement of travel expenses not applying costs to the related subcontract, payments being released without proper identification of labor and equipment costs, incomplete documentation to support services, expenses invoiced, and questionable travel costs practices. In addition, Accounts Payable did not assign subcontractor reimbursement invoices to the correct purchase orders. Audit report FNAL2021-03 highlighted control failures in management not consistently ensuring goods are approved by the requestor, Accounting not revising/updating General Payables and Disbursement Policies to include process changes for approvals resulting from Oracle Approval Management Engine (OAME) implementation, and Finance not consistently ensuring OAME end-users complete training prior to approving invoices using OAME.

There continues to be disconnects between the Finance Section and Internal Audit. Audit findings are not given the appropriate attention which may contribute to the lack of meaningful progress in identifying control deficiencies/weaknesses and implementing corrective actions that permanently resolve outstanding issues.

**Objective 6.2: Provide an Efficient, Effective, and Responsive Acquisition Management System and Property Management System**
**Weight:** 30.00%     **Score:** 1.1     **Grade:** C
**Objective Evaluation:**
Acquisition Management System

Scoring reflects significant systemic deficiencies and continued challenges with acquisition strategy/planning, meaningful procurement organization engagement, market research, selection of subcontract type, evaluation criteria development, exercising/incorporation of subcontract options, and overall subcontract administration. While numerous efforts have been made by DOE to stress the importance of addressing these issues both during the mid-year and on a continuous basis, no meaningful improvement has been made by the laboratory over the course of this fiscal year. Additionally, there continues to be challenges with the quality and substance of procurement packages. Constructive feedback is often challenged, which leads to key issues not being addressed in a timely manner. Also, the laboratory has challenges with incorporating feedback and utilizing lessons learned to make improvements. In reviewing acquisition strategies, DOE continues to observe the omittance of material information, factual misstatements in documents and/or failure of the laboratory to include input from all relevant functional specialists in acquisition plans as it relates to risks, subcontract type, subcontract options, evaluation criteria, market research, and estimated cost/price. Acquisition strategies are rarely based on the full breadth of information available. The laboratory continues to struggle with using market research to support acquisition planning and decision-making processes to ensure that the most suitable approach for procuring supplies/services is pursued. DOE has observed inadequate use of market research to inform acquisition decisions, requirements, source selections, and solicitation timing. Requirements development has also been problematic, often leading to numerous subcontract change orders and/or cost overruns. The LBNF/DUNE Near Site Conventional Facilities (NSCF) Near Detector Complex acquisition plan failed to provide accurate information and omitted critical details regarding potential challenges/risks associated with meeting all applicable aspects of 10 CFR 851 [Editor's note: Worker safety and health requirements]. This issue was discovered by DOE after Head of Contracting

Activity (HCA) approval of the acquisition plan. The laboratory failed to communicate and disclose important concerns to FSO in a timely manner.

Effective contract administration has also been a challenge. There have been recurring FRA Internal Audit and Office of Management and Budget (OMB) Circular A-123 audit findings related to compliance with subcontract payment terms and conditions. [Editor's note: DOE complies with OMB Circular A-123, Management's Responsibility for Enterprise Risk Management and Internal Control, which provides guidance for internal control and risk management requirements. OMB Circular A-123 also establishes the requirement to produce an agency Risk Profile as part of the implementation of an Enterprise Risk Management (ERM) capability coordinated with strategic planning, strategic review, and internal control processes.] DOE has also observed multiple occurrences where the laboratory's failure to act in a timely manner resulted in either an adverse outcome or required intervention or action from DOE. The Mu2e Solenoid procurement has had significant issues related to subcontractor performance, quality, and cost/price control since subcontract award. Per the laboratory, these quality issues have required the implementation of additional quality assurance and oversight measures by laboratory engineers. While the Mu2e Solenoid subcontractor has failed to comply with subcontract terms and conditions, the laboratory's failure to take decisive actions to remedy subcontract non-performance and quality issues throughout the life of this subcontract has been problematic and resulted in substantial DOE involvement at all levels.

In addition, the laboratory has had difficulties complying with their DOE approved procurement policies/procedures and DOE requirements. Sole Source Justifications routinely do not adhere to laboratory Approved Procurement Policy and Procedures Manual. There is inconsistent use of the approved template (FL-57) and often insufficient rationale for other than full and open competition. Also, as a result of a September 2021 Buy American Act (BAA) exemptions data call, the laboratory informed DOE that it failed to execute or obtain BAA exemption approval for 17 procurements from FY 2018- FY 2020, prior to purchase, eight of which required DOE approval.

The laboratory has had minor issues with records management as well. It was not able to locate two Master Agreements but continued to issue task orders against those vehicles. While the Agreements were eventually located (several months later), this highlighted that procurement personnel were not always cognizant of the Master Agreement scope, ordering period of performance, max and min ordering amounts, ceiling amount, or cumulative dollar value of task orders issued to date prior to task order issuance.

Property Management System

[…]

**(Objective 6.2) Notable 1:** Develop and implement a comprehensive property stewardship program that includes an accurate inventory of deployed and stored personal property and expectations for accountability around the stewardship by divisions and sections of this personal property.

Outcome: Through the utilization of the Sunflower property management system and the Self-Service Property application (SSP), the laboratory has been able to ensure an accurate inventory of all deployed and stored personal property and a comprehensive property stewardship program. The laboratory has improved property passes by transitioning to a fully electronic process and has developed an electronic storage justification form to replace the paper forms previously utilized.

**Goal 8.0 Sustain and Enhance the Effectiveness of Integrated Safeguards and Security Management (ISSM) and Emergency Management Systems**

*This Goal evaluates the Contractor's overall success in safeguarding and securing Laboratory assets that supports the mission(s) of the Laboratory in an efficient and effective manner and provides an effective emergency management program.*

**Fermi Site Office**
**Score:** 2.8     **Grade:** B
**Goal Evaluation:**
[…]
The laboratory submitted a Site Security Plan (SSP) in December 2020 which was approved with contingencies. The laboratory reported the implementation of corrective actions for all 27 findings and recommendations from the 2019 DOE-SC-4 Safeguards and Security Assessment, however, an August 2021 DOE Effectiveness Review of the corrective actions found six out of 15 instances of ineffective/partially ineffective corrective actions in preventing reoccurrence.

The laboratory was required to update and submit a new Authority to Operate (ATO) package; however, the document received was not of a quality level that warranted DOE review and was returned to the laboratory for additional work.

In cyber security there have been numerous projects completed and actions taken throughout the fiscal year, though the journey was difficult and highly reactive. Many of these completed projects were as a result of DOE direction to the laboratory requiring specific actions be taken. Additionally, a high level of DOE involvement was required throughout the year to ensure programs were proceeding appropriately.

**Objective 8.2: Provide an Efficient and Effective Cyber Security System for the Protection of Classified and Unclassified Information**
**Weight:** 40.00%     **Score**: 2.2     **Grade:** C+
**Objective Evaluation:**
The laboratory has accomplished several projects over the past year. There has been a big push for Personally Identifiable Information (PII) scanning, moving resources behind authentication, and reviewing web pages that are not needed. The laboratory created a SharePoint site that allows collaboration with the site office to track items that are on their Implementation Plan. The laboratory also moved towards a new security plan that uses National Institute of Standards and Technology (NIST) 800-53r5 baselines. There was a large amount of work completed in response to the Peer Review, new Office of Science Cyber Security Program Plan (CSPP), and several Letters from the Site Office to FRA.

While there have been numerous projects completed and actions taken throughout the fiscal year, the journey was difficult and highly reactive. Instead of the laboratory being proactive about Cyber Security, many of these completed projects were as a result of DOE direction to the laboratory requiring that specific actions be taken. Additionally, a high level of DOE involvement was required throughout the year to ensure programs were proceeding appropriately.

The laboratory was required to update and submit a new Authority to Operate (ATO) package to FSO. The document received was not of a quality level that warranted DOE review and was returned to the laboratory. For example, there were numerous misspellings, references to outdated NIST documentation, outdated OMB memos, and DOE orders used that were not up to date. There were also inconsistent terms used, old data that should had been revised, and some areas that needed stronger language to match the new SC Cyber Security Program Plan (CSPP) requirements.

Over the past year, numerous reviews identified additional areas for the laboratory to focus on and prioritize. As a result of these reviews, the SC CSPP release, and the Cyber Security Executive Order, there was a large list of items that needed to be completed to ensure an effective program. The laboratory turned this list into a "Cyber Security Implementation Plan." This plan identified the task, what the laboratory needed to complete, how the task would be implemented, and evidence of implementation. The Cyber Security Implementation Plan that is now in SharePoint is of interest to other groups at the laboratory, as it has been a huge success in tracking items to completion that require DOE oversight and approval.

An area that needed additional attention was providing consistent and relevant information to ensure the Authorizing Official (AO) was aware of risks and can see progress being made in the Cyber Security Office. The laboratory is now using a "Quad-Chart" that is presented to the AO at least once a month. The "Quad-Chart" contains important information in dashboard form, such as data call submissions, cyber security incidents and events, risk assessment status, upcoming audits, task status, and vulnerability trending information.

Within the past year, there was a Cyber Security researcher that had communicated with the laboratory about a specific vulnerability over a course of time. DOE was not informed of the ongoing interaction in a timely fashion, and because of this communication, a tech blog published a story about this vulnerability and interaction. DOE issued a guidance letter to the laboratory including numerous areas that required immediate attention. The laboratory has made significant progress in scanning its websites for specific types of information and has made important modifications in policies governing information made available to the public. Only information specifically intended for the public was made available without requiring authentication.

The laboratory had a penetration testing firm called "Attack Research" evaluate internal and external systems for potential compromises. The company spent several weeks trying to enter the Fermilab network as an external user and then as a laboratory collaborator. After the assessment was complete, there were no successful entries that were gained, but the assessment did result in some observations for improvement. The laboratory is currently working on a corrective action plan using the observations and will submit the plan to the site office for review. Other areas where the laboratory had improved their security stance was moving all Lab user subnets behind a default deny firewall. The laboratory has also reduced their Internet footprint by instituting default blocks and reducing the use of high-risk protocols, such as FTP, X-Windows, SSHD, Postgres, MySQL, MariaDB, and MongoDB. To verify these results, the laboratory increased the frequency of both privileged and non-privileged vulnerability scanning.

All the work completed is a step in the right direction, risks have been reduced and security increased. However, laboratory implementation of these security upgrades has been reactionary and resulted predominantly from DOE engagement.

**Objective 8.3: Provide an Efficient and Effective Physical Security Program for the Protection of Special Nuclear Materials, Classified Matter, Classified Information, Sensitive Information, and Property**
**Weight:** 40.00%     **Score:** 3.0     **Grade:** B
**Objective Evaluation:**
Overall, the Integrated Safeguards and Security Management (ISSM) and Emergency Management (EM) Systems continue to evolve in response to recent assessments and DOE guidance. A DOE Corrective Action Plan (CAP) Effectiveness Review was conducted in August 2021 to determine the effectiveness of corrective actions taken by the laboratory pursuant to the 2019 DOE Survey. Six (6) out of fifteen corrective actions were identified as ineffective or partially effective in preventing reoccurrence of weaknesses or findings. The assessment team concluded that while work remains for the corrective actions rated as ineffective/partially ineffective there is a high degree of assurance that the probability of recurrence for a majority of the 15 areas of non-compliance is greatly minimized and the effectiveness review is closed satisfactorily.

The Safeguards and Security Program (S&S) has demonstrated a desire to address identified findings/recommendations/guidance from DOE, however the laboratory remains challenged in terms of S&S program elements integration and organization. The laboratory submitted a revised draft of the Site Security Plan (SSP) as the S&S program and program elements continue to evolve, with a specific emphasis on protecting restricted areas and facilitating public engagement on-site. The laboratory should continue to evaluate the S&S needs for the site and evaluate whether opportunities for efficiencies may be lost with managing a decentralized organization.

The laboratory has continued to make improvements in the Physical Security posture. Replacement of obsolete or ineffective Physical Access Control Systems (PACS) with highly efficient and scalable systems continues. Additionally, the laboratory has employed a technology-focused solution to effectively secure the site and integrate emerging security technologies and systems. The laboratory has obtained initial approval for the eventual use of unmanned aircraft system (UAS) to aid in both Safeguards and Security and Emergency Management purposes. This has the potential to provide coverage of the 6800-acre site significantly and more effectively. The laboratory should continue efforts to evaluate the site boundaries and optimize security measures to prevent trespassing. The laboratory has continued to be effective with FS-10 requests and justifications and responsive to "shovel ready opportunities" as offered by DOE.

Conversely, the Foreign Visits and Assignments (FV&A) functional area has lagged in the drafting of specific FV&A procedures which support the Site Security Plan. These include lab procedures to address specific processing of Foreign Nationals, the entry and use of the FACTS system and integration with the Site Access and Badging process. Contributing to this weakness is the recent release of DOE O 142.3B midway through the calendar year. Laboratory compliance with their approved Site Security Plan (SSP) remains unclear as it relates to Property Protection Area (PPA) and General Access Areas (GAA), as full implementation of the SSP, including Wilson Hall controls, was required by March 31, 2021. Progress is ongoing, but FRA has not yet completed actions necessary to facilitate a Facility Clearance License (FCL). The FCL is required in order to appropriately begin the HSPD-12 transition project and ultimately the Access Authorization/Security Clearance process.

The laboratory has implemented a Site Access and Badging (SAB) project to standardize the process and simultaneously ensure required site training is tracked. This project will incorporate Real ID Act compliance, and eventually communicate with the systems used for personnel/employee management. The laboratory has been working towards full Real ID compliance.

The Information Security program continues to evolve; however, it still needs improvement. The program is cyber-focused while not adequately addressing the non-cyber areas of Information Security such as proper document marking, handling, transmission, and storage. The August 2021 DOE Corrective Action Effectiveness Review identified Information Security weaknesses, primarily involving document reviews which uncovered incorrectly marked documents. The overarching Information Security program is underway and must identify all forms of information, both physical and electronic, and properly mark, handle, and store respective records. Subsequent to the Corrective Action Effectiveness Review, a draft framework of a new Information Security Plan has been created to include more robust mediums of information.

The Material Control and Accountability (MC&A) Program continues to be a laboratory strength. The MC&A Program is well implemented and is effectively working in collaboration with DOE experts. This collaboration was highlighted during the initiative to address the vessels of Deuterium (D2) currently stored on-site. DOE is working with the laboratory to help identify the scope of the D2 issue and determine courses of action to properly dispose of the gas. This effort is on-going but will require laboratory involvement and effort to complete successfully.

**(Objective 8.3) Notable 1:** Promptly implement a revised Site Security Plan (SSP) and any additional required updates as approved by DOE. Complete directed security actions per DOE approved timelines.

<u>Outcome</u>: The laboratory's SSP was approved by the FSO Site Office Manager in December 2020 and implemented by the laboratory. This approval was contingent upon actions and deliverables which were adequately addressed by the laboratory.

**(Objective 8.3) Notable 2:** Develop a revised comprehensive ATO package(s) (based on an updated set of risk assessments, security plans and security control assessments) to align Fermilab systems as directed by DOE to the new CSPP and DOE risk expectations. Complete actions per DOE approved timelines.

<u>Outcome</u>: The laboratory submitted an ATO package on August 31 which was returned to due to quality and compliance standards not being met. When the package was reviewed by the Site Office, numerous misspellings existed, reference NIST documents, OMB memos, and DOE orders were not updated, inconsistent terms and old data were used, and some areas needed stronger language to match the SC CSPP requirements. As a result, FSO requested an updated version by September 24, 2021. The laboratory resubmitted the package by the requested date, though work is still needed to clearly document implementation of NIST 800-53r5 controls.

# Analysis of 2023 PEMP (B-/C+ Equivalent Grade when including $1M fine)

| ITEM | DESCRIPTION | OWNER | RATING |
|---|---|---|---|
| **Goal 1.0** | **Provide for Efficient and Effective Mission Accomplishment (30%)** | **Director** | **A-** |
| Objective 1.1 | Provide Science and Technology Results with Meaningful Impact on the Field | CRO | A- |
| Notable Outcome | By July 2023, develop and submit a plan to increase the involvement of Fermilab scientific staff on LBNF/DUNE including timelines of when increases will occur and activities that will receive the increases. | | Achieved |
| Objective 1.2 | Provide Quality Leadership in Science and Technology Results that Advances Community Goals and DOE Mission Goals | CRO | A- |
| Notable Outcome | Contribute to establishing the synergistic research program and deliver impactful science from the FNAL-led QIS Center, as measured by the FY 2023 trimester reports, annual report, common goals and milestones report, research publications and highlights, and participation in periodic conference calls. | | Achieved |
| **Goal 2.0** | **Provide for Efficient and Effective Design, Fabrication, Construction and Operations of Research Facilities (45%)** | **Director** | **B+** |
| Objective 2.1 | Provide Effective Facility Design(s) as Required to Support Laboratory Programs (i.e., activities leading up to CD-2) | CPO | B+ |
| Notable Outcome | Effectively manage and safely execute the assigned LCLS-II-HE project scope in accordance with DOE Order 413.3B. Performance will be assessed based on the assigned project management responsibilities and cryomodule work planned and accomplished during FY 2023. | | Achieved |
| Objective 2.2 | Provide for the Effective and Efficient Construction of Facilities and/or Fabrication of Components (execution phase, post CD-2 to CD-4). | CPO | B+ |
| Notable Outcome | Effectively manage and execute the assigned PPU project scope for magnet fabrication in accordance with DOE Order 413.3B to safely accomplish the planned work per the approved Performance Baseline. Performance will be assessed based on the work planned and accomplished during FY 2023. Deliver chicane magnets 1 and 2, and injection dump dipole to ORNL by June 30, 2023. | | Achieved |
| Objective 2.3 | Provide Efficient and Effective Operation of Facilities | Director | B+ |
| Notable Outcome | Develop a needs assessment on future housing in South Dakota for FRA employees, subcontractors, and collaborators. Based on the identified need, options should be evaluated and documented regarding methods for accomplishing housing near SURF. Coordinate the development of these options with SDSTA to integrate their local knowledge and their unique flexibilities. Submit the report by July 2023. | | Achieved |
| **Goal 3.0** | **Provide for Efficient and Effective Science and Technology Program Management (25%)** | **Director** | **B+** |
| Objective 3.1 | Provide Effective and Efficient Strategic Planning and Stewardship of Scientific Capabilities and Program Vision | Director | B+ |
| Notable Outcome | Submit a staffing analysis of the technical, engineering, computing, scientific, and project management skills of the current staff and the outyear needs for that staff for the next 5 years. | | Achieved |
| Objective 3.2 | Provide Effective and Efficient Science and Technology Project/Program/Facilities Management | Director | B+ |
| Objective 3.3 | Provide Efficient and Effective Communications and Responsiveness to Headquarter's Needs | Office of Communication | B+ |
| **Goal 4.0** | **Provide Sound and Competent Leadership and Stewardship of the Laboratory** | **Director** | **A-** |
| Objective 4.1 | Leadership and Stewardship of the Laboratory | Director | A- |
| Objective 4.2 | Management and Operation of the Laboratory | COO | B |
| Objective 4.3 | Advancing Laboratory Diversity, Equity, Inclusion and Accessibility | COO | A- |
| Notable Outcome | FSO: In order to position Fermilab to meet future success and deliver on mission, FY 2023 must be a year of cultural and functional change. Implement an agreed to set of facility and systems changes in FY 2023 that assure fundamental change in the Laboratory footing for FY 2024. | | Achieved |
| Objective 4.4 | Leadership of External Engagements and Partnerships | COO | A- |
| Objective 4.5 | Contractor Value-added | COO | A- |
| **Goal 5.0** | **Sustain Excellence and Enhance Effectiveness of Integrated Safety, Health, and Environmental Protection [30%]** | **CSO** | **B-** |
| Objective 5.1 | Provide an Efficient and Effective Worker Health and Safety Program | CSO | C+ |
| Objective 5.2 | Provide an Efficient and Effective Environmental Management System | CSO | B+ |
| **Goal 6.0** | **Deliver Efficient, Effective, and Responsive Business Systems and Resources that Enable the Successful Achievement of the Laboratory Mission(s) [30%]** | **COO** | **B+** |
| Objective 6.1 | Provide an Efficient, Effective, and Responsive Financial Management System | CFO | B+ |
| Objective 6.2 | Provide an Efficient, Effective, and Responsive Acquisition Management System and Property Management System | CSO | B |
| 6.2 Part 1 | Acquisition Management System | CFO | |
| 6.2 Part 2 | Property Management System | Facilities Eng. Services | |
| Objective 6.3 | Provide an Efficient, Effective, and Responsive HR and Talent Management Systems | WDRS | B+ |
| Objective 6.4 | Provide Efficient, Effective, and Responsive Contractor Assurance Systems, including internal Audit and Quality | WDRS | B+ |
| 6.4 Part 1 | Contractor Assurance Systems and Quality | COO | |
| 6.4 Part 2 | Internal Audit | FRA Internal Audit | |
| Objective 6.5 | Demonstrate Effective Transfer of Knowledge and Technology and the Commercialization of Intellectual Assets | OPPT | A- |
| **Goal 7.0** | **Sustain Excellence in Operating, Maintaining, and Renewing the Facility and Infrastructure Portfolio to Meet Laboratory Needs [25%]** | **COO** | **B+** |
| Objective 7.1 | Manage Facilities and Infrastructure in an Efficient and Effective Manner that Optimizes Usage, Minimizes Life Cycle Costs and Ensures Site Capability to Meet Mission Needs | Facilities Eng. Services | B+ |
| Objective 7.2 | Provide Planning for and Acquire the Facilities and Infrastructure Required to Support the Continuation and Growth of Laboratory Missions and Programs | Office of Campus Strategy & Readiness | B+ |
| **Goal 8.0** | **Sustain and Enhance the Effectiveness of Integrated Safeguards and Security Management (ISSM) and Emergency Management Systems [15%]** | **Director** | **B+** |
| Objective 8.1 | Provide an Efficient and Effective Emergency Management System | CSO | B+ |
| Objective 8.2 | Provide an Efficient and Effective Cyber Security System for the Protection of Classified and Unclassified Information | CIO | A- |
| Objective 8.3 | Provide an Efficient and Effective Physical Security Program for the Protection of Special Nuclear Materials, Classified Matter, Classified Information, Sensitive Information, and Property | CSO | |

| 8.3 Part 1 | Part 1: Physical Security Program for the Protection of Special Nuclear Material and Property | CSO | B+ |
|---|---|---|---|
| 8.3 Part 2 | Part 2: Classified Matter, Classified Information, and Sensitive Information | CIO | |

NOTE: Whereas the overall rating for the 2013 PEMP was B+, based on clauses B.3 *Performance and Other Incentive Fees* and Clause I.94-DEAR 970.5213-3 *Conditional Payment of Fee, Profit and Other Incentives – Facility Management Contracts*, DOE reduced the fee for FRA's performance by ~$1M, i.e. from the "earned fee" of $4,491,123.00 to $3,491,848.00,

**Goal 4.0 Provide Sound and Competent Leadership and Stewardship of the Laboratory**
*This Goal evaluates the Contractor's Leadership capabilities in leading the direction of the overall Laboratory, the responsiveness of the Contractor to issues and opportunities for continuous improvement, and corporate office involvement/commitment to the overall success of the Laboratory.*
Office of Science (SC)
- FRA has continued to deliver important physics results while building major new projects. The biggest highlight of the year was the new results on Muon g-2 experiment, which delivered a more precise measurement of the anomalous muon magnet moment that continues to be in tension with the Standard Model of Particle Physics.
- […]
- Response to incidents such as the fall at the PIP-II Linac complex have put a spotlight on Environment, Safety and Health (ES&H) challenges at the Laboratory. Leadership must ensure operations at the Laboratory are aligned with Integrated Safety Management (ISM) and safety culture principles to enable operational excellence.
- It is noted that the Contractor Assurance System (CAS) has continued to undergo change including the formation of a new office with full time leadership, however substantive benefit will only come through disciplined use and implementation throughout the Laboratory. Outside assessments have provided insightful recognition of the degree of calibration and benchmarking needed to improve execution. The Laboratory needs to emphasize and make forward progress on the Contractor Assurance System (CAS) program to enhance the robustness of laboratory programs.
- […]

**Objective 4.2: Management and Operation of the Laboratory**
Site Office
The Contractor Assurance System (CAS) has continued to undergo change including the formation of a new office with full time leadership, however substantive benefit will only come through disciplined use and implementation throughout the Laboratory. Outside assessments have provided insightful recognition of the degree of calibration and benchmarking needed to improve execution. Realization of the value in pursing robust extent of condition determinations and extended opportunities to learn and improve from lessons learned elsewhere and in tangential areas is improving. Business as well as operational events benefited from these initial efforts. While this feedback is repetitive from FY 2022, it is restated to ensure understanding of the current state remains the same. Forward progress must be made on CAS implementation and the leveraging of the CAS program to enhance the robustness of Laboratory programs.

Procurement and acquisition processes have continued to have operational impacts to the Laboratory, however many of those impacts are as a result of addressing compliance gaps and resetting historical norms.

Response to incidents such as the fall at the PIP-II Linac complex have put a spotlight on ES&H challenges at the Laboratory. Leadership in response to those challenges have not always aligned with an organization that embraces the principles of Integrated Safety Management (ISM), and in many instances have lacked appropriate formality and safety culture in lieu of project execution milestones. Lessons learned and extent of condition reviews continue to need additional attention should the Laboratory aspire to operational excellence and principles of a learning organization.

Creation of a taxonomy matrix in FY 2023 will enable FRA to clearly identify services provided and application of policies and procedures to individuals accessing Fermilab.

**Goal 5.0 Sustain Excellence and Enhance Effectiveness of Integrated Safety, Health, and Environmental Protection**

*This Goal evaluates the Contractor's overall success in deploying, implementing, and improving integrated ES&H systems that efficiently and effectively support the mission(s) of the Laboratory.*

**Site Office**
- While acknowledging that more focused attention has been placed on specific Environment Safety and Health (ES&H), challenges remained over the course of the year, including a lack of effective program implementation, informality of operations, and weaknesses in site system/process maturation and effectiveness.
- Work Planning and Control (WP&C) activities span multiple systems such as Worklist, FAMIS, and IMPACT tools, which can create confusion around approval authorities and who authorizes the final work package and reviews the hazards. In addition, operational deviations are not consistently addressed and reviewed upon discovery in the field and prior to commencement of work.
- Hazard analysis documentation has been subpar, un-signed, or missing, with troubling inconsistencies across Laboratory divisions and directorates.
- Safety assessments are not being consistently completed as scheduled with delays of years resulting in operational compliance uncertainties, additional administrative controls, and higher operational risk.
- While the Radiation Protection Program (RPP) documentation has improved, FRA continues to struggle with effective implementation of program elements as identified through the two DOE RPP Program assessments, FSO oversight observations, and operational events. FRA continues to be challenged in providing appropriate levels of radiation protection through As Low As Reasonably Achievable (ALARA) principles for both personnel and members of the public.
- Continued ineffective RPP implementation and a lack of Program formality has resulted in regular status meetings between FSO and FRA staff and management to assure continued program improvements are being made.
- FRA completed significant updates and improvements to Occupational Safety and Health program documentation, including Construction Safety, Radiation Protection, Work Planning and Control, Accelerator Safety, and Event Response Program.
- Intervention and direction was necessary to address issues and risk associated with injecting tritiated water into the Industrial Cooling Water (ICW) system which is the source of fire protection water supply across the complex.

**Objective 5.1: Provide an Efficient and Effective Worker Health and Safety Program**

FRA continues to identify improvements and work toward improving the quality and oversight of their subcontractor safety program. The Environment, Safety and Health (ES&H) subcontractor safety group provides oversight to construction projects throughout all phases of the work and continues to try and expand the quantity and quality of their oversight. FRA provided diligent oversight of excavation tasks for culvert repair jobs around the Laboratory which were completed on time without health and safety incidents. However, the incident at the excavation job to install ground cables and a concrete dewar foundation at IB4 highlights the need for more rigor in the Laboratory's work planning and control process. While the quality of Hazard Analysis documentation has shown some improvement early in the fiscal year, FSO is still identifying significant shortcomings with the quality of WP&C documentation. There have been numerous instances of inconsistent implementation in subcontractor expectations involving work planning and control and the flow down of contract requirements to sub-tier contractors. FRA also needs to clearly define roles and responsibilities for staff charged with subcontractor worker health and safety oversight and ensure that training and qualifications are commensurate with areas of responsibility. These concerns have been identified and communicated to the Laboratory as a result of FSO oversight as well as the DOE PIP-II Accident Investigation Report. This incident highlighted program issues with the flow-down of contractual requirements, verification of subcontractor training, inadequate implementation of hazard analyses/work planning and control, and others as identified in the accident investigation report. Efforts to restart work after the pause due to this PIP-II incident were not well coordinated or documented and resulted in substantial DOE involvement after the fact.

FSO has observed Laboratory work where the hazard analysis documentation has been subpar, un-signed, or missing. The WP&C control documents for pure hydrogen storage and use are inconsistent across various Laboratory divisions. FSO formal assessments and targeted surveillances in Laser Safety, Use and Storage Hydrogen Gas Storage and Pyrophoric Materials show that the Laboratory has well written programs to protect workers and the environment from unmitigated exposures to known hazards. However, an independent Issues Management Assessment found that institutional requirements for issues management are not uniformly implemented across the Laboratory.

At the LBNF-Far Site (FS), FRA continues to ensure that WP&C efforts of its subcontractors develop thorough plans and effectively manage work in the field. The design review process involves all stakeholders, however, as evidenced by events at LBNF, this review process has been shown to have gaps that could at times lead to deficiencies in WP&C when implementing approved designs. FRA responded to this trend by refocusing efforts and engaging their subcontractors to review build-plans and HA's for large scopes of work that have a risk of being altered sufficiently to warrant a new analysis and accurately capturing the work being done and methods being used to properly address the associated hazards properly mitigated. While the initial work planning and control documentation for the NuMI target move during July 2023 lacked a level of rigor necessary to evaluate high hazard high risk activities during off hours on a weekend, the Laboratory eventually performed a comprehensive risk analysis and appropriate work planning and control measures which resulted in a successful move of this highly radioactive NuMI component.

The Laboratory's successful transportation of the Short Baseline Near Detector (SBND) as well as the NUMI horn, and target demonstrated solid work planning and control for some high-risk activities.

FRA continues to monitor and proactively respond to the Nitrogen Dioxide (NO2) sampling data obtained to ensure the As Low As Reasonably Achievable (ALARA) principle is maintained for exposures to personnel. This data has been utilized in identifying and resolving issues with equipment that was starting to have mechanical issues leading to increased emissions. FRA continues their rigorous sampling and working on controlling the hazard.

FRA's Radiation Protection Program (RPP) continues to experience numerous operational challenges, identified as part of FSO's operational oversight.

The new DOE Order O420.2D Safety of Accelerators was issued in September 2022 and was added to FRA's contract. FRA developed a comprehensive implementation plan that was approved by FSO. Working groups were formed to review all areas impacted by the revised Order, including the Accelerator Safety Envelope (ASE), Safety Assessment Document (SAD), Unreviewed Safety Issue (USI) Process, Radiation Generating Devices (RGD), Accelerator Readiness Review Process (ARR), Contractor Assurance System (CAS) and Reviews, Training, Fermilab Environmental, Safety & Health Manual (FESHM) and Fermilab Radiological Control Manual (FRCM) Updates.

During FY 2023, FRA submitted a revision to the ASE and several SAD chapters in preparation for the upcoming ARR for the Spinquest experiment and restart of the NM beamline. During the FSO document review, several items were identified that required action prior to approval of the ASE and concurrence of the SAD Chapters ahead of the Spinquest ARR. As currently written, the SAD and ASE do not have the level of detail needed to fully understand all the credited controls and authorized operating limits for the accelerator complex. The comments FSO provided to FRA addressed several issues such as the following:
1. Clearly define credited controls and required maintenance process
2. Clearly define minimum shielding requirements
3. Clearly state and define administrative credited controls
4. Inclusion of radiation monitors as credited controls
5. Inclusion of oxygen deficiency hazard (ODH) controls as credited controls

The comments provided to FRA also addressed necessary high-level changes that were needed to rework the ASE and SAD for operations of the Fermilab accelerator. FSO charged the Laboratory to develop and externally validate a new SAD and ASE.

An external ARR was conducted to validate the updated ASE and SAD documentation. The ARR team determined significant work is needed in order for the updated SAD and ASE to fully address FSO comments and to be compliant with DOE Order420.2D. The team identified areas of concern including the use of compensatory measures in the ASE without properly evaluating the control in the SAD, the SAD chapters did not adequately develop and identify credited controls, the USI process needed improvement, and the use of thirty-year-old outdated shielding assessments as a reference in the ASE was problematic.

**Goal 6.0 Deliver Efficient, Effective, and Responsive Business Systems and Resources that Enable the Successful Achievement of the Laboratory Mission(s)**
*This Goal evaluates the Contractor's overall success in deploying, implementing, and improving integrated business systems that efficiently and effectively support the mission(s) of the Laboratory.*
**Site Office**
- FRA Procurement timelines, backlog, and delays were reduced during FY 2023, resulting in notable improvements in support of the mission and site execution. Building upon this improvement, the Procurement Department should continue its renewed focus to improve documentation and compliance in the areas of acquisition strategy and planning, market research, selection of subcontract type, evaluation criteria development, source selection, exercising and incorporation of subcontract options, and subcontract administration.

- While FRA achieved some success in talent management, the effects of not having a senior director for Human Resources (HR) in place throughout the course of FY 2023 was impactful. This vacancy leads to a missed opportunity to develop a clear agenda for HR and the invaluable support that the HR organization provides to the Laboratory.
- With the establishment of a new Contractor Assurance office, contractor should be the architect of cultivating a cross cutting communication environment, ensuring responses to undesired events minimize reoccurrences, assessing corrective action effectiveness, tracking performance metrics, etc. This system should have reportable metrics across the operations and science functions that can convey the Laboratory Agenda and Enterprise Risks clearly to FRA and DOE management.
- Technology transfer activities continued to advance the relationships with industry partners with several new initiatives announced. International agreements will provide the foundation for the collaborations needed for the major international experiments.

**Objective 6.2: Provide an Efficient, Effective, and Responsive Acquisition Management System and Property Management System**

Acquisition Management System

DOE commends ongoing activities to make improvements to the Procurement organization structure, policies and procedures, and compliance. Procurement was proactively responsive to the FY 2023 Procurement Evaluation and Reengineering Team (PERT) Report demonstrating a willingness to make improvements to the procurement system that will drive compliance. This is an unprecedented level of self-reflection and desire to improve that is well received by DOE.

The Procurement Department underwent a reorganization and showed improvement through onboarding an interim Chief Procurement Officer, in addition to filling 27 positions leading to notable improvements in quality, compliance, mission support and customer needs.

The Procurement Department adopted a people, process and technology improvement strategy that is geared towards attracting, developing, and retaining a skilled workforce.

FRA was able to reduce a backlog of requisitions by ~600 in FY 2023 Quarter three and Quarter four, developed and deployed a requisition tracking report for procurement staff and customers, deployed a Buy American Act policy, implemented a Purchase Card central record repository, and initiated the development of a PERT corrective action plan (CAP) to address the FY 2023 PERT final report findings.

Twelve out of thirteen open internal audit findings were successfully closed.

FRA increased eMarketplace purchases by $2.2M in FY 2023, increasing customer service with embedded compliance and efficiencies. Allowing the procurement staff to focus on more complex procurement actions and the customer to perform self-service requests has aided in reducing wait time and increasing customer satisfaction.

Going forward, the Procurement Department should strive to improve documentation and compliance. Procurement will need to emphasize deploying new procedures through a coherent change management process. Change management must include training for the affected audience. DOE is anticipating progress in the areas of acquisition strategy and planning, market research, selection of subcontract type, evaluation criteria development, source selection, exercising and incorporation of subcontract options, subcontract administration, and the development of a policy on oral presentations for DOE review.

Property Management System

FRA's property management continues to execute all corrective actions identified from the FY 2022 triennial audit, and all work is scheduled to be completed on schedule to meet the required CAP deadlines. There has been increased training and development of property staff (e.g., ten employees completed a High-Risk Property Workshop led by DOE's Export Compliance Assistance Program (ECAP)). Collaboration and coordination with the Radiation Safety team to create new processes and established new e-waste and scrap metal sales contracts to allow for release of scrap after a two-year hold. FRA has coordinated with DOE Consolidated Service Center (CSC) and FSO to develop new contracting mechanisms for equipment sales (e.g., GovDeals). Projects have been executed for removal of hazardous materials that have been stored at the Laboratory for decades (e.g., removal of Deuterium and Nevis Blocks).

# Appendix C – Table of Content of the DOE's RFP for M&O Bidder

We have extracted here below a Table of Content of the REQUEST FOR PROPOSALS (RFP) NO. 89243123RSC000083 FOR THE SELECTION OF A MANAGEMENT AND OPERATING CONTRACTOR FOR THE FERMI NATIONAL ACCELERATOR LABORATORY (FNAL) announced by DOE on January 3, 2024.
https://science.osti.gov/Acquisition-Management/M-and-O-Competitions/Procurement-Information/Request-for-Proposals?utm_medium=email&utm_source=govdelivery

This contract shall be effective up to and including December 31, 2029, unless sooner terminated according to its terms. The contract transition period is from the award date through December 31, 2024. The Contractor will assume full operational control of the Laboratory on January 1, 2025. The contract's maximum period of performance shall not exceed 20 years and 3 months.

For the first three evaluation years, the Contractor must receive at least a "B+" for both Science and Technology and Management and Operations in year 1, and at least an "A-" for Science and Technology and a "B+" for Management and Operations for years 2 and 3. Beyond year 3, in addition to the latter requirement, the Contractor must also obtain a minimum rating of at least a "B+" for each individual Science and Technology goal and a "B-" for each individual Management and Operations goal.

**PART I**

SECTION A – SOLICITATION, OFFER AND AWARD

SECTION B – SUPPLIES OR SERVICES AND PRICES/COSTS

SECTION C – DESCRIPTION/ SPECIFICATIONS/ STATEMENT OF WORK (25 pp)

C.1 INTRODUCTION
C.2 IMPLEMENTATION OF DOE'S MISSION FOR FNAL

C.3 CORE EXPECTATIONS

**(a) General**
**(b) Program Development and Mission Accomplishment**
**(c) Laboratory Stewardship**
**(d) Operational and Business Management**

C.4 STATEMENT OF WORK
**(a) General**

**(b) Research and Development (R&D)**
1. Mission Accomplishment
    (i) SC High Energy Physics (HEP)
    (ii) SC Basic Energy Sciences (BES)
    (iii) Other Programs
        (A) Providing of facilities to Users
        (B) R&D for non-DOE sponsors consistent and complementary to DOE's mission
        (C) Dissemination/publication of unclassified scientific and technical data and operating experience
        (D) Traning
2. Research Facilities and Major Scientific User Facilities
    (i) HEP User Facility: Booster Neutrino Beam; Muon Campus; Neutrino at the Main Injector (NuMI); PIP-II
    (ii) HEP Scientific Areas and Experiments: MicroBooNE; ICARUS; SBND; Muon g-2; Mu2e; NOvA; DUNE; Quantum Science; MAGIS-100.
    (iii) HEP International Collaborations
    (iv) Accelerator Research and Development: 1. Accelerator and beam physics; 2. Advanced accelerator concepts; 3. Particle sources and targetry; 4. Radio-frequency acceleration technology; and 5. Superconducting magnets and materials; FAST/IOTA.
3. Scientific Program Management

**(c) Protection of Workers, the Public, and the Environment**
The safety and health of workers and the public and the protection and restoration of the environment are fundamental responsibilities of the Contractor. The Contractor shall establish an environment, safety and health (ES&H) program operated as an integral, but visible, part of how the organization conducts business, including prioritizing work and allocating resources based on risk reduction. A key element is implementation and sustainment of an Integrated Safety Management System to ensure all work activities are performed in a manner that prevents disruption of the Laboratory's missions by preventing fatalities, minimizing injuries and illnesses, minimizing exposures to hazardous substances and materials, preventing environmental releases in excess of established limits, implements as-low-as-reasonably-achievable releases and exposures, and preventing property loss.

The Contractor shall maintain an organization that supports effective ES&H management by ensuring appropriate levels of ES&H staffing and competence at every level within FNAL. Specifically, the Contractor shall assure that employees are trained, qualified, and involved in aspects of the organization's activities, including providing input to the planning and execution of work, and identification, mitigation, or elimination of workplace hazards. The Contractor shall, similarly, assure that subcontractor employees are trained and qualified on job tasks, hazards, DOE and FNAL safety policies, expectations, and requirements, and shall flow down applicable ES&H requirements down to subcontractors. The Contractor shall, as appropriate, consider ES&H performance in selection of its subcontractors and incorporate ES&H requirements into subcontracts.

The Contractor shall perform all activities in compliance with applicable health, safety, and environmental laws, orders, regulations, national consensus standards, governing agreements and permits executed with regulatory and oversight government organizations.

Incorporating integrated line management, the Contractor shall put in place a system that clearly communicates the roles, responsibilities, and authorities of line managers. The Contractor shall hold line managers, including direct reports, accountable for implementing necessary controls for safe performance of work in their respective area of responsibility. The Contractor shall establish effective management systems to identify deficiencies, resolve them in a timely manner, ensure that corrective actions are implemented, (addressing the extent of conditions, root causes, and measures to prevent recurrence) and prioritize and track commitments and actions.

**(d) Management and Operation of the Laboratory**
1. Strategic Planning
2. Business Management
    (i) Human Resources Management (HR)
    (ii) Financial Management
    (iii) Purchasing Management
    (iv) Property Management
    (v) Legal Services
    (vi) Information Technology Management
    (vii) Other Services
3. Project Management
4. Environmental Management
    (i) Environmental remediation and facility deactivation
    (ii) Construction and maintenance of facilities and infrastructure
    (iii) Tritium management
5. Community Involvement
6. Safeguards and Security (S&S)
7. Cyber Security
8. Emergency Management
9. Waste Management
10. Laboratory Facilities and Infrastructure
11. Sustainability

**(e) University and Science Education Program for Workforce Development**
1. Operation of the U.S. Particle Accelerator School
2. Joint experimental programs with colleges, universities, and nonprofit research institutions
3. Interchange of college and university faculty and Laboratory staff
4. Student/teacher educational research programs at the pre-collegiate and collegiate level
5. Post-doctoral programs
6. Arrangement of regional, national, or international professional, meetings or symposia
7. Use of special Laboratory facilities by colleges, universities, and nonprofit research institutes, or
8. Provision of unique experimental materials to colleges, universities, or nonprofit research institutions or to qualified members of their staff.

C.5 PLANS AND REPORTS

SECTION D – PACKAGING AND MARKING

SECTION E – INSPECTION AND ACCEPTANCE

SECTION F – DELIVERIES OR PERFORMANCE

SECTION G – CONTRACT ADMINISTRATION DATA

SECTION H – SPECIAL CONTRACT REQUIREMENTS (67 pp)

## PART II

SECTION I –CONTRACT CLAUSES (302 pp)

## PART III

SECTION J –LIST OF ATTACHMENTS
Appendix A – Advance Understanding on HR
Appendix B – Performance Evaluation and Measurement plan
Appendix C – Special Financial Institution Agreement
Appendix D – Contractor's Commitments
Appendix E – Key Personnel
Appendix F – Reserved
Appendix G – Purchasing System Requirements
Appendix H – Small Business Subcontracting Plan
Appendix I – DOE Directives/ List B
Appendix J – Treaties and International Agreements/Waived Inventions
Appendix K – Reserved
Appendix L – Performance Guarantee

# PART IV

SECTION K – REPRESENTATIONS, CERTIFICATIONS, AND OTHER STATEMENTS OF OFFERORS (44 pp)
SECTION L – INSTRUCTIONS, CONDITIONS, AND NOTICES TO OFFERORS (65 pp)

**L.1 INSTRUCTIONS FOR THE SUBMISSION OF PROPOSALS AND UNIFORM CONTRACT**
(c) Offerors shall submit three separate volumes as follows:
    (1) Volume I (no page limitations) – Consists of the Uniform Contract. Any contract awarded as a result of this RFP shall contain Part I – The Schedule, consisting of Section A and Sections B through H; Part II – Contract Clauses, consisting of Section I; and Part III – List of Documents, Exhibits, Attachments, consisting of Section J. Volume I shall also contain Part IV – Section K.
    (2) Volume II (150 pp maximum) – Capabilities and Approach Proposal. The Capabilities and Approach Proposal, along with the oral presentation, shall clearly demonstrate the Offeror's understanding of the Government's requirements and the Offeror's capability to perform the prospective contract. The Offeror must provide the written information, required by Provisions L.2 through L.6 of this RFP, in Volume II, which fully demonstrates the Offeror's capability, knowledge, experience, and understanding with respect to the evaluation factors described in Section M Provision entitled "Capabilities and Approach Evaluation Factors" of this RFP. Provisions L.2 through L.6 of this RFP are designed to provide information so that the non-cost evaluation factors specified in Section M Provision entitled "Capabilities and Approach Evaluation Factors" can be evaluated.
    (3) Volume III (no page limitations) – Cost, Fee, Financial and Other Information Proposal. All cost information is to be included in this volume. Requirements for submission of cost information are contained in Section L Provision entitled "Volume III – Cost, Fee, Financial and Other Information Proposal – Instructions" of this RFP. Pursuant to paragraph (b) of Section L Provision entitled "FAR 52.215.1 – Instructions to Offerors-Competitive Acquisition", Offerors shall acknowledge receipt of amendments to this solicitation. In addition to completing Block 14 of the SF 33 to be included in Volume I, Offerors are to provide a written acknowledgement that the Offeror received each amendment to this RFP in Volume III. The written acknowledgement should include identification of each amendment number and the date each amendment was received.

**L.2 SCIENCE VISION AND IMPLEMENTATION PLAN**
(a) Science Vision
The Offeror shall provide documentation that describes its vision for the Laboratory that creates the conditions to enable achievement of the DOE mission, transformational and breakthrough science, and the delivery and optimization of FNAL's world class scientific facilities; enhances the Laboratory's leadership in the national and international arena for research and development; fosters its central role in the international research ecosystem to deliver breakthrough science results; attracts, develops and retains a highly skilled workforce; cultivates and sustains a diverse, equitable, inclusive, and accessible Laboratory culture; and, effectively coordinates activities within the DOE complex, nationally, and internationally.

(b) Implementation Plan

The Offeror shall describe its approach for the implementation of its Science Vision. The proposed approach should demonstrate an understanding of the Laboratory's current major research programs, identify any new directions that the Laboratory might take, and discuss how the Laboratory could most effectively contribute to meeting the challenges facing the major programs in **Section C** entitled "Description/Specifications/Statement of Work". The Offeror shall describe its overall approach to implementing its Science Vision, and shall, at a minimum, include a discussion of its:

• Planned approach to enable achievement of the DOE mission and leverage programs to foster transformational and breakthrough science;
• Planned approach to bring on-line a robust implementation of the LBNF/DUNE and PIP II projects and approach for their future operations and possible upgrades; • Planned approach for maintaining, enhancing and developing cooperative and collaborative partnerships with universities and industry, including emerging research institutions, to enhance the Laboratory's leadership in the national and international arena for research and development;
• Planned approach for fostering the Laboratory's central role in the international research ecosystem to deliver breakthrough science results;
• Planned approach for attracting, developing, and retaining a highly skilled workforce of existing and new scientific personnel with high stature in their disciplines; plan for joint appointments (if applicable); and how it would use the resources of the Laboratory to help develop and educate the next generation of scientists and engineers;
• Planned approach for cultivating and sustaining a diverse, equitable, inclusive, and accessible Laboratory culture;
• Planned approach to support technology transfer and enhance the Strategic Partnership Projects portfolio; and
• Planned approach to leading and coordinating scientific activities at FNAL and within DOE, nationally and internationally; and the Offeror's approach to maintain engagement and positive relations and communications with DOE and other interested stakeholders.

**L.3 LABORATORY OPERATIONS**

The Offeror shall demonstrate a thorough understanding of Laboratory Operations necessary to successfully accomplish **Sections C.4(c) and C.4(d)** of the Statement of Work. The Offeror shall describe its approach for achieving excellence in all areas of operations and business management while maintaining compliance with DOE and other applicable requirements. Areas to be addressed include, but are not limited to:
● Integrated ES&H programs and processes that demonstrate a commitment at all levels within the Laboratory to the safety and health of workers and the public, as well as the protection and restoration of the environment. [See C.4(c)].

• An integrated management system capable of producing implementation-level plans, programs and procedures for the management and operation of the Laboratory.
• A robust, broad scope contractor assurance program to self-assess overall performance and drive continuous improvement of Laboratory operations and management.
• Business management systems [see Section C.4(d)(2)] and how they will be applied to ensure efficient and effective operation, protection and maintenance of the Laboratory's assets, and ability to function as a DOE laboratory.
• Systems for the efficient and effective management of all Laboratory facilities and infrastructure, safeguards and security, cyber security, emergency operations, waste operations, sustainability, and Laboratory strategic planning.
• Involvement of small business in meaningful contract performance, including the extent, variety, and complexity of the work to be performed.
• Developing and maintaining positive community relations and communications with DOE and other interested stakeholders.
• Approach to build, develop, maintain, and encourage a diverse workforce that addresses, at a minimum, the Offeror's approach for promoting diversity through (1) the Contractor's work force, (2) educational outreach, (3) community involvement and outreach, (4) subcontracting, (5) economic development (including technology transfer), and (6) the prevention of profiling based on race or national origin.
• Approach to successfully deliver the Laboratory's project portfolio that details the resources, organization, interfaces, and other elements the Offeror considers necessary.

**L.4 OFFEROR ENGAGEMENT**
<u>Key Personnel</u>
<u>Organizational Structure</u>
<u>Governance Approach and Corporate Assurance</u>
<u>Offeror's Commitments</u>

**L.5 PAST PERFORMANCE** (within last 3 years)

**L.6 TRANSITION PLAN** (3 months maximum)

**L.7 VOLUME III – COST, FEE, FINANCIAL AND OTHER INFORMATION PROPOSAL – INSTRUCTIONS**
[…]

**L.12 ORGANIZATIONAL CONFLICT OF INTEREST (OCI)**
The statement of work includes certain work activities that may present an OCI. ==A contract shall not be awarded to any Offeror having an unresolved OCI.== (See Section K Clause entitled "DEAR 952.209-8 – Organizational Conflicts Of Interest Disclosure- Advisory and Assistance Services".) […]

SECTION M – EVALUATION FACTORS FOR AWARD

| Capabilities and Approach Evaluation Factors ||
|---|---|
| Factors | Description |
| A. | Science Vision and Implementation Plan |
| B. | Laboratory Operations |
| C. | Offeror Engagement |
| D. | Past Performance |
| E. | Transition Plan |

Factors A and B are of equal importance to each other, and are individually of more importance than Factors C, D, and E individually. Factor C is of greater importance than Factor D. Factor D is of greater importance than Factor E. Collectively, these Capabilities and Approach Evaluation Factors are significantly more important than the total evaluated price.

# Appendix D – Unplanned Radiation Dose Incident

The following is an excerpt of a draft Consent Order between the DOE and Fermilab concerning the incident in May 2022:

"In early May of 2022, as part of a research and development project, an employee modified the operating parameters of a radiation generating device (RGD) in the Proton Source Test Stand (PSTS) area of the Linac Building at the Fermi National Accelerator Laboratory. Subsequently, there were indications that the dose rates in the area were elevated (e.g., supplemental dosimeter readings off scale, elevated radiation surveys, and application of shielding to reduce survey readings). Despite the unexpected and unanalyzed dose rates resulting from the modified parameters, the worker and supporting radiological control technician (RCT) did not report or record the conditions and completed the work without stopping. On August 25, 2022, FRA reviewed the second quarter dosimetry results and discovered the worker had received an unplanned dose of 530 millirem. FRA's investigation concluded it was an actual exposure (as opposed to an erroneous reading) that was likely due to ionizing radiation from the RGD.

On September 13, 2022, FRA initiated an internal investigation of the unplanned radiation dose. The laboratory's investigation team focused on Human Performance Improvement and conducted a thorough review of the factors associated with the incident. The team approved and issued the Unplanned Radiation Dose Event Review Report Summary on October 18, 2022. The laboratory's investigation revealed programmatic weaknesses in work planning and control that resulted in less than adequate documentation, review, and authorization of PSTS operations. The investigation also identified weaknesses with escalating issues appropriately, following supplemental personal dosimetry monitoring and reporting expectations, and operations oversight."

FRA self-identified the following potential non-compliances, which were reported in the DOE Noncompliance Tracking System (NTS) under report NTS-SC-FSO-FRA-FERMIBOP-2022-0010490, Unplanned Radiation Dose, on October 18, 2022:
    10 C.F.R. Part 835, Subpart E, Monitoring of Individuals and Areas:
    § 835.401, General Requirements;
    10 C.F.R. Part 835, Subpart H, Records:
    § 835.701, General provisions;
    § 835.703, Other monitoring records;
    § 835.704, Administrative records;
    10 C.F.R. Part 835, Subpart K, Design and Control:
    § 835.1001, Design and Control;
    § 835.1003 Workplace controls.

On February 6, 2023, the Office of Enforcement notified FRA of its decision to investigate the facts and circumstances associated with the unplanned dose and potential deficiencies in the implementation of 10 C.F.R. part 835, Occupational Radiation Protection. The Office of Enforcement conducted an onsite investigation April 24 through 27, 2023. On August 22, 2023, DOE transmitted the Investigation Summary report to FRA and an enforcement conference was held on September 20, 2023.

# Appendix E – A Textbook Case Study in Human Rights

My name is Erica Smith and from 2016-2023 I was a postdoctoral fellow at Indiana University, a Fermilab User, and a member of the NOvA and DUNE Experiments. My interactions with the Fermi Research Alliance and various members of Fermilab leadership have been frustrating at best and harmful at worst. Their actions have left the impression on myself and others that they prioritize maintaining a positive reputation by trying to make problems go away at the expense of victims over cultivating a safe working environment for Fermilab employees and Users. This is my experience.

In May 2018, I reported another Fermilab User and member of the NOvA and DUNE experiments to the Neutrino Division's Human Resources Representative, which, at that time, was the only avenue to report misconduct, including sexual misconduct. The HR representative brought in the assistant general counsel almost immediately, and they began an investigation that would continue over the next several months and was handled poorly in a number of ways.

In the month following my report, I had multiple phone calls with the HR representative and assistant general counsel, emailed them a written summary of my report, and sent pdfs of text chats with friends that had detailed descriptions of the incidents. They asked me to schedule an in-person meeting with them; when I tried, they told me they were going on vacation and they would follow up when they returned. After a month with no contact, I followed up again and they told me we no longer needed to meet. The investigators did not contact me again until scheduling our final meetings months later; I received updates only because my supervisor reached out to the person who was COO in 2018 to check the status of the investigation.

Before they stopped contacting me, the assistant general counsel and HR representative assured me they would inform me when they were going to interview the subject of the report so I would know when he became aware of the report, as we were still working on the same experiments together. They did not; I found out that they interviewed him long after the interview took place.

At no point during the investigation was I told who would know about the investigation or how much information they would receive. I found out the 2018 COO was involved only because someone I reported to at a different institution told me he had been contacted by the COO. I was not told that the general counsel would be involved until he attended our final meeting in September 2018 without ever having been mentioned to me or copied on emails with me. I heard from others that the head of the Neutrino Division, a person that I had interacted with professionally, knew of the report, but they were unsure how many details he knew. Imagine giving a professional talk and not knowing who in the room had read graphic descriptions of a traumatic experience that you shared only with close friends, never thinking that anyone else would read them; this is the result of having investigators with no training in trauma-informed investigations.

When the investigation ended in September 2018, they told me: "the result of our investigation is that FRA is unable to conclude that there was a violation or threat to the Fermilab community or site for which FRA can take personnel or site access action, recognizing that FRA does not manage the NOvA collaboration and FRA does not employ or contract with either you or [him]." I asked if they determined misconduct did or did not take place and received the same statement in an email from the 2018 COO with no further clarification.

Because FRA refused to give an actual finding from the investigation, NOvA leadership was not able to take action to protect me at work; to keep doing my job, I was forced to interact with him during our multiple weekly phone meetings, via our experiment's public Slack channels, and during in-person meetings. I had at least one panic attack at every in-person meeting that we both attended for the next year, and had sporadic panic attacks during our phone meetings. During this year, working group leadership assignments changed, and I found myself reporting directly to him for a short time. I was diagnosed with unspecified trauma disorder and saw a trauma therapist at least weekly – multiple times a week leading up to in-person meetings – just to be able to show up at work. I struggled deeply with my mental health and lost months of productivity vital to securing a permanent position in academia.

At the end of 2019, I requested that NOvA leadership find an external investigator as was allowed by the NOvA Code of Conduct, thinking that because FRA did not actually present a finding of fact as part of their investigation, I should be afforded an investigator who would. At that time, NOvA leadership agreed and in early 2020 I was told that they identified an external consultant to investigate and would contact the DOE to find funding for it. After the subject of the original report was notified that I requested an external investigator he weaponized Fermilab HR against me, reporting me for "making defamatory statements." His report was not treated as retaliation, and the new COO as of 2020 emailed me saying that "FRA is not the 'trier of fact' in determining whether [my] statements are false and defamatory and FRA is the wrong forum to seek relief in your ongoing grievance with each other. Therefore, FRA declines to take up [his] complaint, and considers the matter closed." She also stated, regarding the 2018 report, that "FRA concluded, based on the facts identified in the investigation, that no further action was warranted." After I received this email, NOvA leadership told me the line "no further action was warranted" was enough of a finding from the original investigation and they would not move forward with an external investigator.

After FRA chose to abdicate all responsibility regarding this situation for the second time, the subject of the report began a campaign of online harassment against me. Over the course of several months, he impersonated me on Twitter, posted nude pictures of other women pretending to be me, and gave my phone number to men who contacted the fake profiles so they would send me nude pictures. He signed me up for hundreds of websites including porn sites and Nazi-affiliated websites. He used my home address to sign me up for subscription services and even had someone come to my address with a baby-proofing service. He sent anonymous threats of violence via text and Twitter DM, and threatened to deceive police into sending an armed response to my address. I contacted my local law enforcement multiple times to ensure they had a note to confirm any reports of violence at my address before sending an armed response. I bought a portable lock for my door and kept a baseball bat near my desk because I was afraid of where he might be publishing my home address. In mid-2020 I hired a lawyer to find proof of his harassment and bring a lawsuit against him. We won the lawsuit in October 2022; you can find the press release from my lawyers here:
https://www.brettwilson.co.uk/blog/press-release-academic-pays-50000-damages-to-colleague-falsely-touted-online-as-a-sex-worker/ and an article in The Guardian written about the situation here:
https://www.theguardian.com/education/2022/oct/12/former-ucl-academic-to-pay-damages-after-harassing-colleague-for-months .

During the harassment campaign and the subsequent lawsuit, he continued to interact with me in public Slack channels and gave unprompted feedback on my work during weekly phone meetings. I realized I could no longer tolerate being forced to interact with him, and for my mental health, I left the working group I had been in for nearly 4 years. I picked up a new analysis in a working group in which he was not active, effectively abandoning all my previous work and putting me extremely behind in terms of showing prospective search committees I was capable of a complete physics analysis, an essential part of being hired as tenure track faculty.

After the lawsuit was public, I was contacted by the new Fermilab director and new CRO to have a meeting. In November 2022 I met with them, the same Fermilab general counsel from 2018, and my lawyer, and told them everything I've said in this testimonial. Their general counsel even agreed with me when I told them I felt I had been "ghosted" by the original investigators. During this meeting, I received a lot of promises that the system would change, but I've seen no evidence that it has.
I will note that the Fermilab reporting system has changed since I reported, and one now makes reports through a third-party system. However, I've listened to multiple presentations on the topic and while the reporting system itself has changed, reports are still investigated by "subject matter experts" at Fermilab. Despite asking for clarification on what this means, particularly in the case of sexual harassment and assault – who at Fermilab is an "expert" in investigating that? – there has never been a clear answer other than to say that they are employed by Fermilab, i.e., people beholden to protecting the interests of Fermilab and FRA, not employees or Users. This is a problem that has been pointed out, at the very least, to Fermilab's general counsel, but to my knowledge no changes have been made to the investigative process.
The problem starts at the top, with FRA and the Fermilab directorate. Everyone who hears about my experience agrees that it was terrible, including the people with the power to ensure it doesn't happen again, yet, all I've seen is leadership sitting on their hands while claiming their hands are tied. Things will never change until Fermilab learns the meaning of institutional courage and stops prioritizing protecting bad actors to avoid legal retaliation over creating and maintaining a safe working environment.

There is no happy ending to this testimonial. I left the field in 2023 as the lost productivity through the majority of my postdoc affected my ability to get a permanent position. I've been unemployed for months, trying to recover from all that has happened to me. None of this had to happen; if FRA had acted appropriately in 2018, I might still be a physicist today. I'll never know. There is one silver lining that came from my experience, which was coauthoring a 54 page contributed paper regarding the climate of the field where my focus was experiment codes of conduct and concrete suggestions to labs and funding agencies to make the changes needed to effectively remove bad actors, which can be found here https://arxiv.org/abs/2204.03713 , and support for these suggestions was included in the P5 report. I may no longer be part of the field, but perhaps my experiences will mean others in the future don't have to go through what I went through; that will have to be enough.

# Appendix F – Fermilab Concern Reporting System

As presented on December 9, 2022, within an all-hands meeting at Fermilab.

## Ways to Report Concerns at Fermilab

We encourage employees and Fermilab users/affiliates to use any of the following avenues for reporting concerns of improper conduct; harassment or discrimination; unsafe conditions; conflicts of interest; fraud, waste or abuse; violations of law or policy; or other irregularities:

- Managers or Directorate/Division/Department Leadership
- HR Partners or Human Resources Leadership
- The Office of General Counsel
- The Fermilab Security Department
- The **Fermilab Concerns Reporting System**, a third-party provided hotline/website hosted by **Integrity Counts** where reporters may either self-identify or remain anonymous.
  https://app.integritycounts.ca/org/fermilab
  Phone Hotline: 866.921.6714 (USA), 00-800-2002-0033 (Switzerland).

In addition, FRA employees and Fermilab users may report directly to the Department of Energy (DOE) as follows:
- DOE Employee Concerns Program at the Chicago Integrated Support Center – Hotline (800) 701-9966, or scfieldecp@science.doe.gov, or employeeconcernsprogram@doe.gov
- DOE Fraud, Waste, and Abuse Hotline – (800) 541-1625 or IGHotline@hq.doe.gov
- DOE Office of the Inspector General – (202) 586-4073, or IGHotline@hq.doe.gov, or https://www.energy.gov/ig/complaint-form. Individuals who contact the DOE/OIG Hotline may self-identify or remain anonymous.

**FRA policy prohibits retaliation against individuals who make a report through any of these channels.**

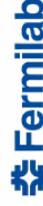

48   12/9/2022     Merminga, Fleming, Tingey | Thank you for 2022 | All Hands

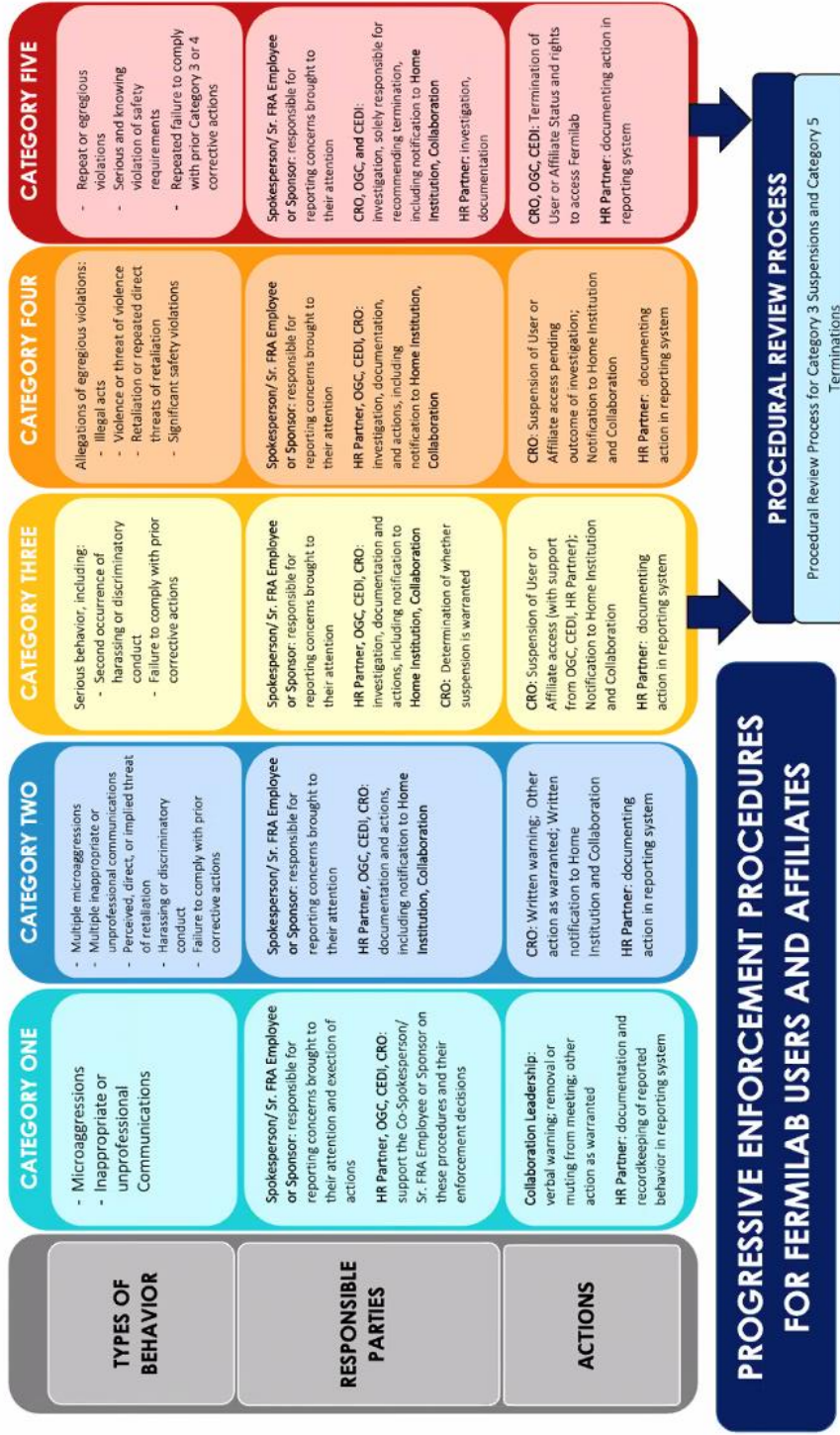

# Lab-Wide – Intake Metrics

- **Integrity Counts** was implemented in October 2020
- Reported concerns at Fermilab have increased from FY 2021 (30) to FY 2022 (53)
- Concerns are reported in two ways: self-identified and anonymously
  - Self-reported concerns increased by 8% from FY 21 to FY22
  - Anonymous reports account for 36% of all concerns received
- 78% of concerns are reported by Employees, 18% by Users/Affiliates, and 4% unidentified

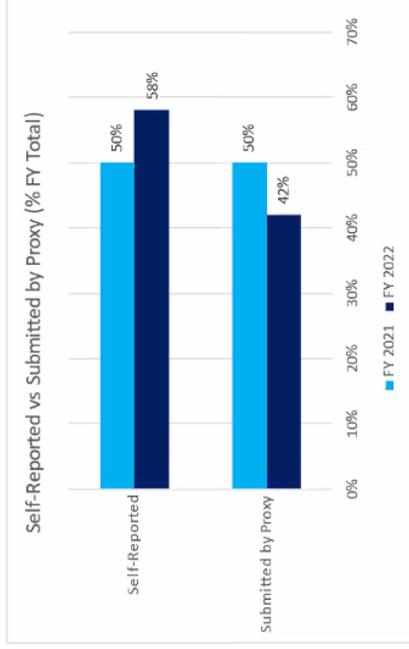

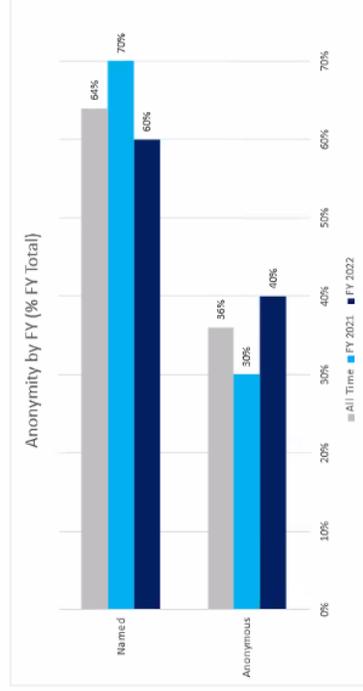

**Note:** Proxy concerns are concerns reported into the system that have been raised through all other channels of reporting concerns. See this website for more information on these channels.

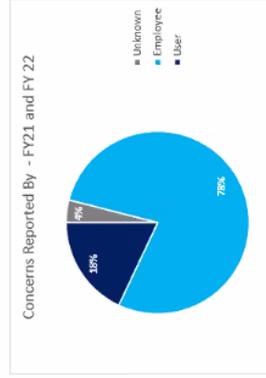

50   12/9/2022   Merminga, Fleming, Tingey | Thank you for 2022 | All Hands

# Lab-Wide – Concern Types and Outcomes

- The top five case types received and reviewed include:
    - Hostile Work Environment,
    - Equity, Diversity, or Inclusion,
    - Other,
    - Inappropriate / Unprofessional Conduct, and
    - Harassment

- Over 50% of concerns are resolved within 40 days.

- The top outcomes for substantiated reports include:
    - Guidance Provided
    - Corrective / Remediation
    - Written Warnings
    - Suspended Without Pay

- When a concern is substantiated, action is taken by the Laboratory. In addition to addressing the matter directly with individuals involved, the Lab:
    - Shares data with Senior Leadership Team
    - Metrics Committees reviews for trends
    - Reporting to DOE Employee Concerns Program
    - Notification of substantiated concerns are made to home institutions and collaborations pursuant to the User/Affiliates Progressive Enforcement Procedures

## Average Case Closure Times

| All Reports | FY 2021 | FY 2022 |
|---|---|---|
| 77 days | 106 days | 56 days |

**Fermilab**

# Appendix G – Analysis of "Fermilab Concern Reporting System"

The following analysis was provided to Fermilab deputy Director Bonnie Fleming in March 2023 on her slides from Appendix F. <u>The analysis was specific to employees.</u>

1. According to the first slide, complaints can presently be placed with half a dozen different entities. Unless these different entities transfer the information to a single responsible unit, this is not conducive to a good control of the narrative and of the context, especially for the more complex cases that occur over an extended period of time. At least until Summer 2021, HR accepted complaints by anybody, outside of any context. Some HR personnel go as far as verbatim stating in email communications that "they do not want the context."
2. It is not clear at all what is the present role/function of the so-called "third-party". <u>Within this allegedly "new" system, Fermilab lawyers continue to be involved in "investigations"</u>. Because of the demonstrated conflict of interest by the lab HR/Legal Counsel system, complaints about abuses by supervisors (as defined by the NLRD) should be <u>handled and investigated by an independent entity. This should be a law firm with labor lawyers</u>.
3. A major risk of such a fragmented system is for instance that of an employee who has placed a prior complaint against a manager or who has a pending investigation with one of the various entities listed, say the General Counsel. If the manager comes back at her with made-up complaints to a yet different entity, say HR, the victim will have little recourse. HR accepting complaints by her own perpetrators against a prospective victim and the General Counsel using it to denigrate the victim allows the system to protect management or any other perpetrator, as in the case described in Appendix E, independently of their behavior. <u>The present system inherently fosters a culture of abuse of power and harassment.</u> The only way to monitor and control retaliation is to use the same rules for all and send any possibly fake complaints about potential victims to one single office or entity.
4. In the "Progressive Enforcement System for Users and Affiliates" shown in the second slide, the "Responsible parties" are the same half-dozen entities listed in Slide 1, and the Senior FRA employee or sponsor. It is antithetical to the definition of "responsibility" to have so many "responsible" parties. Responsibility is clearly defined when there is one entity that is responsible. Further, this is perfectly consistent with the above, i.e. <u>complaints should be managed and controlled by one single entity, whichever that is</u>.
5. In the assumption that the Senior FRA employee is a manager, how is the conflict of interest of their personal agenda in denigrating victims by retaliation as explained in point 3 addressed? Managers can be responsible ONLY if we have a minimal uncertainty that they do not have conflict of interest with the victim, for instance in cases of retaliation against prior accusations. Unfortunately, based on how several managers were groomed at the lab, there are still some in managerial positions that do not have the necessary ethics to put aside their own agenda. In the Table of Slide 2 there is talk of "verbal and written warnings", as well as "documenting actions in reporting system". However, we haven't seen any changes in practice.
6. Point 5 overlaps with the key concept that <u>a toxic environment cannot be healed without first removing the first order causes</u>. At Fermilab, at least those areas which have been engulfed by cronyism have to be renewed. This is the only way to eliminate the deeply grounded culture of nepotism and cronyism that has established a strong authority in the hands of a few, as well as extended unethical human treatment in some areas of the lab. And <u>this cannot be done without the will of the M&O</u>. Until employees see that cronyism is accepted, as opposed to a merit-based approach for growth, no changes will be effective because the core human and professional values are tainted and therefore have no meaning. Also, another very important concept is leaders leading by example. How can we feel safe in an organization that promotes or even just accepts bad behavior in its leaders?
7. <u>HR does not use the same rules for all</u>. So, for instance, when a manager or a crony being groomed for ascension misbehaves, they do not get an official reprimand, which ends up in one's personnel file for the duration of employment. At most they get what is called a "memo", which has no formal consequences. Several individuals, males and females, who were promoted to high/very high positions of responsibility have received dozens of HR complaints. Using the "memo" stunt, their behavior does not constitute an obstacle to their promotion and they end up occupying leadership positions.
8. The need to reset house in an organization, when charged with its renewal, applies to any cause of ineffectiveness in the organization (not just to cronyism.) There are several factors that come into play for undue promotion of ineffective individuals who end up in charge of responsibilities well beyond their capabilities. <u>These ineffective individuals have to be promptly identified and replaced.</u>
9. <u>Lastly, the Legal Counsel/HR system focuses on isolating the individuals submitting a complaint by often imposing a communication veto with any other employee</u>. The formal justification reads like: "It is inappropriate for me or anyone in laboratory leadership to discuss any aspect of anyone else's concern or investigation with you as you are not their representative." The victim has hence no way to talk to anybody at the lab with their version of the events, with the consequence that the lab leadership has access ONLY to the version by HR/ Legal Counsel. There is no legal need for this to happen. Not only is it disrespectful to the victim, but it benefits only the General Counsel itself, as it precludes any of their action to be put under check and balances by an actual third party. The lab would do well in learning about best practices now used in hospitals. Lawsuits by patients against hospitals have dramatically decreased since hospitals' lawyers keep a dialogue open with the victims. [one example among others is https://www.ncbi.nlm.nih.gov/pmc/articles/PMC1071103/]

<u>In short, this allegedly new concern reporting system does not work for employees. This is consistent with being a system that reports to the organization's general counsel, as described in Section II.F.2</u>. The system may have different outcomes when used for Users.

# Appendix H – A Textbook Case of Cover-up[7]

**We are going to describe here below a textbook case of elaborate cover-up. In the following is a summary of the story. More details of the events can be found in the documentation provided by the witnesses, shown later. In this Appendix the names of the witnesses are not shown for their own protection. However, for everybody else we decided to use actual first names and sometimes also last names for accountability. A reason is that any statement described here can be corroborated by a dozen witnesses willing to come forward in a court of law.**

Sometime in 2022, former Operation Maintenance Superintendent Greg at the Infrastructure Services Division (ISD), under Mark T. Jeffers as Senior Director, assigned Randy an unofficial position of Foreman for a large group of electricians and other employees of an ISD shop. Randy, Greg and Jeffers were all former Military personnel or contractors. A few months later, Randy was officially promoted to Foreman without any interview or posting of the position. He was known for threatening/ harassing people by whispering bloody and graphic scenes in their ears, and also claimed to carry a loaded gun at all times, either on his ankle or in his truck. Before the official promotion occurred, a majority of the team in the shop tried to prevent it. An electrician from this team brought Greg a petition signed by 12 employees (names available), over a total of 14 from the shop, asking to remove him from the position. After reading the petition, not only did Greg toss it back to the messenger, but he also broke confidentiality and informed Randy of that petition. Nothing changed when the original electrician brought forward a female electrician and former veteran, to another private meeting on 12/12/2022 with Greg so she could testify to all kinds of harassment from Randy and Dave, another employee from the shop. In this case too, Greg immediately informed Randy what was said in this confidential meeting before the two witnesses even left Building 38 where Greg's office was located. The witnesses were soon informed by colleagues that Randy was after them! At that point, the witnesses had no other recourse than going to HR and were told to leave the premises asap. Security and the Kane County Sheriff were called and they stopped Randy before he left the site. However, despite FNAL policies stating that Security has the right (and duty) to check a vehicle when on site, THEY DID NOT DO IT and let the guy leave.

The female witness had been repeatedly harassed also by peer electrician Dave. The latter became obsessed with her and in Spring of 2022 he backed a 25,000 pound bucket truck at her and another peer in the parking lot near Lab A. She and her peer were nearly crushed by Dave and they got out of this alive only because the peer was able to promptly back their van away from the truck coming at them! The female victim's electrical supervisor, DJ, was present and did not do anything to discipline the perpetrator.

---

[7] We are disclosing here information received directly by the witnesses themselves outside of working hours, and therefore these are not data gained within any work agreement. In addition, we believe in reporting violations of regulations, mismanagement, abuse of authority, substantial dangers to public safety, and in general any other violation under whistleblower protection.

With all these events happening, as noted, HR at some point became involved. The two witnesses met and cooperated with HR, and also provided detailed statements in writing (see also in the following). At this stage the first witness had a meeting with ISD Senior Director Mark Jeffers too. The electrician had the signed petition with him and handed it over to Jeffers. The latter looked at it, but would not take it. In January of 2023, the electrician was called to a meeting with former COO Scott Tingey, current lab Deputy Director Bonnie Fleming, current ISD Senior Director Mark Jeffers, the entire lab legal team, and former HR interim Senior Director Jennifer Gondorchin. While faced by this whole crowd, he was proffered some hollow statements, with no resulting disciplinary action whatsoever for Greg breaking the confidentiality of private meetings and putting lives in danger; and for Dave, despite his attempt at badly hurting others. Despite the presence of several witnesses, Dave did not receive any disciplinary action, and is still employed at the lab. For all of his misdeeds, Randy was given one week of unpaid leave of absence as only disciplinary measure. He is still employed at the lab and was further promoted. And Greg, Randy's abettor, was promoted too and retired only recently.

In fact, the above facts were not the first time that Randy had misbehaved. As a Union Steward for Local 701, the original witness kept all his confidential documents in a locked drawer in his office in Building 39. These documents included various types of votes for submission to the Union. In January of 2022, Randy entered his office one night, broke into his desk and stole all of his confidential documents. He then proceeded to transfer them to the Union. His accomplices at the Union took the documents instead of giving them back to their rightful owner and then fired the Union Steward from his position. He reported this theft to HR and to the ISD Head Mark Jeffers on March 2022, but no disciplinary action was taken even then.

Because of the total lack of protection and of safety for her own life, the female witness was forced to quit the lab sometime in 2023. Similarly, 6 other electricians from this toxic group under Al and Greg have left the lab between January 2022 and May 2023.

In October 2023 the original witness was tasked to take care of the electrical system for a new well pump that had been installed between the master substation and the fire station. A few days after performing that job, the electrician was driving by the well and saw the truck of mechanical lead Omar. He stopped and found the latter inside a hot electrical cabinet 480 Vac 400 Amps. Omar had completely undone his wiring. Mechanical foreman Mike was watching him undo his work while violating most electrical safety policies. The latter did nothing to stop him. The electrician reported the policy violations to his boss Frank and nothing was done.

At the same time, the electrician was also interviewing for the electrical foreman position with Greg and his electrical supervisor Frank in November 2023. Frank had repeatedly told the technician that he was his only choice for the foreman position.

On the Friday before Thanksgiving week, Frank called the electrician and requested him to postpone his vacation week to go see his family in Michigan for a very important job. The actual reason was instead to report to Greg's office to be informed that he had not been given the foreman job.

The electrician was fired when he sent out an upset phone text to his supervisor, mentioning Omar and Mike sabotaging the well, and threatening to report Greg for doing nothing about it. Greg saw the electrician's text by taking Frank's phone.

**The following more detailed information was also part of written documentation provided by the original witness to Heather A. Sidman at HR on Dec. 16, 2022.**

"**Events of 12/12/2022**
Months ago, Randy was unofficially given a Foreman position at our shop and paid as such. Like I said when I spoke with you, I have repeatedly gone to meet with Mr. Gilbert to discuss Randy being permanently promoted into the Electrical Foreman Position. During these discussions with Mr. Gilbert, one in particular that took place in early October, I brought Mr. Gilbert a signed petition of a dozen members of our electrical shop that are fearful of Randy continuing in this position and possibly being permanently assigned to this position. Despite me stating that I had a petition with a dozen electricians' signatures, Mr. Gilbert was unconcerned. He didn't ask me to see it. When I nevertheless handed it to him, he tossed the petition back at me once he read the names. He also never indicated he would be conducting any sort of investigation. <u>I felt Mr. Gilbert didn't take any of my accusations seriously.</u> Upon leaving that meeting that day at 3:30 pm, Randy was somehow informed that I had brought in a petition signed by a dozen of electricians who did not want him to be assigned a permanent position of shop foreman.

During this meeting with Mr. Gilbert, I stated that "Randy is an instigator, relishes promoting drama between fellow employees, tells violent stories of hurting people, and is not very stable." In this meeting with Mr. Gilbert, I explained that the electrical group is extremely concerned about Randy being given the foreman position. Mr. Gilbert did not ask me to go into more detail about our concerns or ask any more specifics about Randy. <u>To my knowledge, Mr. Gilbert has never subsequently interviewed anyone in our shop, nor did he remove Randy from the position of Foreman.</u> It is my belief, that Mr. Gilbert, almost immediately after I spoke with him, violated my confidentiality and informed Randy of our meeting. Roughly one week later I had another discussion with Mr. Gilbert to relate several examples of Randy's deception, lying about several facts, and creating anger in our shop. Again, Mr. Gilbert to my knowledge did nothing about our concerns.

Roughly one month later I was interviewing with Mr. Gilbert, Rick, and John for an open position in our shop. Immediately following the interview, John stated "interview is over, now I have some questions for you." John asked why I had attacked my fellow electrician Frank some weeks ago? When the event in question happened, I had gone to Mr. Gilbert's office that same day and explained this incident to him between Frank and myself. I explained that Randy was the catalyst of the whole misunderstanding between myself and Frank. From that day forward, Frank and I apologized to each other and realized that Randy was responsible for our misunderstanding. After relating the entire story to Mr. Gilbert weeks earlier, why would I have to sit in an interview for a job position and hear a false hostile story presented by John and be forced to repeat the incident when it was already discussed by Mr. Gilbert the day the incident happened. Mr. Gilbert said nothing about how I came to him immediately the day of this altercation, he sat there and said nothing, that forced me to defend myself and my character to a hostile John.

That character "attack" on me from John, which I believed came to them from Randy, forced me into a disagreement of how they viewed the event and caused me to explain we have terrified electricians in our shop, at which time I was FINALLY asked to give them a name. I said that our female colleague is terrified of Randy and Dave. Greg's response was "now I have to interview her." This reluctance on performing an objective investigation is consistent with the fact that Mr. Gilbert did not take the petition when it was originally presented to him which ironically contained the signature of our female colleague among others. I believe Mr. Gilbert and John had made their disdain for the electrical group clear and that <u>they wanted to put someone in a position of power to reinforce their intimidating, dishonest, and horrible culture regardless of the consequences.</u>

In the meeting on 12/12/22 between Greg, Rick, our female colleague and myself, <u>we mentioned to Mr. Gilbert that Randy also brags about having a loaded gun on Fermilab site. Following that statement, my female colleague and I felt Mr. Gilbert viewed our concern as rumor and conjecture rather than a bona fide safety issue</u>. I would like to know if he immediately notified HR of my female colleague and my concerns or did he say nothing and go to the Holiday party instead? I have never been as scared as when it was made clear to her and myself that Randy was aware of us meeting with Greg Gilbert and the accusations that were made. Someone once again has violated the confidentiality of myself, her, and our meetings with Mr. Gilbert. Despite our female colleague reporting to Mr. Gilbert that Randy approached her in the past and whispered in her ear something to the effect of "I'm a Marine I miss killing people," upper management has essentially placed both of our lives and our families lives in a horrible situation. Both myself and my female colleague are in fear for our lives because lab security and the Kane County sheriff's office failed to search his truck, essentially allowing a person who makes racist and violent statements on a regular basis to fellow employees exit the site with the possibility of having a loaded weapon. <u>I now sit at my house on administrative leave with my 21-year-old daughter scared to death knowing he still has that gun and knows exactly who spoke up about him.</u>
This process was filled with leaks and lies and took everything I had left as a concerned employee to keep pressing the serious nature of Randy being promoted. Our "shop culture" is such that if you bring any stated concerns to management and they're deemed inconvenient or seem unbelievable, management tends to sweep them under the rug while simultaneously informing the accused of their accusers resulting in either their careers or personal safety jeopardized.

Chauncey from HR brought up "culture" and why did we wait so long to bring this to the attention of our supervisors? Clearly, this is not the case. I personally, along with several others in our shop, have repeatedly brought the issue of Randy as a dangerous individual to the attention of Greg and John over a 4-month period. <u>Given management's lack of response to any of the previous concerns we did not feel confident in reporting what Randy said he was carrying, fearing that the lab would tell Randy and allow him to get offsite with the weapon.</u> WHICH IN FACT IS WHAT EXACTLY HAPPENED!!! <u>In this day an age of mass shootings I would never have anticipated such a lackluster response to the accusations of an employee possessing a weapon on site!</u> To be clear, Randy is a very capable electrician but unfortunately, he is completely devoid of any ability to interact with his peers in a respectful manner and routinely engages in incendiary racist and other inacceptable rhetoric."

**Some additional detailed information, which was also part of written documentation provided by the female witness to Heather A. Sidman in December 2022, is presented below.**

"**Randy**
First day that I showed up to Fermilab, I met Randy, we talked about how we were both in the military and both in Afghanistan (me as a Marine and him as a contractor), and he explained to me about how he hated all Muslims and that as far as he is concerned, "they should drop a bomb on the entire middle east".

The first time I worked with Randy we were wiring an outlet. He asked me if I knew about the color coordination in relation to the outlets, after he told me I said, "hmm, did not know that". After working with him, he went to DJ's office and told him that I was too dumb to work there and that the only reason I got the job was because I was a woman and I was Latina. He complained about this also at meetings with my peers present and told them that I only got the job because I knew DJ (not sure what he was implying, if it was that we had a relationship or if it was because DJ, others and I did come from Caterpillar). He said I wasn't journeyman material so I shouldn't have ever gotten hired.

One day that I was working with Randy, he asked me what we would say in one of our drill moves (I think it was either a left face or right face) in the military because when he was in they would say "whoosh" which is the sound that would come out of their vaginas (can't remember if he said vagina or pussies) when they would close their legs.

One of the first times I worked with Randy, he described to me a story of a time when Doug and himself had gotten into an argument (this took place before I started at Fermilab). I had been aware of the story and Doug said that it was because Randy had said some nasty things about some of the women that had just made it into office along with AOC (Alexandria Ocasio Cortez). Others were present at the time because it was at the morning meeting. Randy's version was that he didn't say anything wrong and that Doug flipped out on him without knowing what he really meant. Then he continued on with his story and said something about stabbing Doug in the neck with a pen if he ever tried anything… He said "I got a ball point pen and if he ever tries anything…. I'll stab him in the neck and watch him bleed out".

In passing one day I was in my cubicle, he approached and said something along the lines of "I miss killing people". I just looked at him and looked away.

He called our female maintenance planner a "stupid bitch" and a "dumb cunt." "

**The following is a testimonial from a former female Maintenance Planner/Scheduler at ISD. This is a widow who raised many successful children as a single mother.**

"Randy was described to me by most as a ticking time bomb and that he was ready to go off on people. In particular, he had repeatedly expressed hate for Rob and was on the verge of wanting to hurt him. It didn't take long for me to be faced with the same. Once he told me a joke about an old woman being raped. He called people foul names and called me foul names too to my face and he didn't care who else was around. He called his own daughters bitches, as in: "I have two daughters and they can do so much better than you, those bitches… Oh, look they can't believe I'm calling my daughters bitches, well they are, women are bitches by the very nature that they are women".

The most horrific stories that he told us were about prostitutes from his time in the Marine Corps. One story that he was telling everybody was about buying a cheap watch in the Middle East and passing it as an authentic brand and then using it to pay for a prostitute so that he short-changed her. <u>Another story from when he was in the military was about going out one night in the Philippines to look for a prostitute with some military buddies. They apparently found one, but could not stand that she was transgender and, according to him, they all killed her!!!</u>

Also, we all knew that he went shooting with Dave and that he kept a gun with him at all times, either in his truck or on his ankle.

As a former maintenance manager for 25 years at a pharmaceuticals corporation first, and at a medical innovative engineering company next, I had never seen such obsolete equipment, and such lack of correct procedures. When I was hired in 2022, I was not given a working computer for 4 months! The ISD maintenance management system was incredibly antiquated. When I was there, they were replacing it with an allegedly more modern one, which did not work either! The equipment is in horrible shape also because when a work order wasn't done, they would hit "delete" and get it next year. WOW ! It's a horrible way to operate. Not effective at all. That's the norm at the lab. Electrical and mechanical systems were down, fire alarms were down, the equipment was in shamble. Because they were not OSHA compliant, I have seen so many of their electricians operating in totally unsafe conditions. I was told that OSHA is not needed there as they are the Federal government and it doesn't apply! Even some parts of the tunnel were collapsing. It is the worst facility equipment I've ever seen in my 25 year career managing facilities. The organization is not federal, they are just funded by the government and they have to be OSHA compliant. The fact that it is not, makes it incredibly dangerous. Tax payers money to fund such a fraudulent place. No accountability for anything.

In the entire campus the leadership is nonexistent. A bunch of words to make it sound good but as we know it's the actions that matter, and Fermilab is negligent in every way. They just cover it up and keep sociopaths employed. It's the most dangerous and disgraceful place I've ever experienced. For instance, in June 2022 there was a nearly fatal accident that happened to a diver from a Chicago diving company who nearly died in a Fermilab pond. The diver had been hired to check out the water pump in the pump, and ISD had not turned that pump off! Why? Because they did not want to interfere with the fire system! It is unheard of to check or repair an OPERATING pump, especially if it is an underwater one. Fire systems are put on fire test for this kind of repairs. There was no written procedure, no safety checks. Total negligence. Also in this case an "investigation" was allegedly performed but some ISD managers are still laughing about the whole thing.

Fermilab's HR ignored the harassment by Randy, Scott, Lee and my manager Jason. Most people witnessed my harassment. But in the so-called "investigation" they did not talk to any of my peers who had witnessed all these harassing events. They just talked to managers and others who were part of the verbal assault and harassment.

I walked into a very dangerous situation, and I will not ever tolerate being bullied and talked down to with vulgar language. I decided to speak up and stand up for myself, other women and everyone in that Lab. After I did that, I was let go with the following argument: "We don't want to see any woman treated like that." Two women from ISD who contacted the Fermilab Hotline [Editor's note: see "Statement of Fermilab Community Standards" at the bottom of this Appendix] on my behalf were displaced from their jobs at ISD.

Unfortunately Fermilab keeps the sociopaths and they lose great people. Most of the Electricians have left that facility. Al condescendingly asked me "Who do you think you are?" My response was "Who do you think you are?" and I walked out."

**We are now going to provide some additional detailed information by the female electrician in reference to the troublesome man who persecuted her. This was also part of written documentation provided by the female witness to Heather A. Sidman in December 2022.**

**"Dave's incidents**
We had worked together at Caterpillar, he was going through a divorce at the time, a divorce in which his wife was cheating on him with a guy they knew. She and the lover were both teachers in the local area. Dave would complain about the situation with people at the Fermilab Electrical Shop, and he would show me naked pictures of his ex-wife. I did not ask to be shown. Many of the other people who were shown expressed feeling uncomfortable being shown her nude pictures.

One day I had to work with Dave on the main ring, he walked about 12-15 ft away from me, on the other side of his truck. He started talking to me about how someone peed on his equipment right before he had to clean it. Without turning around, I asked him, "are you telling me that because that is what you're doing right now?" he said "yeah, I told you I was going to", I said "no you didn't, and if I knew that I would have walked away" and when he came back to work on the AC unit with me, I told him, "and now you're touching my tools?" he said, "I used my other hand".

There was an incident where I did express some firm resistance against Dave when he was questioning why my husband got to drive a nice car when mine is not. I firmly expressed to him, to shut his questioning down that it was what we wanted and that he deserved it. He apologized a day later saying that he just meant that the car my husband has was one that he was led to believe was for older people.

One Saturday I volunteered to work from 6am to 2:30pm to work on lighting in Wilson Hall (I believe it was the 4th and 5th floor, finishing the fifth and moving on to the 4th). I was assigned to work with Rey. Our work orders only showed we were to work on the 5th floor, I didn't even have the work order number for the 4th floor in my queue. Dave showed up sometime after his shift which ended at 8am, he was to work from 12am to 8am, and then 4hrs of OT from 8am to 12pm. The fifth floor required a sign in on the Hazardous Analysis form. When we first began, we had a hard time finding it, we were just going to stay in the sides that didn't require a signature. Pretty shortly after deciding that, Rey found where it was. We signed it and continued working in that area, one person on the ladder and one handing the bulbs. Dave showed up sometime after 8 and closer to 9 on the fifth floor. He asked for the HA form and I told him that we didn't find it. He walked around for a bit (in this time Rey and I were just working and not paying too much attention to him) and said that he found it, as soon as he started to sign it I told him that we did find it but it was on the wall (because the one he was trying to sign was for the contractors). I went over to show him where, and while I was near him I smelled alcohol, and I said, "were you drinking last night?", he said maybe it was the hand sanitizer, and I said ok. I had told him twice, "we are almost done here, you can move to the 4th floor and we will pick up there after lunch". I said that because Dave hated Rey, and other weekends when the three of us were put to work together it was very awkward for me because I was in the middle with, "tell him this tell him that, what's he doing, why is he working so slow" from Dave. Dave agreed to go to the 4th floor, he took one case of bulbs that Rey and myself had brought up for the job. After he left, we realized there was a conference room with no windows that needed relamping and one other room that needed to be finished before we could leave for lunch. Rey and I had left at about 10:20 am for lunch. First we got rid of the used bulbs. Afterwards we went to site 39, Frankie was on shift and was scheduled to be the duty electrician from 08:00 to 16:00. We talked to Frankie till about 10:40, left to pick up lunch, and in that time Dave called, but the reception was lost, I tried calling him back but he didn't answer. Rey and I went to lunch as normal, when we came back Frankie told us that Dave asked where I was and Frankie told him something along the lines of, "she's not with you? She must be with him" and that Dave reacted as if he was upset and walked away.

Sunday when there was a changeover (Josh was scheduled to work on call from 4pm to midnight) between Josh and Dave, Dave went on to raise his voice, complaining and telling Josh that I had lied to him, that I had deceived him and that I was a liar and that I couldn't be trusted while using the phrases "how can I work with her if I can't trust her." "She is a liar". Josh called me early in the morning before I was even awake and said that I needed to talk to Dave, and that he is flipping out etc. I tried all weekend to call him to apologize and to explain what my side of the events were, he ignored my calls on his personal phone and his work phone, till Monday morning I messaged him and asked if we could talk. He asked me to give him a few minutes to use the bathroom (this was about 7:20am before the morning meeting). When we finally talked he explained to me that he still felt that I purposely deceived him when I told him that I didn't know where the HA form was and that I didn't answer his phone call (when the reception was lost), and that I tried acting as if I was his boss by telling him that he could go to the 4th floor and that it seemed like I was trying to get rid of him. I still apologized and told him that a lie is something told with the intention of deceit, and he said a lie is a lie and that it didn't matter. Frankie and Randy were with Dave, and I was told that when I was on the phone with him apologizing, Dave tried making a scene by standing up and raising his voice very loudly, and that the impression he was giving them was that "he was trying to put me in my place to establish his dominance over me". He also said, "she doesn't even know me, if she knew me she'd know I don't drink". How it was explained to me was that he got very irritable when telling his story to Josh and Frankie even after they had tried to explain to him that it was a misunderstanding. He also went to DJ, our former electrical supervisor, about this situation. He showed up to the meeting in sunglasses and crossed his arms and didn't move or speak.

As the days went by, he continued to badmouth me to other coworkers. While explaining to Frankie about how upset he was he had tears in his eyes (I was told by Frankie that there were visible tears in his eyes) and explained that he was crying because he was so worked up over the whole situation. Dave talked to the coworkers for months after this about how I lied to him and how I deceived him. The guys at the shop explained to me that maybe because he was cheated on by his wife and that because I am a woman, that maybe he sees me like his wife. He had been heard joking about me being his work wife.

Some time in April 2022, shortly after that we had worked overtime again on a Saturday, but this was for a shut down, Rey and I had worked together, not by choice, but DJ had paired us together. We were doing high voltage cleaning. I left for lunch before Rey did, and in that time Rey had found something that needed to be replaced. DJ had Dave drive the bucket truck out to the site after lunch. After it was parked in the vicinity, Rey drove me out there in the work van and pulled up behind the bucket truck (and at the time I did not have a work van because mine was under repair). <u>Dave started to drive in reverse, slowly at first and when Rey started to go in reverse as well, then Dave "gunned it" and accelerated.</u> I was yelling and getting panicky at Rey telling him to hurry up. I am not sure how he did it, but Rey was able to back our van away before the truck could hit us. At some point Dave slammed the brakes and got out of the truck, and DJ told him "what's wrong buddy, need a hug?" and tried to joke about it and Dave yelled about how we weren't moving fast enough. Eric was there and said that he told him that he can't do that and explained the dangerousness of it. Dave then sat on the edge of the truck and crossed his arms with his sunglasses on and didn't respond to people for the rest of the day, as if he was the victim. Eric said that afterwards he tried talking to DJ about how he should take some action on that and also explaining how dangerous that is. However, my supervisor DJ did nothing about it."

"I can still remember the dread and anxiety that I felt every day. Despite those terrible events, not only nothing was done, but on the contrary JP and Greg went out of their way to pair people who didn't get along together. I once heard them telling each other how funny that was. Tom, Frankie and Rick instead were really good at keeping Dave away from me. Still, the latter had told me of times when he had broken people's windshields in road rage incidents. He said that the police were often called. I told HR that this kind of violence did not help my concerns. Clearly, that meant nothing to them."

**In the story above, one can tell how the guy's sickness escalates slowly and relentlessly. At the end it reads like a horror story. Some thoughtful additional words to HR from the colleague who had tried to help the victim are in the following.**

"<u>I now want to address who are we supposed to trust at the Lab?</u> How many times has HR come down to our shop and pulled us aside, or just made their presence known in person and said, "HR is here for you" or asked any single one of us "how are things going?" Never. I have never met anyone from HR in our shop nor have I ever seen your presence in our shop, not once in almost 5 years. That alone has left us to feel isolated. We all know we can go to HR anytime, but what about HR reaching out to us to see how things are going? I believe this is a two-way street and that would have helped to build trust between HR and our groups. <u>Maybe occasionally someone should pull aside a female employee in a shop filled with men and ask her "is everything ok?"</u> I did this repeatedly and finally that is how our female colleague was finally able to tell me the truth. <u>I have 4 daughters and want them to exist in a world that doesn't push them around or put their work qualifications down and intimidate them with horrible whispers.</u> Men and any employee for that matter should not be allowed to take "possession" of them at work like Dave did to our female colleague.

He was allowed to harass her and when she pushed back, he called her a liar and a deceiver! So, my point here is, we need to know we are safe at our place of employment. We need to know we are safe when we come forward with an uncomfortable situation. <u>Our management has failed us in that regard and not created a culture of trust but one of fear for coming forward and speaking out!</u> Our good old boys club of managers in our shop make us feel guilty, scared, and apprehensive to ever want to come forward and have dialogue for fear of being implicated as a snitch.

<u>It is my serious hope that this investigation will lead to a positive change in our culture, demonstrated by our HR Department and how you handle this situation. None of us deserve to be unsafe at work</u>, especially when as electricians we are working with the risk of injury or death from a mistake because our minds are filled with fear and worry about our job security, personal security, and we are not able to focus 100% attention on our work tasks!"

**As we know by now, the lab and FRA are not interested in positive change and instead decided to cover everything up. How? Let's start with the gun holder. The official narrative by the lab was that there was just one lone paranoid guy complaining about Randy carrying a gun. They claim that an investigation was performed and successfully concluded. They call it successful because a consequence of these horrible events is a policy change on inspecting vehicles on site, i.e. now security does not have to inspect any vehicle, they just call law enforcement to do it for them (clearly without a warrant required).**

**For the above cover-up to be successful, and specifically to be able to claim that there was just one lone wacko complainer, EVERYBODY in the chain of command and at HR had to pretend that there was no petition signed by 12 people who knew Randy well. This is why the line managers were so terrified to even just touch that petition and why the HR representatives NEVER asked for its signatories. They knew that if they had, several of those signatories would have confirmed Randy's threats of carrying a gun.**

**For the cover-up to be successful, the lab also had to somehow twist the lame truth that they had not been able to inspect the vehicle. In short, they let that gun slip through their fingers. And exactly as in movies with corrupt cops covering up their own misdeeds, they decided to blame the witness.**

**Everybody knows that when masquerading the truth with lies, another will be needed, and then another, etc. with no end. One of these lies is that they had to convince themselves that there was no gun in that car.**

**Perhaps the most dire consequence of the previous fabrication is that they had to give the benefit of the doubt to the bad guy. All these deceptions completely clouded the standard approach for safety of using a cost-risk-benefits analysis. So the lab decided that it was safer to believe a troublesome individual who had**

been very likely carrying a gun as he had always claimed to everybody, rather than getting rid of him.

And eventually, once the loose cannon was not just kept on the job, but further promoted, what other choice was there other than cutting any loose ends? Which was done by firing the good guy some time in 2023 under some behavioral pretext.

**Compare all of the above with the "Statement of Fermilab Community Standards" at the bottom of this Appendix, and talk about leaders leading by example, or at least implementing their organization's advertised policies!**

Unless… in the same way that Security was justified in not performing their job because, poor them, they are not armed, some in upper management and at HR were scared of the bad guys. After all, what if they went after them with the gun??? That would be much different than the bad guys going after two diligent employees, wouldn't it? If this were the case, however, it would mean that the lab has become a street jungle where the strongest street gang wins. We pray that this is not the case.

Similarly, Randy's own claim to several employees of killing a transgender prostitute abroad has gone unchecked! By closing an eye on a possible murder by assuming that it cannot be true, the lab is violating any conceivable safety policy, to say the least.

And what about DOE? How much did DOE know about the gun on site? We interviewed the lab's FSO on this subject and it was clear that he had at least bought the official narrative.

Finally, how did the lab cover up what happened to the female electrician? This was an easy case (for the lab). In addition to the usual avoiding like the plague any witness' interview on the bucket truck incident when it happened, in December 2022 they told the victim that she had reported the April 2022 events too late! (The assault by the obsessed electrician was at the very least a misdemeanor, if not a felony, and the statute of limitations is at least 18 months.) Then they just waited it out for her to feel completely overwhelmed and leave. Easy peasy with all that harassment going on all the time.

**In conclusion, one wonders how this can be called a "culture of safety" and how many know about this. What is certain is that all of this happened under FRA's purview. FRA is controlled by the University of Chicago. We hope that the DOE will consider what we presented in this report to restore Fermilab as a true leader in HEP science in the US and the world.**

## Statement of Fermilab Community Standards

Fermilab's status as a world-class, scientific-research destination and its reputation as an institution of choice requires a community that promotes professionalism, mutual respect, inclusion and that is free from discrimination. Fermilab welcomes a diverse populace with varying backgrounds and experiences including international members. Fermilab expects that all community members commit to the same high standards of ethics and behavior.

**Intention**

The purpose of this Statement of Community Standards is to establish and communicate the set of expectations that all members of the Fermilab community shall follow. Disruptive or harassing behavior shall not be tolerated regardless of race, color, religion, disability, age, gender, veteran status, sexual orientation, gender identity, and/or nationality.

**Applicability**

This Statement applies to all members of the Fermilab community: employees, users, subcontractors, visitors, and guests.

These expectations apply for all interactions including but not limited to: • on the Fermilab site, including the Village; • at Fermilab installations or Fermilab-supported events that are off-site, including Sanford Lab in South Dakota, CERN, and off-site conferences; • online within the Fermilab domain; and • when representing or associated with Fermilab off-site or online, including in-person interactions and on social media.

**Responsibility**

The responsibility for proactively facilitating a safe and welcoming environment -- including voicing concerns -- rests with each individual member of the Fermilab community. Leading by example, line management, beginning with the lab director, shall hold themselves and their teams accountable.

The following basic principles apply to all members of the Fermilab community:

1. Build trust and credibility

Fermilab's success depends on the trust and confidence we develop with one another and with our stakeholders and partners including the U.S. Department of Energy, the American public, 1 and the global scientific community. As we engage in business on behalf of the Laboratory, we each are ambassadors of Fermilab. Fermilab's reputation is determined by our words and our actions. We gain credibility by fulfilling on our commitments and, when appropriate, acknowledging when we will not and taking responsibility for the consequences.

2. Communicate openly and honestly

Every member of the Fermilab community is welcome, and each encouraged to communicate their ideas and/or concerns. We seek to promote open communication which emphasizes listening, creativity, growth and development, collaboration, and inclusivity. This culture encourages respectful discussion and debate for mutual benefit.

3. Respect one another

Members of the Fermilab community shall be able to work in a safe and welcoming environment where everyone is treated with dignity and respect. Peer-to-peer or supervisory relationships that interfere with an individual's research or work performance, limit access to educational experiences and career opportunities, or adversely impact an individual's well-being are unacceptable.

Mutual respect is developed by: • Valuing others and their points of view including those who don't easily find their voice. • Being open to correction. • Recognizing the skills and expertise of others in the Fermilab community. • Sharing the credit. • Being courteous. • Welcoming new perspectives and ideas. • Continuing to grow and develop and helping others to grow and develop.

**How to report concerns**

All members of the Fermilab community are responsible for reporting concerns of improper conduct; harassment or discrimination; unsafe conditions; conflicts of interest; fraud; waste or abuse; violations of law or policy; or other irregularities. ethical standards, contractual requirements or laws. You are encouraged to raise concerns about any violation of this Statement of Community Standards. Confidentiality shall be maintained and reporting should proceed without fear of harassment or retribution.

Contact either of the following:
- Managers or Directorate/Division/Department Leadership
- HR partners or Human Resources Leadership
- The Office of the General Counsel
- The Fermilab Security Department
- The Fermilab Concerns Reporting System, a third-party provided hotline/website hosted by Integrity Counts where reporters may either self-identify or remain anonymous. https://app.integritycounts.ca/org/fermilab
  Phone Hotline: 866.921.6714 (USA), 00-800-2002-0033 (Switzerland).

**FRA policy prohibits retaliation against individuals who make a report through any of these channels.**

Fermilab leadership is committed to addressing all incidents promptly and thoroughly. Violations can result in disciplinary action up to and/or including termination of employment, contract and/or site-access privileges.